\documentclass[%
 reprint,
superscriptaddress,
nofootinbib,
nobibnotes,
bibnotes,
 amsmath,amssymb,
prd,
]{revtex4-2}
\usepackage{hyperref}
\usepackage{graphicx}
\usepackage{dcolumn}
\usepackage{bm}
\usepackage{color}
\usepackage{xcolor}
\usepackage{xspace}
\usepackage{float}
\usepackage[caption=false]{subfig} 

\newcommand{\hn}[0]{\ensuremath{\hat n}}
\newcommand{\lcdm}{$\Lambda$CDM\xspace}
\newcommand{\nlth}{\ensuremath{N_L^{(3/2)}}\xspace}

\newcommand{\be}{\begin{equation}}
\newcommand{\ee}{\end{equation}}
\newcommand{\bea}{\begin{eqnarray}}
\newcommand{\eea}{\end{eqnarray}}

\newcommand{\mnras}{MNRAS}
\newcommand{\aap}{A\&A}
\newcommand{\physrep}{Physics Reports}
\newcommand{\jcap}{JCAP}

\newcommand{\Nth}{$N_L^{(3/2)}\ $}
\newcommand{\NthXY}{\ensuremath{N_L^{\rm{(3/2)}}}\xspace}

\newcommand{\delensalot}{\texttt{delensalot}}
\newcommand{\lenspyx}{\texttt{lenspyx}}

\newcommand{\plancklens}{\texttt{plancklens}}

\newcommand{\SB}[1]{{\color{olive}{SB: #1}}}

\begin{document}

\preprint{}

\title{Non-Gaussian deflections in iterative optimal CMB lensing reconstruction}

\newcommand{\Geneve}{Universit\'e de Gen\`eve, D\'epartement de Physique Th\'eorique et CAP, 24 Quai Ansermet, CH-1211 Gen\`eve 4, Switzerland}
\newcommand{\Cardiff}{{School of Physics and Astronomy, Cardiff University, The Parade, Cardiff, Wales CF24 3AA, United Kingdom}}
\newcommand{\ICTP}{Instituto de F\'isica Te\'orica da Universidade Estadual Paulista and ICTP South American Institute for Fundamental Research,
R. Dr. Bento Teobaldo Ferraz, 271, Bloco II, Barra-Funda - S\~ao Paulo/SP, Brasil}
\newcommand{\CCA}{Center for Computational Astrophysics, Flatiron Institute, 162 Fifth Avenue, New York, NY, 10010, USA}
\newcommand{\KICC}{Kavli Institute for Cosmology Cambridge, Madingley Road, Cambridge CB3 0HA, UK}
\newcommand{\IOA}{Institute of Astronomy, Madingley Road, Cambridge CB3 0HA, UK}
\author{Omar Darwish}
\email{omar.darwish@unige.ch}

\affiliation{\Geneve}

\author{Sebastian Belkner}
\email{sebastian.belkner@unige.ch}
\affiliation{\Geneve}

\author{Louis Legrand}
\affiliation{\Geneve}
\affiliation{\ICTP}

\author{Julien Carron}
\affiliation{\Geneve}

\author{Giulio Fabbian}
\affiliation{\IOA}
\affiliation{\KICC}
\affiliation{\Cardiff}
\affiliation{\CCA}

{}

%

\date{\today}

\begin{abstract}
The gravitational lensing signal from the Cosmic Microwave Background is highly valuable to constrain the growth of the structures in the Universe in a clean and robust manner over a wide range of redshifts.
One of the theoretical systematics for lensing reconstruction is the impact of the lensing field non-Gaussianities on its estimators. 
Non-linear matter clustering and post-Born lensing corrections are known to bias standard quadratic estimators to some extent, most significantly so in temperature.
In this work, we explore the impact of non-Gaussian deflections on Maximum a Posteriori lensing estimators, which, in contrast to quadratic estimators, are able to provide optimal measurements of the lensing field. 
We show that these naturally reduce the induced non-Gaussian bias and lead to unbiased cosmological constraints in $\Lambda$CDM at CMB-S4 noise levels without the need for explicit modelling. 
We also test the impact of assuming a non-Gaussian prior for the reconstruction; this mitigates the effect further slightly, but generally has little impact on the quality of the reconstruction.
This shows that higher-order statistics of the lensing deflections are not expected to present a major challenge for optimal CMB lensing reconstruction in the foreseeable future.

\end{abstract}

\maketitle


\section{\label{sec:level1} Introduction}
Gravitational lensing of the cosmic microwave background (CMB) is one of the leading cosmological probes of the next generation of CMB polarization surveys such as Simons Observatory (SO) \cite{so} and CMB-S4 \cite{cmbs4}. Maps of the CMB lensing potential and its summary statistics can provide clean and robust probes of the large-scale structure (LSS) in the Universe and allow to constrain cosmological parameters that govern the LSS growth and to which CMB anisotropies alone are weakly sensitive (for example neutrino mass, dark energy properties, curvature) \cite{lewis2006,2009PhRvD..79f5033D,1999MNRAS.302..735S,Sailer_2021}. On the other hand, galaxy surveys in different wavelengths probing the same LSS distribution (through galaxy clustering, weak lensing or intensity mapping) are also expected to extract complementary cosmological information on the LSS growth, or on primordial non-Gaussianities. Through galaxy-CMB lensing cross-correlations we will be able to marginalize over some observational systematics (e.g. shear multiplicative bias) \cite{vallinotto2012,2017PhRvD..95l3512S,2020PhRvD.101f3509C}, or modeling uncertainties (e.g. galaxy bias, magnification bias) \cite{vallinotto2013,roman,euclid}. Masses of high-redshift galaxy cluster samples can also be calibrated with the highest sensitivity through stacking of CMB lensing maps at location of the clusters. and later used to probe cosmology through their mass function \cite{lewis-king,hu-dedeo,2017JCAP...08..030R}. Additionally, CMB lensing maps can also be used to predict and subtract the lensing-generated $B$-mode signal of CMB polarization to enhance constraints on inflationary physics achievable through precise measurements of the primordial $B$-mode signal on large angular scales \cite{2004PhRvD..69d3005S,smith2012} \\*
Given the importance and diversity of the science case connected to CMB lensing, it is crucial to understand all the properties and shortcomings of statistical estimators employed to reconstruct maps of the CMB lensing potential from the observed maps of the CMB anisotropies. The most commonly employed technique used for this purpose is the so-called quadratic estimator (QE) \cite{hu-okamoto}. This uses the breaking of statistical isotropy introduced by the projected matter potential field into the observed CMB to reconstruct the lensing modes through weighted couplings between pairs of harmonic modes of the observed CMB itself. Such technique has been used to derive the most sensitive measurements in the field so far from the Planck satellite data or ground-based CMB polarization experiments such as ACTpol, SPTpol, Polarbear and BICEP \cite{plancklensing2018,plancklensingdr4,actlensingdr6,sptpollensing,pblensing2020,biceplensing, Sailer:2022jwt}. 

Extensive effort has been recently carried out to evaluate the sensitivity to the QE to noise anisotropies, instrumental systematics, galactic and extragalactic foregrounds, and suitable modification have been proposed to minimize their impact in the final reconstructed lensing map and power spectrum for future experiments \cite{Maniyar:2021msb,Carron:2022edh,Beck:2020dhe,2019PhRvL.122r1301S,mirmelstein2021,Sailer:2022jwt,darwish2023}. But for very deep surveys such as CMB-S4, the QE is expected to be suboptimal since it only accounts for the lensing coupling at linear order. Several additional methods that account for the full lensing information have been proposed. These advocate either sampling techniques \cite{Millea:2017fyd,Millea:2020cpw,Millea:2021had}, iterative maximum likelihood \cite{Hirata:2003ka} or Maximum a-Posteriori (MAP) estimates \cite{maplensing} to reach the lowest possible reconstruction noise for future measurements of the lensing potential and of its power spectrum. MAP estimators in particular offer the advantage of a reduced computational cost compared to sampling or iterative spectrum reconstruction methods and have so far been employed successfully on data covering large sky fraction. Very deep polarization observations of Polarbear were used to validate for the first time the performances of MAP estimates for joint lensing reconstruction and delensing analyses \cite{pbdelens}. Similar results have then been achieved by sampling methods on the SPTpol data \cite{millea2021}. Recent work also showed how to achieve unbiased measurements of the CMB lensing spectrum from MAP estimates accounting for residual noise, normalization biases and mean field effects induced by noise anisotropies and masking, even in presence of mischaracterization of the data statistical properties \cite{legrand2022,legrand2023}, so that it is now possible to start investigating biases induced by foregrounds and theoretical assumptions employed in the design of this new class of lensing estimator. \\*
In this work we focus our attention in particular on the assumption of the Gaussianity of the CMB lensing field for iterative MAP CMB lensing reconstruction. As shown in previous work of \cite{bohmBiasCMBLensing2016,beckLensingReconstructionPostBorn2018a,bohmEffectNonGaussianLensing2018}, non-Gaussian effects induced by the non-linear evolution of the LSS and post-Born lensing corrections due to multiple photon deflections \cite{pratten2016} can bias the QE reconstruction (\Nth bias) and in turn affect the constraints achievable on cosmological parameters, in particular on the total mass of neutrinos \cite{beckLensingReconstructionPostBorn2018a}. Such bias becomes more important when correlating CMB lensing with external LSS tracers \cite{fabbianCMBLensingReconstruction2019a}.

In Sec. \ref{sec:theory} we review the properties of MAP estimator and $\nlth$ for the QE. In Sec.~\ref{sec:results} we measure the $\nlth$ bias in the MAP estimator, comparing it to the QE results and assess its importance for future surveys. In Sec.~\ref{sec:params} we evaluate the impact $\nlth$ on cosmological parameters and cross-correlation science and propose mitigation strategies. \\*
In the following we use the standard convention and denote CMB lensing multipoles $(L, M)$ and CMB multipoles with $(\ell, m)$. We will also mention in the discussion the CMB lensing convergence $\kappa$, defined in function of the CMB lensing potential $\phi$ in real and harmonic domains as 

\bea
\kappa(\hn) &=& -\frac{\nabla^2}{2}\phi(\hn)\\
\kappa_{LM} &=&  \frac 12 L(L + 1)\phi_{LM} \label{eq:kappa_phi_rels}
\eea

\section{Lensing reconstruction and non-Gaussian deflections}  \label{sec:theory}
\newcommand{\va}[0]{\ensuremath{\boldsymbol{\alpha}}}
\newcommand{\Da}[0]{\ensuremath{\mathcal D_{\va}}}
\newcommand{\Beam}[0]{\ensuremath{\mathcal B}}

In this section, we provide background information on the lensing reconstruction methods we employ and offer a brief recap of the origin of the non-Gaussianity induced bias in the CMB lensing estimates.

First, we summarize the Maximum A Posteriori (MAP) CMB lensing estimator. 
For more details, we refer the reader to \citep{carronMaximumPosterioriCMB2017, Belkner:2023duz} for the specific implementation we use.

\subsection{Lensing reconstruction from the MAP estimator}
To obtain the MAP estimator \citep{hirataAnalyzingWeakLensing2003, hirataReconstructionLensingCosmic2003, carronMaximumPosterioriCMB2017} we will start by modelling the observed CMB as

\begin{equation}
    X^{\rm dat} = \Beam \Da X + n,
\end{equation}

where \Da~is the lensing operator mapping the primordial CMB $X$ into the lensed CMB through the deflection $\va$ \citep{hirataAnalyzingWeakLensing2003}, $\Beam$ a linear response matrix that includes the beam, and $n$ the noise, that we assume to be uncorrelated with the CMB; in this work we will ignore foregrounds, and consider only experimental noise. 

The goal of the MAP estimator is to derive a maximum a posterior estimate of the CMB lensing field, given the data, by maximising the posterior

\begin{equation}
    p(\va|X^{\rm{dat}}) \propto p(X^{\rm{dat}}|\va)p_{\va}(\va)\ ,
\end{equation}

given the likelihood $p(X^{\rm{dat}}|\va)$ and a prior $p_{\va}(\va)$. For simplicity, due to the monotony of the natural logarithm, from now own will take the natural logarithm of probabilities, and so maximing it will be equivalent to maximizing the original quantities.

We will assume a Gaussian log-likelihood for the data \cite{carronMaximumPosterioriCMB2017}

\begin{equation}
    \ln p(X^{\mathrm{dat}}|\va) = -\frac{1}{2}X^{\rm{dat}} \cdot \mathrm{Cov}_{\va}^{-1} X^{\rm{dat}} - \frac{1}{2}\det \mathrm{Cov}_{\va}+ \rm const
\end{equation}

with a covariance

\begin{equation}
    \mathrm{Cov}_{\va} = \langle X^{\mathrm{dat}}X^{\mathrm{dat},\dagger}\rangle = \Beam \Da C^{\mathrm{unl}}\Da^{\dagger}\Beam^{\dagger} + N \ ,
\end{equation}

where $C^{\rm unl}$ and $N$ are the covariance of the unlensed CMB and that of the noise respectively, and we explicitly specify that the covariance has a structure that depends on the realization of the deflection angle \va.

For now we will ignore the curl of the deflection field \va, and we consider only its gradient mode. We can parametrize with $\kappa$, such that the final log-posterior for $\kappa$ can be written as

\begin{equation} 
\begin{split}
-2 \ln p(\kappa|X^{\mathrm{dat}}) =  X^{\rm{dat}} \cdot \mathrm{Cov}_{\va}^{-1} X^{\rm{dat}} + \det \rm{Cov}_{\va} \\
 +\sum_{LM} \frac{\kappa_{LM}\kappa^\dagger_{LM}}{C_{L}^{\kappa\kappa}}
\end{split} \label{eq:posterior}
\end{equation}

up to irrelevant constants.

The CMB lensing estimator is then derived by requiring the gradient of the posterior, Eq.~\eqref{eq:posterior}, to vanish. To solve for this, we will follow \citep{carronMaximumPosterioriCMB2017} and employ an iterative scheme starting from the standard quadratic estimator to finally obtain an estimate of the lensing potential field.

In particular, the calculation of the gradient of the likelihood gives a term that is quadratic in the data, which we will refer to as $g^{\mathrm{QD}}$, and is given by the product of an inverse variance weighted field, and a \textit{deflected} Wiener-filtered one \citep{carronMaximumPosterioriCMB2017}

\begin{align}
    g_{\mathrm{QD}}(\hat{n}) &=  -\Big(\Beam^{\dagger}\mathrm{Cov}_{\va}^{-1} X^{\mathrm{dat}}\Big)(\hat{n})  \label{eq:qdlike}\\
    &\cdot\Big(\Da\eth C^{\mathrm{unl}}\Da^{\dagger}\Beam^{\dagger}\mathrm{Cov}_{\va}^{-1} X^{\mathrm{dat}}\Big)(\hat{n}) \ , \nonumber
\end{align}

where $\eth$ is the spin-raising operator.

The gradient of the likelihood also introduces a mean field term which, in the absence of other sources of anisotropies, represents the anisotropy introduced delensing the noise map by the estimated $\va$ in the quadratic gradient \citep{carronMaximumPosterioriCMB2017}. This term predominantly produces dilations (convergence-like) rather than local anisotropies (shear-like) terms, hence generally very small for polarization-only estimators~\citep{Belkner:2023duz}. However, it is larger for temperature estimators, possibly contributing up to $10-20 \%$ to the cross-spectrum with the input field. For further exploration of this impact, a brief discussion is presented in Appendix~\ref{sec:meanfieldapp}, while a more detailed study is deferred to further work.

\subsection{Lensing reconstruction from the QE estimator}

The standard quadratic estimator (QE) \citep{hu-okamoto} can be easily obtained from this likelihood perspective. If one writes the gradient  (`$g_{\va}$') to linear order in \va, and forces it to vanish, one obtains an estimate $\hat \va$ defined by:

\begin{equation}
g_{\va} \approx g^{\mathrm{QD}}_{\va=0}-H_{\va=0}\:\hat{\va} = 0.  \label{eq:system} 
\end{equation}
In this equation, $H$ (the Hessian) is minus the second derivative of the log-likelihood function. Taking instead of the realization-dependent curvature its average of data realizations, $\hat \alpha$ will be quadratic in the data. This average is a Fisher matrix calculated at no deflection, and is identical to the quadratic estimator response function calculated in the standard manner~\cite{hu-okamoto}.

In this case, the unnormalized estimate $g^{\mathrm{QD}}_{\va=0}$ of the CMB lensing field is given by the product of an inverse-variance-weighted field, and an undeflected Wiener-filtered one

\begin{align}
   & g^{\mathrm{QE}}(\hat{n}) =\\
   &- \Big(\Beam^{\dagger}\mathrm{Cov}_{\va=0}^{-1} X^{\mathrm{dat}}\Big)(\hat{n}) \Big(\eth C^{\mathrm{unl}}\Beam^{\dagger}\mathrm{Cov}_{\va=0}^{-1} X^{\mathrm{dat}} \Big)(\hat{n})\ .
\label{eq:qelike}   
\end{align}

This is similar to the quadratic part of Eq.~\eqref{eq:qdlike}, but for the absence of all deflections: the gradient part in~\eqref{eq:qdlike} is deflected, because the iterative estimate works by capturing the residual lensing at the unlensed position, and then remapping back to give the estimate at the observed locations~\citep{Hanson_2010}. 

\subsection{Non-Gaussian deflections effect on the reconstruction}

The autospectrum of an estimated normalized QE CMB lensing map then results in

\begin{equation}
    C_L^{\hat{\phi}\hat{\phi}} \approx C_L^{\phi\phi} +N_L^{(0)}+N_L^{(1)} +N_L^{(3/2)} +... \ ,
\end{equation}

where the total estimated power spectrum $C_L^{\hat{\phi}\hat{\phi}}$ has contributions from the signal of interest itself $C_L^{\phi\phi}$, from chance Gaussian fluctuations $N^{(0)}$ \citep{lewis2006}, from additional secondary lensing contractions $N^{(1)}$ \citep{Kesden_2003}, and non-Gaussian contributions $N_L^{(3/2)}$. These are induced by non-zero higher order statistics in the CMB lensing potential, arising from non-linear growth of structure and post-Born lensing \citep{pratten2016}.

The study of these effects on the estimate of the CMB lensing potential has been thoroughly studied for the quadratic estimator \citep{bohmBiasCMBLensing2016, bohmEffectNonGaussianLensing2018, beckLensingReconstructionPostBorn2018a, fabbianCMBLensingReconstruction2019a}. They mainly arise due to a non-zero response in the auto-spectrum to the presence of a bispectrum term of the lensing potential.

For next generation surveys, such as CMB-S4 \citep{cmbs4}, the $N_L^{(3/2)}$ bias, if unaccounted for, can lead to $1$-$2\sigma$ induced biases in cosmological parameters \citep{beckLensingReconstructionPostBorn2018a} such as the sum of neutrino masses, or have a large impact on cross-correlations with external matter tracers such as galaxy clustering or lensing \citep{fabbianCMBLensingReconstruction2019a}. \footnote{See Section \ref{sec:params} for an updated discussion on the bias induced on the sum of neutrino masses.}
While methods to mitigate these effects has been proposed, such as using polarization data only (by excluding temperature information that often leads to large biases), or alternative quadratic estimators, such as bias hardened ones \citep{Namikawa_2013, fabbianCMBLensingReconstruction2019a}, these will not be able to fully harness the statistical power of upcoming CMB data as the MAP estimator does. 
Indeed, for low noise and high resolution experiments, the MAP will be able to reconstruct the lensing potential to the highest significance \citep{carronMaximumPosterioriCMB2017, legrand2022}, and therefore it is important to assess its potential in dealing with known lensing QE biases.

In this paper, we will focus on the quality of reconstruction of the MAP estimator, specifically examining its performance in the presence of non-Gaussian contributions. 
Our primary focus will be on the lensing auto-spectrum, which serves as the main observable for CMB lensing analyses. We additionally present results for the lensing cross-spectrum with the input lensing potential. 
We plan to investigate the effect on cross-correlations for MAP with large-scale structure surveys in future studies (as it was done for the QE case \citep{fabbianCMBLensingReconstruction2019a}).

Finally, we also present simulations results demonstrating the robustness of the bias hardening technique \citep{Namikawa_2013} to the non-Gaussian bias.

\section{Measurement setup}


\subsection{Experimental setup}

We consider acrudely CMB-S4 wide-field-like experimental setup \citep{cmbs4}. 
The assumed observed sky fraction is $40\%$, and the noise is modelled for simplicity as a homogeneous and isotropic white noise, with a noise level after component separation of $N_{\rm lev}^{\rm T}$ ($N_{\rm lev}^{\rm P}$) = $1(\sqrt{2}) \mu\mathrm{K-arcmin}$ for the temperature $T$ (polarization $P$), and a beam modelled as a Gaussian with full width at half-maximum of $1$ arcmin. 
Our focus is to assess the fundamental performance of the MAP to the non-Gaussianity of CMB lensing, though it is worth noting that, at the small CMB scales considered here, extragalactic foregrounds will play an important role in temperature data.

\begin{figure}[h!]
    \centering
    \includegraphics[width=\columnwidth]{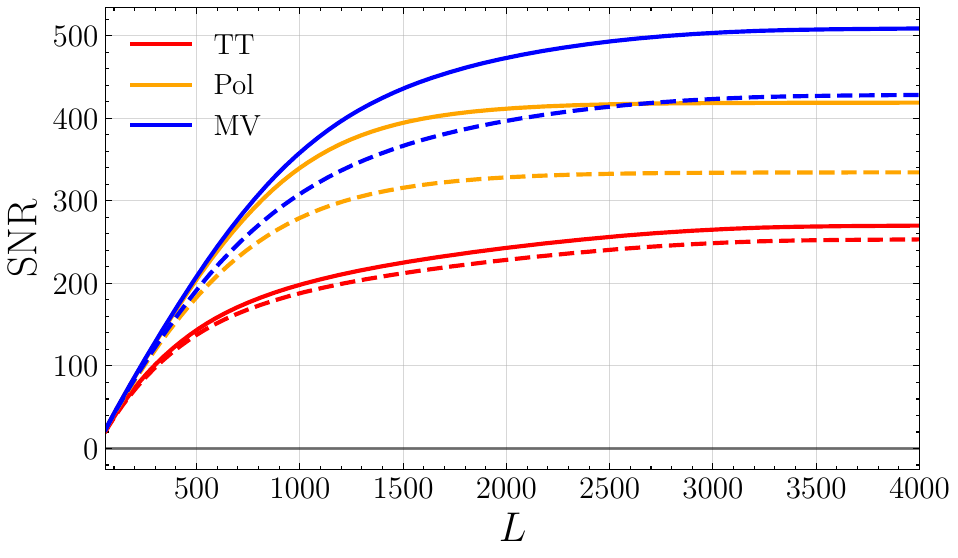}
    \caption{Predictions of the signal to noise ratio, defined in Equation \ref{eq:snr}, on the lensing auto spectrum for different lensing estimators, for our baseline CMB-S4 wide-like configuration, with $\ell_{\rm max} = 4000$. Dashed is the standard QE and solid the MAP estimator. The temperature-only, polarization-only and minimum variance estimators are shown in red, orange and blue respectively.}
    \label{fig:snrplot}
\end{figure}

These will still play a key role for future CMB lensing measurements like Simons Observatory~\cite{so}, and are potentially the most sensitive reconstruction channel at even lower noise levels, provided scales deep enough in the damping tail of the CMB spectrum can be used, where the lensing effect is large.
Figure \ref{fig:snrplot} shows forecasts\footnote{The MAP curves were calculated using 5 iterations, with theoretical reconstruction noise calculated using the \texttt{plancklens} code.}  for MAP estimators involving different data combinations in our baseline configuration, assuming CMB modes up to $\ell_{\rm max} = 4000$ are used in the reconstruction. The polarization data on its own can reach a nominal precision of 0.25\% on the spectrum using $L \leq 1500$ with the iterative approach, and provided all sources of foregrounds and complications are under control, one can push to slightly less than $0.2\%$.

To extract the \Nth bias we will consider different kind of simulations, classified based on the nature of the input convergence field. To remove unphysical effects that could come from finite number of particles in the N-body simulation used in this work, we keep input lensing modes up to an $L_{\mathrm{sim}}=5120$.  Given then an input CMB lensing map, we use \lenspyx\ \cite{reinecke2023improved} to generate lensed CMB maps deflected with $\va = \nabla\phi$.\footnote{\url{https://github.com/carronj/lenspyx}} Finally, we convolve the CMB simulation with the beam, and add a noise realization. The total map $X^{\rm{obs}}=X^{\rm{cmb}}+n$ will be our observed map.

We now proceed to describe the sets of deflection fields used to lens the CMB.

\subsection{Simulated data sets}

In order to isolate the relevant effects and test the impact of the assumption of the prior on $\kappa$ for the lensing reconstruction, we made use of different sets of simulations, summarized below:

\begin{itemize}
    
    \item $\kappa^{\rm tot} \equiv \kappa^{\rm{LSS+PB}}$, CMB lensing convergence simulations that include the total effect from LSS non-linearity and post-Born (PB) effects.
    \item $\kappa^{\rm{LSS}}$, CMB lensing convergence simulations that include only the effect from LSS non-linearity.
    \item $\kappa^{\rm{X,R}}$, with  $X \in \{\rm{tot, LSS}\}$ where we randomize the phases of the CMB lensing convergence simulations that include the total effect from LSS non-linearity and PB effects, or only LSS if $X=\rm{LSS}$.
    \item $\kappa^{\rm{G}}$, Gaussian CMB lensing simulations that include the exact power spectrum of the raw input convergence field $C_L^{\kappa^{\rm{X}}\kappa^{\rm{X}}}, X \in \{\rm{tot, LSS}\}$. \footnote{Though we generate Gaussian simulations with CMB lensing power spectra matching the total and LSS only cases, we note that the impact of post-Born corrections on $C_L^{\kappa\kappa}$ has been proven to be negligible at the level of accuracy considered in this work, at the $0.25\%$ level at $L = 8000$~\cite{pratten2016,fabbian2018}.
    Therefore one might only use only Gaussian simulations with LSS only spectra.}
    
\end{itemize}

In addition, in Section \ref{sec:lognormalsims} we will also use log-normal simulations of $\kappa$ with a custom third moment $\gamma$, and they will be indicated by $\kappa^{\rm{log}}$, and their randomized version by $\kappa^{\rm{log,R}}$.

Below we provide more details on each type of simulations.

\subsubsection{Fully non-linear observables}

To model the realistic effect of the non-linear LSS clustering of matter and the effect of post-Born corrections on lensing observables we used the full-sky maps of the CMB lensing and curl potential of \cite{fabbian2018}. These were constructed using a multiple-lens plane raytracing algorithm \cite{hilbert2009} with lensing planes constructed from a $\Lambda$CDM simulation of the DEMNUni suite. This was designed to study the impact of massive neutrinos on the universe evolution and its interplay with different dark energy models \cite{carbone2016, castorina2015}. The \lcdm simulation used a Planck 2013 cosmology with massless neutrinos
\begin{eqnarray}
\label{eq:cosmodemuni}
&&\{\Omega_{\rm cdm},  \Omega_{\rm b}, \Omega_{\Lambda}, n_{\rm s}, \sigma_8, H_0,M_{\nu},\tau
  \} = \\ \nonumber
&&\{0.27, 0.05, 0.68, 0.96, 0.83, 67 \: \rm{Km / s /
    Mpc}, 0, 0.0925 \},
\end{eqnarray}
\noindent
and sampled the matter distribution with $2048^3$ dark matter particles in a volume of 2Gpc/$h$ between $z=99$ and $z=0$. The mass resolution of the simulation at $z = 0$ is $M_{\rm CDM} = 8.27 \times 10^{10} M_{\odot}/h$. The details of the full-sky lightcone construction from the finite volume of the N-body simulation are provided in Refs.~\cite{carbone2008, calabrese2015}. The output of the lightcone construction consists in 62 surface mass density planes $\Sigma^{\theta(k)}$ including the mass contained in spherical shells of comoving thickness $\Delta\chi\approx150\ {\rm Mpc}/h$ that are used to construct k-th CMB lensing potential convergence plane as
\bea
\Delta^{(k)}_{\Sigma} &=& \Sigma^{\theta(k)} /\bar{\Sigma}^{\theta(k)} -1. \\
\kappa^{(k)}_{\chi_{CMB}} &=& 4 \pi  G\frac{D_a(\chi_{CMB}-\chi_k)}{D_a(\chi_{CMB})}\frac{(1+z_k)}{D_a(\chi_k)}\Delta^{(k)}_{\Sigma} 
\label{eq:kappa1}
\eea
where $D_a$ is the angular diameter distance and $\chi_{CMB}, \chi_k$ are comoving distances to the CMB and to the $k$-th lensing plane respectively. The simulation neglects the matter distribution at $99< z \leq 1089$ without any effective loss in accuracy. The lensing convergence planes can then be summed together to obtain the lensing potential in Born approximation ($\kappa^{LSS}$) following \cite{fosalba2008} or used to propagate the full lensing distortion tensor beyond the Born approximation as discussed in \cite{fabbian2018} and obtain also maps of CMB $\kappa^{\rm tot}$ and lensing rotation $\omega$ which arises from coupling between subsequent lensing events. The fact that both Born and post-Born lensing maps are derived from the exact same matter distribution allows to disentangle the impact of each of the specific term as anything depending on $\kappa^{\rm tot} - \kappa^{LSS}$ will isolate the effect of post-Born corrections alone.  The pipeline is general and can be adopted for lensing planes located at $\chi_s$ different from the CMB. 
We refer the reader to  \cite{fabbian2013, fabbian2018} for more technical details of the raytracing procedure used here. 

\subsubsection{Log-normal simulations \label{sec:lognormalsims}}

The main relevant effect for our study is the bispectrum signal introduced in the N-body derived convergence map. For comparison, it will be useful to generate simpler maps that introduce a connected higher than two order correlation function, such as log-normal simulations. These are much cheaper to produce and better understood analytically compared to full N-body simulations, while allowing to tune the skewness of the generated map to match the value found in the full N-body simulation.

We follow the methods of~\cite{xavierImprovingLognormalModels2016a} to produce our log-normal simulations.\footnote{As implemented in \url{https://github.com/Saladino93/fieldgen}.} We model the CMB lensing with a shifted lognormal field:

\begin{equation}\label{eq:klog}
    \kappa^{\rm{log}}(\hat{n}) = e^{Z(\hat{n})}-\lambda
\end{equation}

where $Z$ is a Gaussian random field with mean $\mu$ and variance $\sigma^2$, and $\lambda$ a shift parameter. These values are calculated in such a way to match a desired power spectrum and skewness calculated from a band-limited map. We relate the moments of the desired map $\mu_{\kappa^{\rm{log}}}$, $\sigma^2_{\kappa^{\rm{log}}}$, and $\gamma_{\kappa^{\rm{log}}}$ to the parameters $\mu$, $\sigma$, and $\lambda$ to generate the log-normal simulation (see e.g. \cite{xavierImprovingLognormalModels2016a})



\begin{figure}[htp]
\centering
\includegraphics[width = 0.45\textwidth]{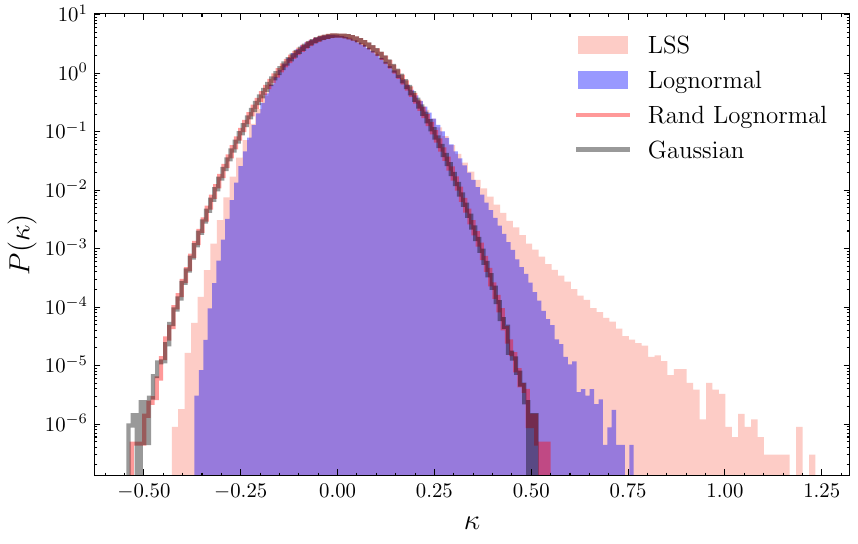}
\includegraphics[width = 0.45\textwidth]{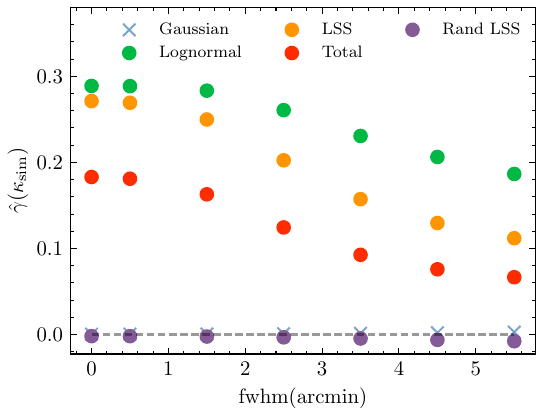}
\caption{
\emph{Upper panel}: binned histogram of several input convergence maps. We can see that the histograms from the LSS and lognormal convergence maps are slightly skewed compared to the Gaussian and randomized ones. We do not plot an histogram of a convergence map including LSS+Post Born effects, though this will have a reduced skewness compared to the LSS only one \citep{fabbian2018}.
\emph{Lower panel}: The skewness for different input convergence maps in function of the FWHM of a Gaussian smoothing. The lognormal case shown here is built to match the log-normal parameter $\lambda$ of the $\mathrm{LSS}$ map at $\rm{fwhm}=0$. The values are calculated considering lensing modes up to $L_{\mathrm{sim}}=5120$ where the CMB lensing modes used to lens the CMB simulations are cut at.
}

\label{fig:thirdmoments}
\end{figure}

To obtain the desired log-normal field, we first calculate the correlation function of the convergence field, and obtain from it the correlation function of the Gaussian field $Z$ according to~\cite{xavierImprovingLognormalModels2016a}

\begin{equation}
    \xi_Z = \ln\left(\frac{\xi^{\kappa}}{\alpha^2}+1\right)
\end{equation}

where $\alpha=\mu_{\kappa^{\rm{log}}}+\lambda > 0$. We then generate the zero mean Gaussian field $Z-\mu$, from the power spectrum  $C_l^{Z}$, obtained from $\xi_Z$,\footnote{We use the package \url{https://cltools.readthedocs.io/flt/} to perform Legendre transforms to quickly switch from/to correlation functions/angular power spectra.} using \texttt{Healpy} \cite{2005ApJ...622..759G, Zonca2019}.\footnote{\url{healpy.readthedocs.io/}} Finally, we go back to real space and obtain

\begin{equation}\label{eq:klog2}
  \kappa_{\rm{log}}(\hn) = e^{Z(\hn)}-\lambda,
\end{equation}

where $e^{\mu}=(\mu_{\kappa^{\rm{log}}}+\lambda)e^{-\sigma^2/2}$, with $\sigma^2 =  \xi_Z(0)$ is the variance of the Gaussian field $Z$. The generated maps are saved at the same resolution $\mathrm{NSIDE}=4096$ of the N-body convergence simulation, and are then processed with the same pipeline. In Figure~\ref{fig:thirdmoments} we show on the top panel an histogram of the non-Gaussian input and log-normal maps, and on the lower panel the third moment for the different simulations used in this work. As we will discuss in detail later in the text, the bulk of non-Gaussian biases is related, to leading order, to the bispectrum of the lensing field and recent work has shown that shifted-lognormal simulations like the ones adopted here can provide a moderately accurate approximation to weak lensing fields' higher-order statistics \cite{hall2022}.

\subsection{Lensing potential reconstruction setup}


Unless indicated otherwise, we reconstruct the lensing potentials for $L\in (2, 5120)$ from the generated simulations, using CMB modes in the range $\ell_{\mathrm{T,CMB}}\in (10, 4000)$ for $T$, $E$ and $B$.

We use \delensalot~\cite{Belkner:2023duz}\footnote{\url{https://github.com/NextGenCMB/delensalot}} for the MAP or QE reconstruction of the lensing potentials (the QE reconstruction of \delensalot~follows \plancklens). Unless stated otherwise, throughout the paper the presented MAP results performed at least 5 iterations, where we find converged results, in the sense of negligible changes in the derived spectra of the reconstructed fields.\footnote{In the case of temperature, convergence of the results can be reached earlier though we still keep a larger number of iterations.}

For the MAP estimates, reconstruction of lensing potential and the Wiener filtered CMBs are attempted up to an $\ell_{\rm{unl, max}}=5120$ at each step. 
The step size used in the Netwon-Raphson method to updated the iterative estimates is $\lambda = 0.5$, and the tolerance to check for convergence of the conjugate gradient method used for the inverse variance operation is set to $\texttt{cg\_tol}=10^{-7}$, running $7$-$10$ iterations per reconstruction.

We renormalize the reconstructed maps as follows: we do not employ analytical expressions, but use reconstructions based on observed CMB simulations lensed with the Gaussian CMB lensing simulations set ($\kappa^{\mathrm{G}}$), and we calculate a normalization factor given by

\begin{align*}
    \mathcal W^{\mathrm{emp, XY}}_\textit{L} = \frac{\langle C_\textit{L}^{\hat{g}^{XY}\kappa} \rangle_{\rm sims} }{\langle C_\textit{L}^{\kappa\kappa}\rangle_{\rm sims}}\,,
\end{align*}

where $\hat{g}^{XY}$ represents an unnormalized CMB lensing potential reconstruction, and $\kappa$ the corresponding input used to lens the CMB simulation.\footnote{We checked that for the QE, we get consistent results both with this method and by using a response function given by CMB gradient spectra, following \citep{fabbianCMBLensingReconstruction2019a}.}

We then use this isotropic factor to normalize the lensing potential estimates from all simulation sets ($\kappa^{X},\ \kappa^{\rm X,R}\ | \ X \in \{\rm LSS, \rm T, \rm log\}$  )

\begin{align}\label{eq:renorm}
    \hat{\kappa}_{LM}^{XY} = \frac{\hat{g}_{LM}^{XY}}{\mathcal W^{\rm{emp, XY}}_\textit{L}}
\end{align}

from which we calculate the raw power spectrum

\begin{equation}
   C_L^{\hat{\kappa}^{XY}\hat{\kappa}^{XY}} = \frac{1}{2L+1}\sum_{M}\hat{\kappa}_{LM}\hat{\kappa}^{\dagger}_{LM}\ .
\end{equation}

Finally, to assess the quality of the reconstructions, it is useful to define the cross correlation coefficient $\rho_L$
\footnote{By `quality' of the reconstruction we mean here its information content on the true lensing potential, irrespective of potential biases (the cross-correlation coefficient is invariant under arbitrary multiplicative biases in the reconstructed field (small additive ones can also be cast as multiplicative biases))}.



\begin{align*}
    \rho_L = \dfrac{C_L^{\hat{\kappa}^{XY}\kappa}}{\sqrt{C_L^{\hat{\kappa}^{XY}\hat{\kappa}^{XY}} C_L^{\kappa \kappa}}}\,,
\end{align*}

\section{$\NthXY$ bias} \label{sec:results}

With the notation now established, we are ready to calculate the biases associated with the presence of non-Gaussianity in our CMB lensing fields. The $\nlth$ biases are determined through simulations, specifically by cross-correlating the estimated CMB lensing field with another tracer, such as the input CMB lensing potential itself, or by examining the CMB lensing auto-correlation. This methodology, as outlined in (\citep{beckLensingReconstructionPostBorn2018a, fabbianCMBLensingReconstruction2019a}) allows us to quantitatively assess and account for the impact of non-Gaussian features in our lensing fields. This is achieved by calculating the difference from the scenario where no non-Gaussianity is present.

The biases, denoted as $\NthXY$, are computed through the following expressions:

\begin{align} \label{eq:N32calc} \nonumber
  \mathrm{Total:\ } \NthXY &= \langle \hat{C}_L^{\hat{\phi}^{XY}\phi^{\rm ext}}[\kappa^{\rm tot}]-\hat{C}_L^{\hat{\phi}^{XY}\phi^{\rm ext}}[\kappa^{\rm{tot,R}}]  \rangle_{\rm CMB}
\\ \nonumber
    \mathrm{LSS:\ } \NthXY &= \langle \hat{C}_L^{\hat{\phi}^{XY}\phi^{\rm ext}}[\kappa^{\rm{LSS}}]-\hat{C}_L^{\hat{\phi}^{XY}\phi^{\rm ext}}[\kappa^{\rm{LSS,R}}]  \rangle_{\rm CMB}
\\
    \mathrm{PB:\ } \NthXY &= \langle \hat{C}_L^{\hat{\phi}^{XY}\phi^{\rm ext}}[\kappa^{\rm tot}]-\hat{C}_L^{\hat{\phi}^{XY}\phi^{\rm ext}}[\kappa^{\rm{LSS}}]  \rangle_{\rm CMB}\ .
\end{align}


We denote $\phi^{\rm ext}$ either the input lensing field or the reconstructed field $\hat{\phi}^{XY}$ itself. The angle brackets signify the average over lensed CMB simulations. To gauge the reliability of our measurements, we will compute the scatter in the biases from these simulations to obtain the uncertainty on the mean measurement. Our subtraction method ensures that simulations which are differentiated in these equations share identical primordial CMB and noise, mitigating realization-dependent biases and cosmic variance effects.

In Figures~\ref{fig:n32bornauto} and~\ref{fig:n32borncross} we show the \Nth bias for the CMB lensing autospectrum and cross-spectrum with the input, respectively. The shaded areas are computed from the scatters on the mean of the simulations, while grey area represent statistical error bars as calculated from the diagonal of the covariance matrix

\begin{equation}
    \mathbf{C}_{LL'} =  \frac{2\delta_{LL'}}{(2L +1) f_{\rm sky}} \left(C_{L}^{\rm \phi\phi} + N_L^{(0)} +  N_L^{(1)}\right)^2 \ .
\end{equation}

In Figure \ref{fig:n32bornauto}, it is evident that, for the lensing autospectrum, the \Nth bias of the MAP estimator shows a lower absolute value compared to the QE estimator for $L \leq 1800$. Specifically, at the scales presented, the LSS effect in the QE tends to suppress power in the reconstructed potential. This effect is alleviated in the MAP for all estimators except on the smallest scales. The situation is similar, but with the opposite sign, for the PB bispectrum effect. This results in partial cancellation between the two effects for the total contribution for both the QE and MAP. Similar conclusions are achieved with the cross-spectrum with the input lensing potential as depicted in Figure \ref{fig:n32borncross}.

Both MAP and QE estimators share the characteristic that the total bias effect depends on the combination of used data. Specifically, in the minimum variance combination, the polarization estimator dominates at lower modes $L \leq 1500$, while the temperature one dominates at higher $L$'s.

Finally, a distinct feature is the noticeable steep rise in the bias of the MAP estimator for the LSS cases, particularly in the TT and MV combinations, evident in both the auto and cross spectra. While the specific origin of this rise remains uncertain, we validated our bias method by assessing Gaussian simulations with no LSS bias and randomized ones. We checked that the difference is consistent with zero. Therefore this rise at small scales in the MAP might be due non-Gaussanities, or from the choice of normalisation of the maps (obtained from Gaussian simulations). To investigate the choice of normalisation, we consider the cross-correlation coefficient of the reconstructions with their respective inputs

\begin{eqnarray}
    \rho^X = \frac{\langle C_L^{\hat{\kappa}\kappa} \rangle}{ \sqrt{\langle C_L^{\hat{\kappa}\hat{\kappa}}\rangle \langle C_L^{\kappa\kappa}}\rangle}\ \,
\end{eqnarray}

where $\langle . \rangle$ denotes the average over sims. This quantity does not require any normalisation, and it is related to the faithfulness of our reconstruction to the input. We assume that we can directly take the difference of the cross-correlation coefficient with the corresponding input for the LSS non-Gaussian MAP vs the Gaussian MAP estimates, as shown in Figure \ref{fig:crosscorrcoeff}.

As expected, we can see that on large scales the Gaussian case reconstruction is better correlated with the input, for both the QE and MAP cross-correlation differences. While on small scales, $L>1500$, interestingly we see a small rise in the difference with the cross-correlation coefficient of the MAP of the non-Gaussian case. This might be related to additive terms that we test in Section \ref{sec:signflipped}.


\begin{widetext}

\begin{figure*}[h!]
    \centering
    \includegraphics[width=\textwidth]{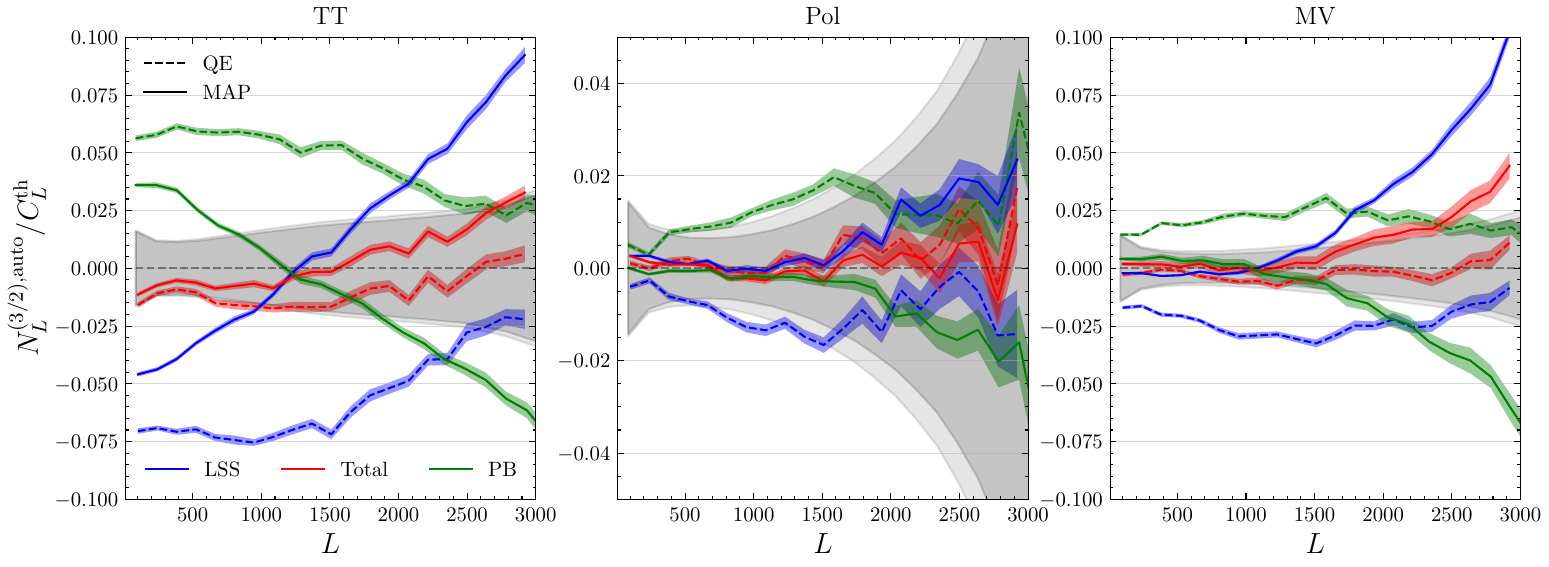}
    \caption{Fractional $N_L^{(3/2)}$ bias on the CMB lensing auto-correlation for a CMB-S4 like configuration, as described in the text. This is calculated by averaging over 64 realization of the input CMB primordial field. We show results for the LSS only, post-Born (PB) and full cases in blue, green and red, respectively. In grey and light grey we show the statistical error bars for MAP and QE respectively. We bin the spectra from $L_{\rm{min}} = 30$ to $L_{\rm{max}} = 3000$ with a wide binning of around $140$.}
    \label{fig:n32bornauto}
\end{figure*}

\begin{figure*}[h!]
    \centering
    \includegraphics[width=\textwidth]{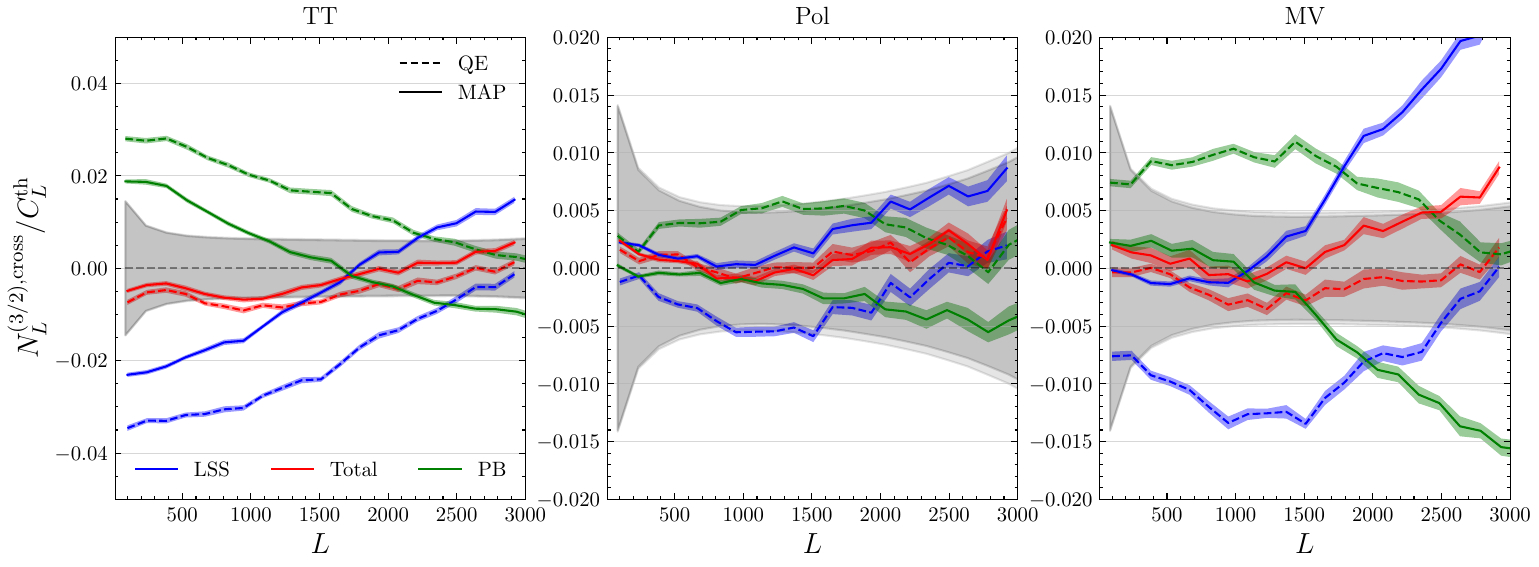}
    \caption{Same as Figure \ref{fig:n32bornauto} but for the cross-spectrum between the reconstruction and the input lensing potential.}
    \label{fig:n32borncross}
\end{figure*}

\end{widetext}

\begin{figure}[h!]
    \centering
\includegraphics[width=0.95\columnwidth]{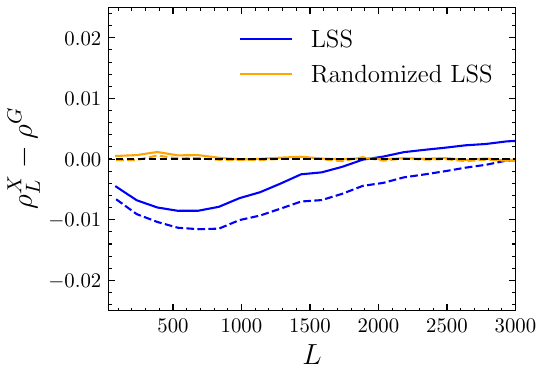}
    \caption{Difference in the cross-correlation coefficients between the LSS non-Gaussian (blue), randomized LSS non-Gaussian (orange) cases with the Gaussian case, for the $TT$ estimator. In solid (dashed) we have the MAP (QE). We can see that the difference with randomized (orange) is consistent with zero. While as expected, the non-Gaussian case is below zero on large scales. Nevertheless, for the non-Gaussian MAP (blue solid line), we can see a small rise after $L>1500$.}
    \label{fig:crosscorrcoeff}
\end{figure}

\subsection{Simple theory calculations \label{sec:simpletheory}}
Given the complexity of beyond-the-QE reconstruction, a comprehensive analytic treatment of the theory curves for the MAP-$\nlth$ bias seems out of reach. Nevertheless, we can proceed by analogy with quadratic estimator theory to obtain some curves that matches reasonably well our findings.

We will focus on the simpler case of the cross-correlation of the reconstructed CMB lensing potential $\hat{\phi}$ with the input lensing potential $\phi$ on the flat-sky:\footnote{We focus on the cross-correlation with the input, and not on the reconstruction auto-spectrum, as the former is easier to predict with respect to the latter. In particular, we perform all of our studies using CMB modes up to $l_{\rm{max}}=4000$. Our baseline perturbative analytical model for prediciting the QE auto-spectrum, based on \citep{bohmBiasCMBLensing2016, fabbianCMBLensingReconstruction2019a}, works well only on the largest scales, where we get the prediction $N^{(3/2),\rm{auto}} \sim 2N^{(3/2),\rm{cross}}$. This will require a more careful treatment for future studies.}

\begin{equation}\label{eq:n32x}
    N^{(3/2), \rm{cross}}_L = \langle \hat{\phi}(\vec{L})\phi^*(\vec{L}) \rangle - \langle  \phi(\vec{L}) \phi^*(\vec{L}) \rangle,
\end{equation}
where we assume the reconstruction $\hat{\phi}(\vec{L})$ is unbiased, in the sense that $\langle \hat{\phi}\rangle=\phi$. The difference between these two terms depends predominantly on the bispectrum of the observed lensing potential $B^{\phi\phi\phi}(L, l_2, l_3)$, where $L$ stands for the external tracer (here $\phi$ itself) multipole and $l$'s for the ones entering through the CMB modes used for CMB lensing reconstruction.

For the standard QE CMB lensing temperature-only estimator, calculations are given in \citep{bohmBiasCMBLensing2016, fabbianCMBLensingReconstruction2019a}. We briefly review the rationale here.



A temperature-based quadratic estimator is of the form

\begin{equation}
    \hat{\phi}(\vec{L}) = A_L^{TT} \sum_{\vec{l}} g(\vec l, \vec L) T^{\rm{dat}}(\vec{L}-\vec{l})T^{\rm{dat}}(\vec{l}).
    \end{equation}
for some normalization $A_L^{TT}$ and weights $g(\vec l, \vec L)$.

 The lensed CMB may be perturbatively expanded in a series with respect to the lensing potential, leading to
\begin{equation}
    T = T^u+\delta T^u+\delta^2 T^u+ \mathcal{O}(\phi^3)\ ,
\end{equation}

where $T^u$ is the unlensed CMB, and $\delta^n T^u$ depends on the lensing potential power $n$ \citep{lewis2006}.

On evaluating the cross-correlation~\eqref{eq:n32x},

\begin{equation}
    \langle\hat{\phi}(\vec{L}) \phi(-\vec{L})\rangle  \sim \langle T(\vec{L}-\vec{l})T(\vec{l})\phi(-\vec{L})\rangle\ ,
\end{equation}
the lensing potential bispectrum will give rise to the $N^{(3/2), \rm{cross}}_L$,

\begin{equation}
    N^{(3/2), \rm{cross}} \sim \langle \delta T^u\delta T^u \phi \rangle + 2 \langle T^u \delta^2 T^u  \phi \rangle\ .\label{eq:n32refsimple}
\end{equation}

The full expression for the temperature case, including to a good approximation non-perturbative lensing remapping effects~\citep{fabbianCMBLensingReconstruction2019a}, is

\begin{align} \nonumber
    N^{(3/2), \rm{cross}}_L  = A_L^{TT} \int_{\vec{l}_1} B^{\phi\phi\phi}(L,l_1,|\vec L - \vec l_1|)\int_{\vec{l}_2} g(\vec{l}_2, \vec{L})  \\  \left( - C^{T\nabla T}_{|\vec {l}_1 - \vec {l}_2|} [(\vec{l}_1-\vec{l}_2)\cdot\vec{l}_1] [(\vec{l}_1-\vec{l}_2) \cdot (\vec{L}-\vec{l}_1)]  \right. \nonumber
    \\ \left. +\: C^{T\nabla T}_{l_2} [\vec{l}_2\cdot \vec{l}_1] [\vec{l}_2\cdot (\vec{L}-\vec{l}_1)]  \right). \label{eq:n32fullcrosstt}
\end{align}
The spectra $C_l^{T \nabla T}$ are here the same lensed gradient spectra that enters the non-perturbative lensing response functions.

Let's discuss how we connect the MAP bias to this representation. Reference \citep{legrand2022} demonstrated through simulations that the converged CMB lensing MAP solution power spectrum (for polarization at least) can be accurately described as a quadratic estimator with partially lensed CMB spectra. These spectra can be obtained using an iterative scheme initially proposed for the $EB$ estimator by~\cite{smith2012}. Hence, it is natural to test this recipe for $N^{3/2}_L$ as well.

In this picture, the lensing potential entering the CMB legs are now given by the unresolved, residual lensing map
\begin{equation}
\hat{\phi}^{\rm res}_{LM} \equiv (1-\mathcal W_L)\phi_{LM}.
\end{equation}

We then make use of the same flat-sky prediction Eq~\eqref{eq:n32fullcrosstt}, but with the substitution
\begin{align} \nonumber
B^{\phi\phi\phi}(L,l_1,l_3)  \rightarrow B^{\phi\phi\phi}(L,l_1,l_3) (1-\mathcal W_{l_1})(1- \mathcal W_{l_3}).
\end{align}
where $\vec l_3 = \vec L - \vec l_1$, and $\mathcal W$ is the Wiener-filter.
The normalization $A^{TT}_L$ is the standard quadratic estimator normalization but calculated with partially lensed gradient spectra, and $C_l^{T \nabla T}$ are also calculated with the partially lensed spectra. 

\begin{figure}[h]
    \centering
    \includegraphics[width=1.0\columnwidth]
    {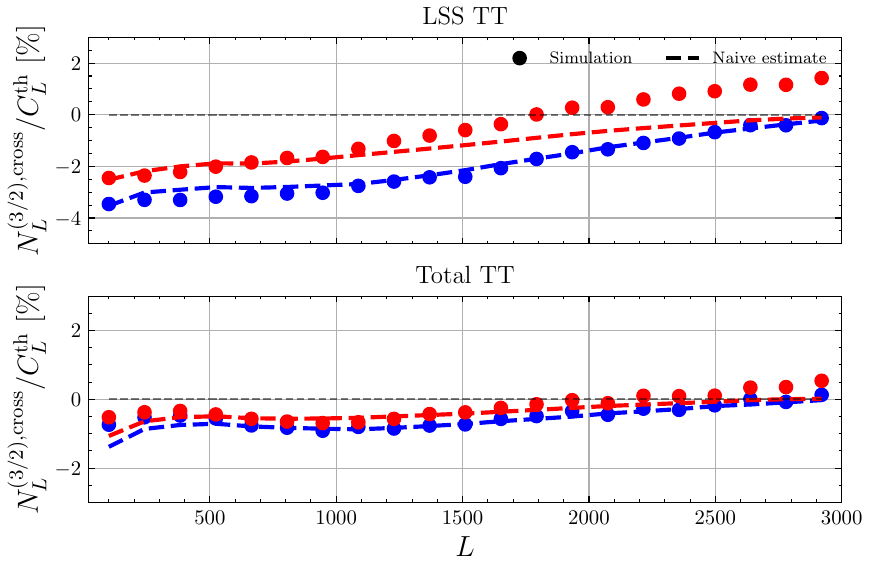}
    \caption{Approximate theory calculation for the $N_L^{(3/2)}$ bias on the CMB lensing cross-spectrum. In blue, we show results for the quadratic estimator, and in red for the MAP reconstruction. The MAP predictions are obtained from the naive analytic prescription given in the text. This is for our CMB-S4-like reconstruction from temperature.}
    \label{fig:simpletheorycurve}
\end{figure}

In Figure~\ref{fig:simpletheorycurve} we show the result of this naive calculation, compared to simulations.\footnote{The code to calculate these biases is based on \url{https://github.com/Saladino93/lensbiases}.}. For simplicity we consider the case of temperature only. On the upper panel we show the LSS-only case, where $N^{3/2}$ is stronger than with the total bispectrum, shown in the lower panel.
It assumes a LSS bispectrum calculated at the tree level in perturbation theory with corrections coming from \citep{gil-marinImprovedFittingFormula2012} (though an improved version can be found in \citep{namikawaCMBLensingBispectrum2019}).
The Post-Born corrections are based on \citep{pratten2016}. See the Appendix of \citep{fabbianCMBLensingReconstruction2019a} for a short review.

We can see that on a wide range of scales this simple predictive scheme is able to recover the estimated biases from simulations, validating the understanding that the bispectrum of the CMB lensing field correctly describe the bias also in the case of the iterative estimator. In the LSS-only case, and on small scales, the prediction deviates significantly from our findings in simulations. In Section \ref{sec:signflipped} we check that its origin is not from higher even CMB lensing connected $n$-point functions.


\subsection{Consistency and robustness tests}
To have a better understanding of our reconstructions and validate our pipeline, we perform a series of consistency checks to explore possible residuals in our estimates, specifically those of higher order than the bispectrum residuals.

\subsubsection{Joint reconstruction with lensing field rotation \label{sec:jointrec}}

In this section we test the impact of post-Born lensing rotation on the MAP reconstructions.

The CMB lensing deflection vector field $\vec{\alpha}(\hat{n})$ can be written thanks to the Helmholtz decomposition in flat-sky notation as
\begin{equation}
  \vec{\alpha}(\hat{n}) = \vec{\nabla}\phi(\hat{n})+  \vec{\nabla} \times \Omega(\hat{n})\ ,
\end{equation}
On the full-sky, the decomposition is compactly written using the spin-weight formalism,
\begin{equation}
	_1{\alpha}(\hn) = - \eth \phi (\hn) - i \eth \Omega(\hn),
\end{equation}
where $\Omega$ is the lensing curl potential, and 
\begin{equation}
\begin{split}
	\omega(\hn) &= -\frac{\nabla^2\Omega(\hn)}{2} \\
	\omega_{LM} &= \frac 12 L (L + 1) \Omega_{LM}
\end{split}
\end{equation}
is the lensing field rotation; the angle by which tiny local images are rotated by lensing. The leading rotation is induced by post-Born lensing, that couples pair of non-aligned shearing lenses at different redshifts, producing in this way a net rotation. For a single deflection or in the Born approximation the rotation is null. Non-zero bispectra of the kind $\kappa\kappa\omega$ or $\kappa\omega\omega$~\citep{pratten2016,fabbian2018} are also generate at higher perturbative orders. The amplitude of the rotation field power spectrum is about 3 to 4 orders of magnitude smaller than the one of $\kappa$~\cite{Hirata:2003ka, pratten2016} on the scales considered here.

The simulations of~\citep{fabbian2018,fabbianCMBLensingReconstruction2019a} naturally include the rotation component of CMB lensing. We use this field in combination with the post-Born convergence map  to estimate the lensing potentials in the presence of a curl-like displacement using either a joint $\phi$-$\Omega$ optimal reconstruction (the differences to the standard $\phi$-only algorithm are minors and described in~\citep{Belkner:2023duz}), or completely neglecting $\Omega$ in the reconstruction.

We can then estimate the effects in the \nlth bias coming from the mixed-bispectrum terms $\kappa\kappa\omega$, $\kappa\omega\omega$ as :

\begin{equation}
\mathrm{{\rm{Mixed}}:\ } \NthXY = \langle \hat{C}_L^{\hat{\phi}^{XY}\phi^{\rm ext}}[\kappa^{\rm{tot}}, \omega]-\hat{C}_L^{\hat{\phi}^{XY}\phi^{\rm ext}}[\kappa^{\rm{tot}}]  \rangle_{\rm CMB},
\end{equation}
where the first term include the lensing rotation in the inputs and reconstruction, and the second does not.
As for the biases obtained in previous sections, here we use 64 CMB primordial realizations, and we run the pipeline up to 10 iterations, though outputs appeared to have converged starting from the fifth one.

\begin{figure}[h!]
    \centering
    \includegraphics[width=\columnwidth]{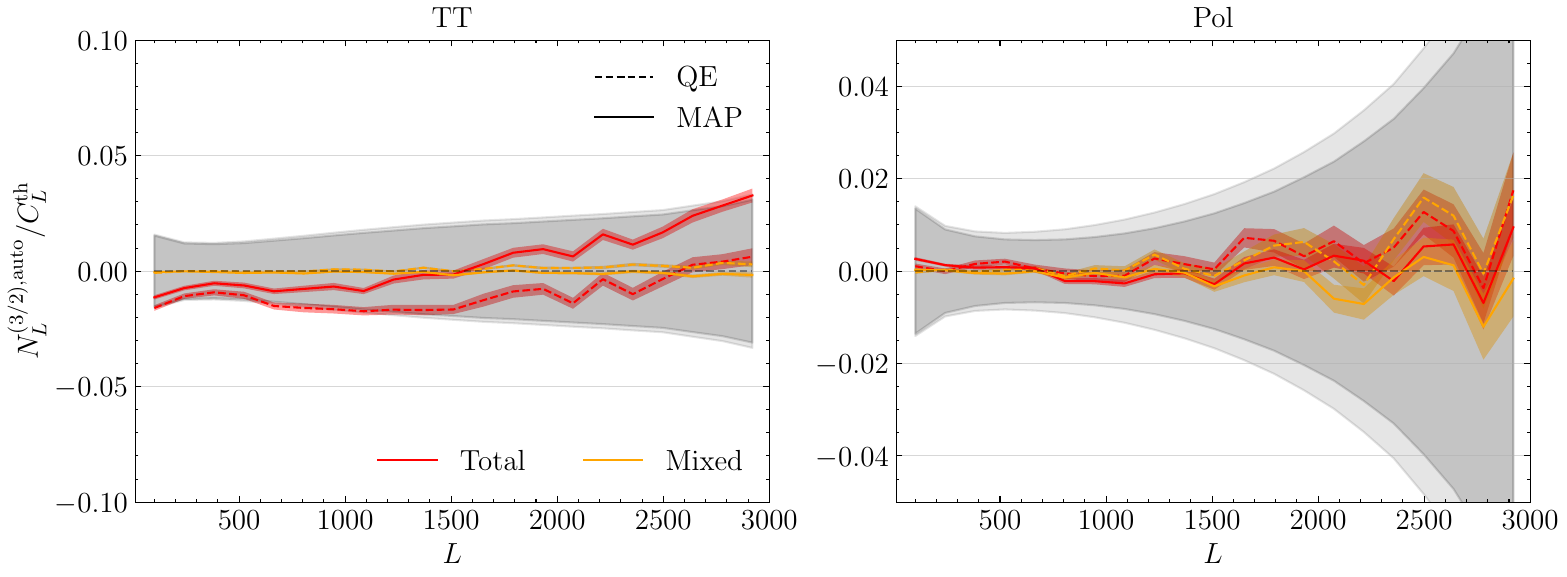}
        \includegraphics[width=\columnwidth]{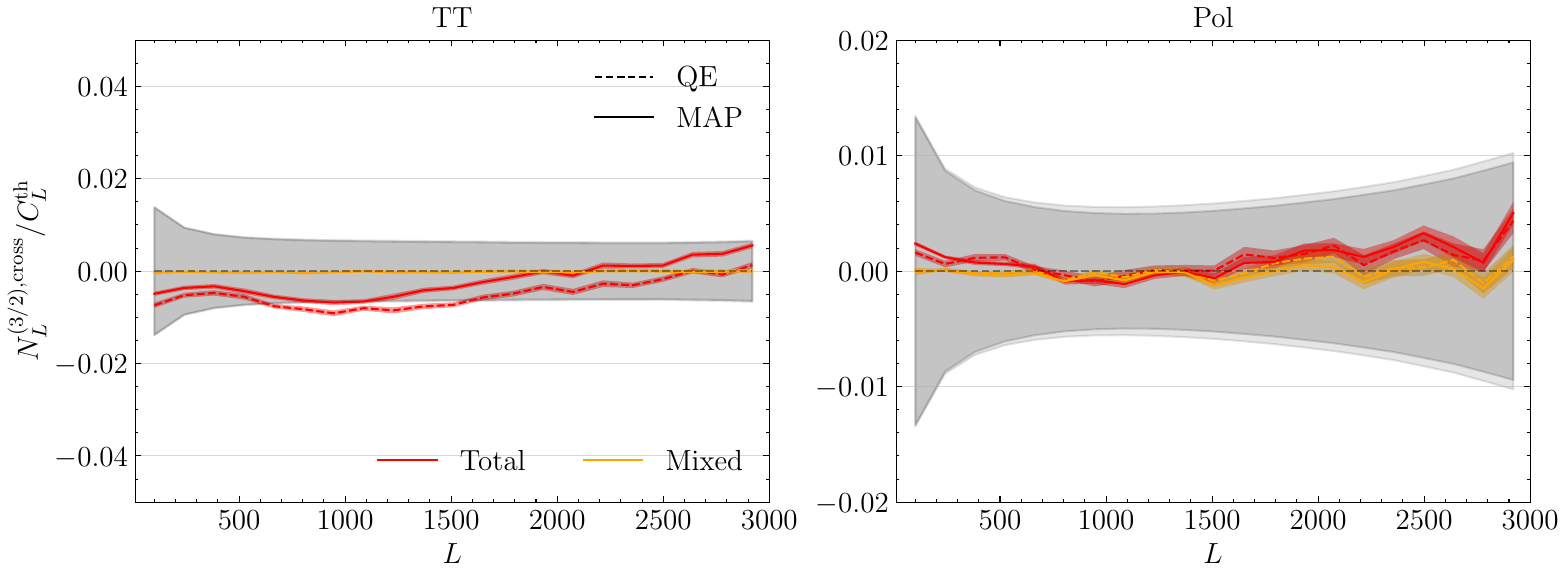}

    \caption{Non-Gaussian $N^{(3/2)}_L$ lensing biases in the estimated lensing potential auto-spectra (top panels) or cross-spectrum to the true lensing (bottom panels), with and without curl-like deflections in the simulations. In red we show the total bias that does not include post-Born lensing field rotation (same as Figure \ref{fig:n32bornauto}), while in orange we show the difference between the biases obtained including post-Born corrections for both the gradient and curl component of the deflections and the red curve. The difference is consistent with zero in both panels.}
    \label{fig:n32autojointcurl}
\end{figure}

We show in Fig.~\ref{fig:n32autojointcurl} the \nlth biases to the auto and cross-spectra in the presence of a full deflection field (including LSS and post-Born effects in the gradient-like deflection component as well as curl-like deflections), compared to a case without curl-like displacement component in the simulations. We see no statistically significant differences.

\subsubsection{Sign-flipped $\kappa$ \label{sec:signflipped}}
At the perturbative level, the $\nlth$ bias is proportional to a projected bispectrum on large scales~\citep{bohmBiasCMBLensing2016}. We will test this assumption in this section. 

Flipping the sign of the lensing potential should flip the sign of the bias, if it depends the bispectrum that is odd in the lensing field. For the QE, this argument breaks down only when using the smallest CMB scales for our reconstructions, where perturbative arguments are less effective (e.g., at $\ell_{\rm{max}} \sim 4000$~\citep{bohmBiasCMBLensing2016}). On the other hand, the MAP always uses the full likelihood information and, therefore, higher-order point functions to reconstruct the lensing field. This also makes the MAP more difficult to assess analytically compared to the QE. Therefore, we turn to simulations to check if the MAP bias behaves significantly differently with respect to the sign of the input lensing map.

For each of the convergence maps ($\kappa^{\rm G}$, $\kappa^{\rm LSS}$, $\kappa^{\rm tot}$) we calculate their flipped version

\begin{align}
    \kappa^{\rm in,\neg}_\textit{LM} = (-1) \cdot \kappa^{\rm in}_\textit{LM}\,,
\end{align}

where $\kappa^{\rm in}_\textit{LM}$ is the reference input CMB convergence lensing field.
We generate then 64 lensed CMB simulations with this flipped lensing potential, and calculate the half-sum and half-differences

\begin{equation}
    \Delta_{\pm}^{XY} = \left\langle \frac{1}{2}\Big(\NthXY[\kappa^{\rm in}_\textit{LM}]\pm\NthXY[\kappa^{\rm in,\neg}_\textit{LM}]\Big) \right \rangle_{\rm{sims}}\ .
\end{equation}

for the $TT$, $EE$, $EE+EB$, and $MV$ estimators, and where the \nlth are estimated as explained in Section \ref{sec:results}.

\begin{figure}[h!]
    \centering
    \includegraphics[width=\columnwidth]{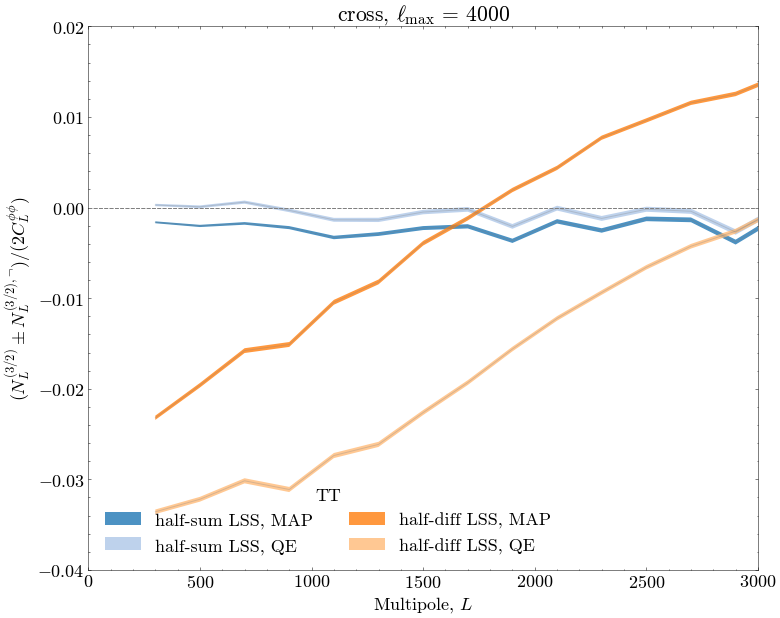}
    \caption{Half-sum and half-differences of $N^{(3/2)}_L$ biases, built from a baseline simulation, and a second one flipping the sign of the input lensing potential, in order to isolate the constributions of even(blue) and odd(orange) $n$-point functions on the total bias. The light (dark) colors are the QE (MAP) reconstructions for $\ell_{\rm max}=4000$, from temperature-only and LSS bispectrum only.}
    \label{fig:hshdttlss}
\end{figure}

Any contribution from contractions which are even in the lensing field will appear in the half-sum of the biases, and similarly for odd terms in the half-differences. Figure~\ref{fig:hshdttlss} shows these half-sum in blue and half-differences in orange, for the cross LSS-only reconstruction bias from temperature. We can see that for MAP and QE reconstructions we have a small residual in the half-sum ($\sim 0.2\%$). In particular, the rise at the highest $L$ observed in the previous section is still most likely to an odd $n$-point function effect.

To quantify the consistency of these results with a signal dominated by the bispectrum, we calculate the probability-to-exceed (PTE) using Welch’s test. This test compares the means of two simulation sets ($\kappa$, sign-flipped $\kappa$) with unequal variances, calculated from simulation scatters. The input data vector is the average over simulation measurements for $21$ multipole bins over the range $5\leq L\leq 3000$. The results are shown in Table~\ref{tab:PTEhalfsum}, where the average PTEs over the $21$ multipoles are shown.  A test is assumed successful if the PTE is greater than $0.05$. With this definition, the QE $MV$, $EB$ cross-spectrum for $PB$ cases fail, though this is acceptable given the number of tests performed.

\begin{table*}[t]
    \centering
    \setlength{\tabcolsep}{12pt} 
        \begin{tabular}{ l c c c c } 
            \hline
            \hline
             Estimator & QE & & MAP & \\
             & auto & cross & auto & cross \\
             & LSS / TOT / PB & LSS / TOT / PB & LSS / TOT / PB & LSS / TOT / PB \\
             \hline
             & \multicolumn{4}{c}{$\ell_{\rm max} = 3000$}\\
            $TT$ & 0.51 / 0.78 / 0.31  & 0.56 / 0.21 / 0.24  & 0.89 / 0.51 / 0.98  & 0.66 / 0.20 / 0.33 \\
            $EE$ & 0.42 / 0.28 / 0.95  & 0.44 / 0.63 / 0.56  & 0.32 / 0.19 / 0.86  & 0.65 / 0.79 / 0.72\\
            $EE + EB$ & 0.11 / 0.17 / 0.69  & 0.07 / 0.61 / 0.04  & 0.58 / 0.96 / 0.25  & 0.10 / 0.28 / 0.10\\
            $MV$ & 0.90 / 0.47 / 0.13  & 0.11 / 0.79 / 0.02  & 0.20 / 0.10 / 0.94  & 0.75 / 0.74 / 0.45 \\
            \hline
            & \multicolumn{4}{c}{$\ell_{\rm max} = 4000$}\\
            $TT$ & 0.97 / 0.87 / 0.88 & 0.74 / 0.71 / 0.75 & 0.38 / 0.43 / 0.37 & 0.25 / 0.09 / 0.36\\
            $EE$ & 0.07 / 0.06 / 0.43 & 0.31 / 0.28 / 0.37 & 0.52 / 0.29 / 0.74 & 0.28 / 0.27 / 0.35 \\
            $EE + EB$ & 0.38 / 0.99 / 0.31 & 0.15 / 0.44 / 0.14 & 0.82 / 0.88 / 0.65 & 0.32 / 0.59 / 0.22 \\
            $MV$ & 0.42 / 0.96 / 0.14 & 0.25 / 0.25 / 0.27 & 0.99 / 0.93 / 0.95 & 0.77 / 0.73 / 0.79\\
            \hline
            \hline
        \end{tabular}
    \caption{Averaged over $L$-bins PTEs for the half-sum tests of the \Nth biases. Shown are the temperature, polarization, their combination, QE and MAP estimators, for \Nth auto- and cross spectra.}        
    \label{tab:PTEhalfsum}
\end{table*}



\section{Alternative CMB lensing estimators}

\subsection{Iterative estimator with a non-Gaussian prior}\label{sec:lognormalsampling}

The conventional iterative estimator for CMB lensing assumes a Gaussian potential field due to its incorporation in the prior distribution. However, the observed CMB lensing field exhibits non-Gaussian characteristics that the standard iterative approach does not directly address. In this context, our objective is to accommodate potential non-Gaussian aspects of the field by proposing an alternative method for estimating $\phi$.

To begin with, we introduce a non-Gaussian prior. Specifically, we characterize the lensing convergence field using a log-normal model, as outlined in section \ref{sec:lognormalsims}. This choice establishes the foundation for our new approach and is motivated by the fact that the log-normal approximation is a sensible one for moderately non-linear fields, e.g. \cite{das2006}, \cite{barthelemy2020}.

The basic idea is that instead of iteratively solving for the CMB lensing potential field directly, we iterate over a posterior that is 
a function of the Gaussian field whose exponential gives the CMB lensing convergence field.\footnote{This allows us to use our baseline MAP pipeline with little modifications.}

In this case, we will assume a prior on the Gaussianized field $Z$, which may be written in harmonic space
\begin{equation}
   -2 \ln p_Z(Z)=\frac{\left( Z^2_{00} - \mu \sqrt{4\pi} \right)^2}{C^Z_{L=0}} + \sum_{L \ge 0, |M| \leq L} \frac{Z_{LM} Z^{\dagger}_{LM}}{C^Z_L},
\end{equation}
up to irrelevant constants. The un-normalized posterior can be written as before

\begin{equation}
    p(Z|X^{\mathrm{dat}}) \propto p(X^{\mathrm{dat}}|Z)p(Z)
    \label{eq:unnormposteriorZcomplete}
\end{equation}

Subsequently, our objective is to determine the optimal estimate of $Z$ based on the available data.\footnote{In principle we could also attempt to reconstruct $Z$ jointly with the parameters $\mu$ and $\lambda$. This is a complication with no relevance for what is being tested here however, and we assume fixed fiducial values of $\mu$ and $\lambda$, and fixed fiducial spectrum $C^Z_L$ for the reconstructions.}

In this case, the total gradient with respect to the Gaussian field is

\begin{equation}
    g^{\rm tot}_Z(\hn) = \frac{\delta \ln p(Z|X^{\mathrm{dat}})}{\delta Z(\hat{n})} = g^{\rm QD}_Z(\hn) - g^{\rm MF}_Z(\hn) + g^{\rm PR}_Z(\hn)\ ,
\end{equation}
This gradient can be calculated with minimal modifications to our $\phi$-based MAP reconstruction code: using the chain rule, we have namely
\begin{align} \nonumber
	& g^{\rm QD}_Z(\hn) - g^{\rm MF}_Z(\hn) = \frac{\delta \ln p(X^{\mathrm{dat}} |Z )}{\delta Z(\hat{n})} \\ \nonumber
	&= \int d^2\hat n' \frac{\delta \kappa(\hat{n}')}{\delta Z(\hat{n})}\frac{ \delta \ln p(X^{\mathrm{dat}} | \phi )}{\delta \kappa(\hat{n}')} \\&= e^{Z(\hn)}\left(g^{\rm QD}_\kappa(\hat n)-  g^{\rm MF}_\kappa(\hat n) \right) 
\end{align}
We have used in the second line
\begin{equation}
    \frac{\delta \kappa(\hat{n}')}{\delta Z(\hat{n})} = \delta_D^{(2)}(\hat{n}'-\hat{n})e^{Z(\hat{n})},
\end{equation}
which follows directly from our definition $\kappa(\hn) = e^{Z(\hn)} - \lambda$.
Finally, the gradients $g_\kappa(\hn)$ are easily obtained in harmonic space from those of $\phi$ which the original MAP reconstruction calculates: from $\kappa_{LM} = \frac 12 L(L + 1) \phi_{LM}$ follows for the gradients
\begin{align}
	g_{\kappa}(\hn)&=  \int d^2n' \frac{\delta \phi(\hn')}{\delta \kappa(\hn)} g_{\phi}(\hn') \\&= \sum_{LM}\frac{2}{L(L + 1)} g_{\phi, LM} Y_{LM}(\hn).
\end{align}



Starting from the quadratic estimator solution we then iterate over estimates of $Z$  in the same manner than the original code.\footnote{A way to see how the starting point is chosen is the following: if we take the gradient in $Z$, $g_Z$, then we require as a first step that: $g_Z \approx g_{Z_0}+F_{Z_0}(Z-Z_0) \approx 0$. By exponentiating the expression that we find for $Z$ we can relate it to $\kappa+\lambda$, and derive that $\hat{Z}=\ln(\hat{\kappa}+\lambda)$. 
}



Once the MAP point $\hat Z$ is reconstructed, we apply the transform to get an unnormalized estimate of the lensing field,

\begin{equation}\label{eq:ZMAP2pMAP}
	 \hat {\kappa}_{\rm log}(\hn) \equiv e^{\hat Z(\hn)} - \lambda.
\end{equation}
which we then normalize in the same way as before, by rescaling with the inverse averaged cross-spectrum to input Gaussian fields, Eq.~\eqref{eq:renorm}.
We then reconstruct the biases in the very same way.

The approach we follow (maximizing for $Z$) does lead to a slightly different map than if we were maximizing directly for $\kappa$, even if using the same lognormal prior on $Z$. 
This is because the relation between the point-estimates \eqref{eq:ZMAP2pMAP} is nonlinear. 
Our approach is arguably more natural than maximizing for $\kappa$, and avoids one issue we encoutered in preliminary work, which is how to enforce the lognormality of $\kappa$ map across iterations: when maximizing for $\kappa$, one must ensure that $\kappa(\hn) + \lambda$ is always positive at each point, since $Z(\hn)$ is its logarithm.

\begin{figure}[h!]
    \centering
    \includegraphics[width=\columnwidth]{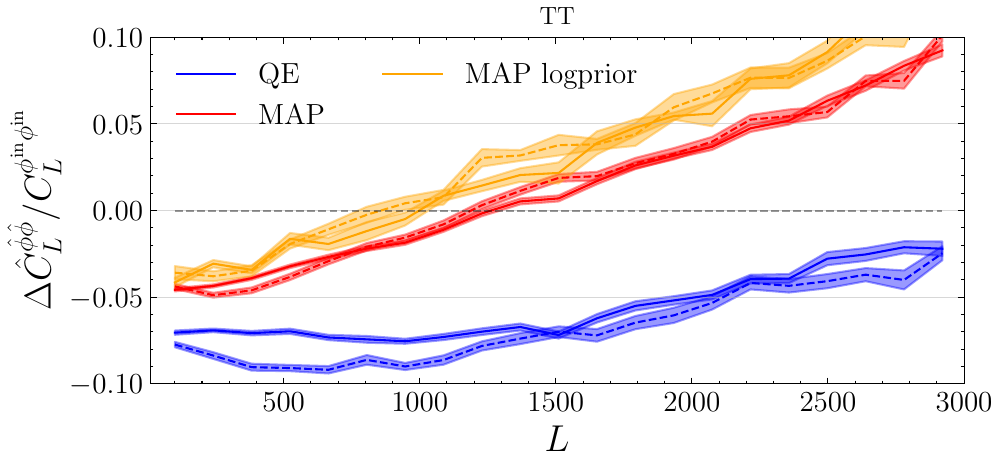}
        \includegraphics[width=\columnwidth]{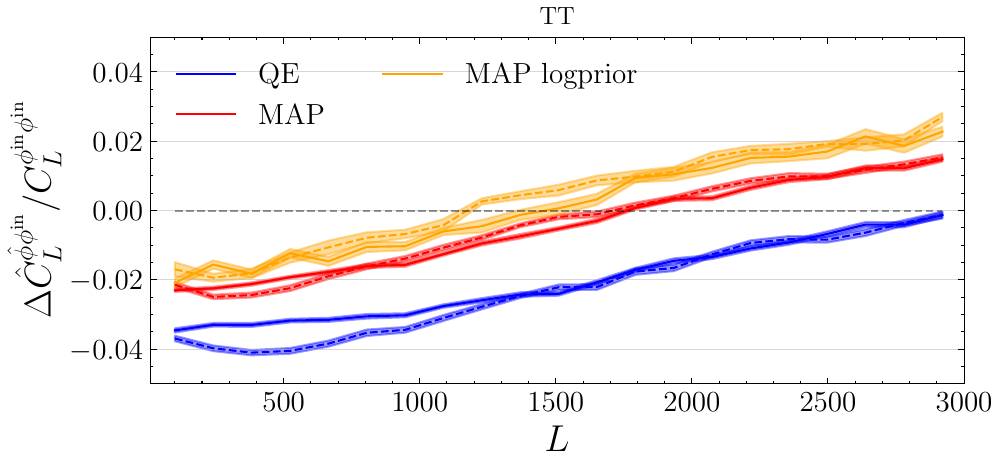}

    \caption{Comparison of the $N^{(3/2)}_L$ bias for different variants of temperature reconstructions. The solid lines use the $N$-body LSS map, while the dashed ones a lognormal map with a similar skewness (at the same input resolution). We show QE results in blue, MAP results with Gaussian prior in red, and MAP results with lognormal prior in orange. We see a small shift upwards between these two choices of priors. The biases found on lognormal maps match well the ones found on the LSS input map. The top shows the bias in the auto-spectrum, and bottom for the cross-spectrum to the true lensing.}
    \label{fig:n32autolognormal}
\end{figure}

In Figures \ref{fig:n32autolognormal} we focus on temperature-only reconstructions, and with the LSS non-linear part of the non-Gaussianity only, which has the strongest signature.  

For each set of curves, the dashed lines show the biases found in maps with the N-body LSS kappa convergence map, and the dotted lines the biases found in our simple lognormal simulated convergence maps with the same skewness. 

In blue and red we have results for the QE and baseline MAP, respectively, where the reconstruction is done as in the previous section, with a Gaussian prior on $\kappa$. Given the simplicity of the lognormal simulations, it is remarkable how well they reproduce the $N^{(3/2)}$ bias, particularly in the MAP case.


In orange, we show the curves obtained from the reconstructions using the lognormal prior as just discussed. Again, there are essentially no difference found between the N-body input (orange dashed) or lognormal input (orange dotted). Both curves shifts up slightly, reducing slightly the bias on most signal dominated scales, but remain qualitatively very similar. 

The fact that the lognormal prior improves only mildly the Gaussian prior is due the fact that we are likelihood dominated, and therefore the MAP reconstruction is mainly data driven.

\subsection{Bias Hardening}

The small-scale CMB temperature is strongly contaminated by foregrounds, limiting the use of more modes to perform CMB lensing reconstruction. Recently, foreground-mitigating lensing reconstruction methods have been developed, allowing more robust and powerful CMB lensing measurements.

 In particular, bias-hardening \citep{Namikawa_2013, Osborne_2014, Sailer_2020}, deprojecting the response of the CMB lensing QE to a foregrounds QE, have been used in recent data analyses to extract cosmological parameters from observations. 

At the likelihood level, we can derive the bias-hardened QE CMB lensing estimator by looking for modulations in the observed map, going beyond the CMB and noise, that could be attributed to point sources. 

We imagine to model our data in pixel space as

\begin{equation}
   X^{\rm dat} = \Beam \Da X + \Beam S+n\ ,
\end{equation}

where $S$ is some source field. From this equation, the pixel-pixel covariance is

\begin{equation}
    C \equiv \langle X^dX^{d,\dagger} \rangle = \mathrm{Cov}_{\vec{\alpha},S^2}=\Beam \Da C^{\mathrm{unl}}\Da^{\dagger}\Beam^{\dagger}+\Beam S^2\Beam^{\dagger}+N
\end{equation}

assuming no cross-terms between CMB, the source, or the noise terms. We can see that at the covariance level the source term induces a variance that is larger than the expected one from just experimental noise.

The log-likelihood becomes then

\begin{equation}
    \mathcal{L} \equiv \ln L(X^{\rm dat}|\vec{\alpha}, S^2) = -\frac{1}{2}X^{\rm dat} \cdot \mathrm{Cov}_{\vec{\alpha},S^2}^{-1}X^{\rm dat}-\frac{1}{2}\det \mathrm{Cov}_{\vec{\alpha},S^2}\ .\label{eq:likeS2}
\end{equation}

To derive the bias-hardened estimator, we adopt a method akin to the standard quadratic estimator. We start by nulling the gradients for both the lensing potential $\phi$ and the source term variance $S^2$, expanding around $(\phi, S^2) = (0, 0)$ using a first-step Newton iteration:

\begin{equation}
    \vec{G}_{\phi,S^2}\approx \vec{G}_{0, 0}+F_{0, 0}\begin{bmatrix}
\phi\\S^2
\end{bmatrix}\approx 0\rightarrow \begin{bmatrix}
\phi\\S^2
\end{bmatrix} \approx F_{0, 0}^{-1} \vec{G}_{0, 0}\ ,
\end{equation}

where the total gradient of the log-likelihood $\vec{G} = [\frac{\partial \mathcal{L}}{\partial \phi}, \frac{\partial \mathcal{L}}{\partial S^2}]^T$ is approximated by its value at $(\phi, S^2) = (0, 0)$, $\vec{G}_{0, 0}$, and a curvature matrix $F_{0, 0}$. The latter, encapsulates the responses $R^{ab}$ of the estimator for $a$ to the presence of $b$, with $a,b \in{\phi, S^2}$.

\begin{figure}
    \centering
    \includegraphics[width=\columnwidth]{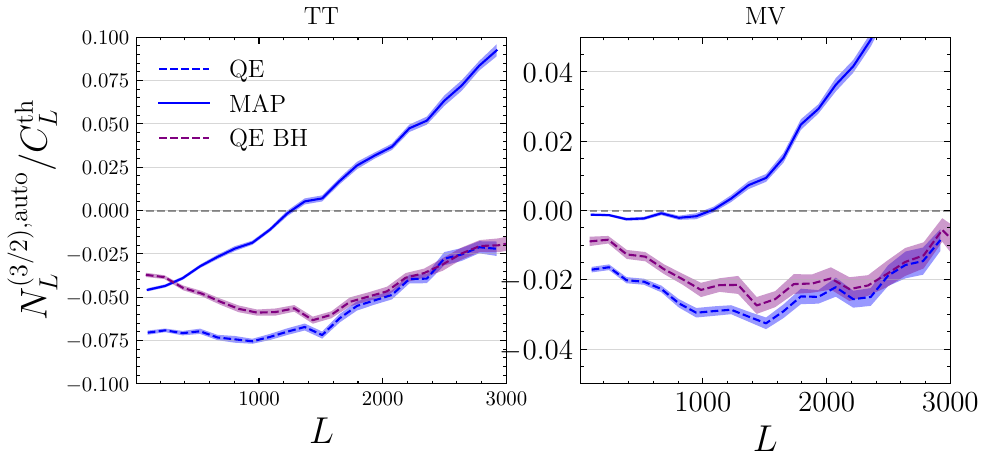}
    \caption{Auto-spectrum $N_L^{(3/2)}$ biases induced by the non-Gaussian LSS-only map on several lensing reconstructions; in dashed we have the QE, in blue, and the bias hardened QE, in purple. The solid line is the MAP. The QE BH is able to mitigate the bias on large scales on a similar level to the MAP estimator, at a small cost in signal to noise compared to the QE. On the left panel, results with temperature data only, while and on the right including polarization too.}
    \label{fig:n32autobh}
\end{figure}

\begin{figure}
    \centering
    \includegraphics[width=\columnwidth]{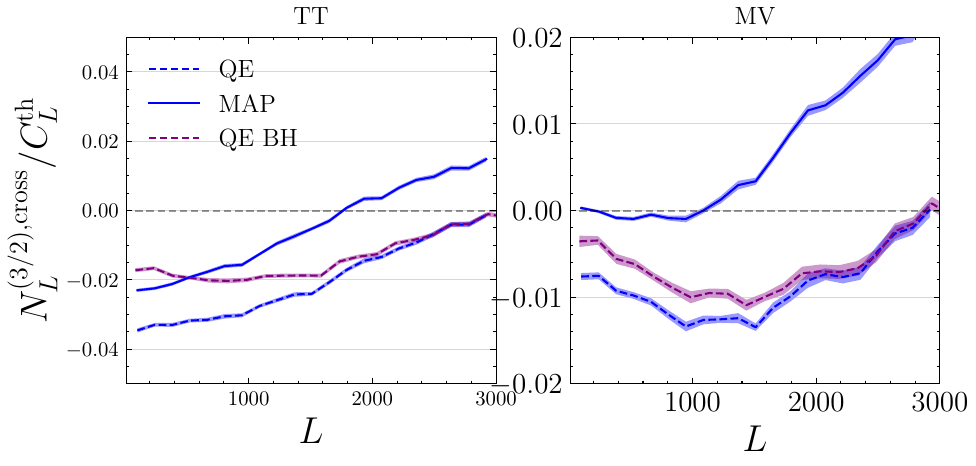}
    \caption{Similar to Figure \ref{fig:n32autobh} but for the bias in cross-correlation to the true lensing. After bias-hardening, the quadratic estimator also shows a somewhat reduced $N_L^{(3/2)}$ bias.}
    \label{fig:n32crossbh}
\end{figure}

In Figures \ref{fig:n32autobh}, \ref{fig:n32crossbh} we show results for the \nlth bias from our simulations using a bias-hardened estimator against a point source. We can see that, compared to the QE, the bias hardening estimator mitigates the impact of the non-Gaussian bias, confirming analytical expectations of~\citep{fabbianCMBLensingReconstruction2019a}. This suggests a bias-hardened MAP-solution might also be helpful, which just adds a prior to Equation \ref{eq:likeS2}, which is work in progress.

\section{Impact of \nlth for cosmological analyses}
We now turn to the impact of the bias for parameter inference.

\subsection{Bias in the lensing amplitude}

In this subsection we estimate the bias in the amplitude of CMB lensing power spectrum due to \nlth. We assume that the covariance matrix of the CMB lensing power spectrum is diagonal and is given by 
\begin{equation}
    \mathbf{C}_{LL'} =  \frac{2 \delta_{LL'}}{(2L +1) f_{\rm sky}} \left(C_{L}^{\rm \phi\phi} + N_L^{(0)} +  N_L^{(1)}\right)^2 \, .
\end{equation}

We report the signal to noise ratio of the CMB lensing power spectrum between multipoles $L_{\rm min}$ and $L_{\rm max}$ as
\begin{equation}
    \mathrm{SNR} = \sqrt{\sum_{L_{\rm min}}^{L_{\rm max}} C^{\phi\phi}_L \mathbf{C}_{LL'}^{-1}  C^{\phi\phi}_{L'}} \ . \label{eq:snr}
\end{equation}

This signal to noise ratio is exactly the square root of the Fisher information matrix on the amplitude of the CMB lensing spectrum. The inverse of this SNR thus gives the expected constraint on the amplitude of the CMB lensing spectrum if we only vary this parameter.

\begin{figure}
    \centering
    \includegraphics[width=\columnwidth]{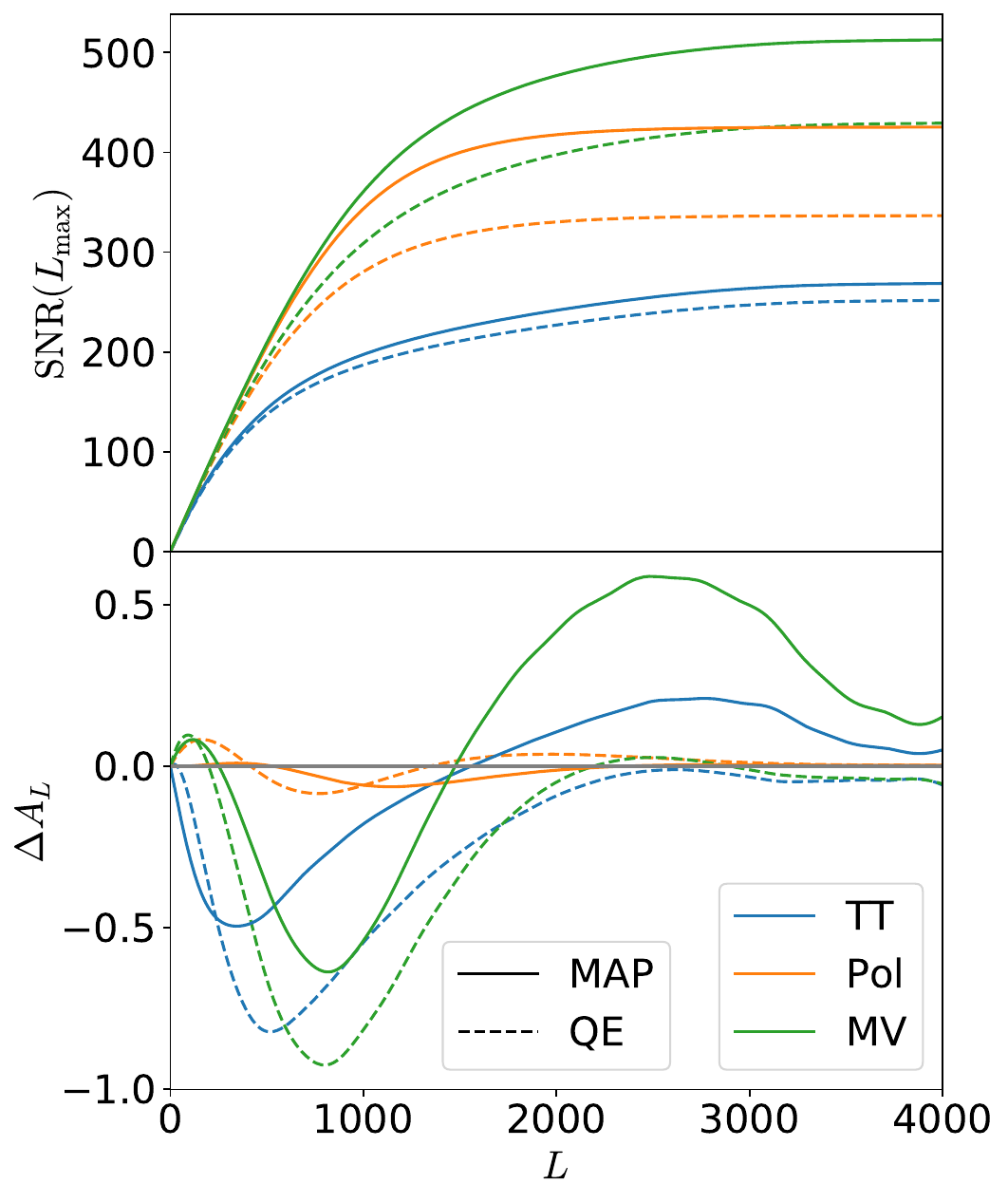}
    \caption{\textit{Upper panel:} Signal to noise ratio on the lensing power spectrum as a function of the maximum scale considered. We show the QE in dashed lines and the MAP in plain lines. The temperature only, polarization only and minimum variance are respectively in blue, orange and green.
    \textit{Lower panel:} Bias on the lensing power spectrum amplitude as a function of scale, for the different estimators considered, including both the LSS and post-Born effects in \nlth.}
    \label{fig:snr_bias}
\end{figure}

We show on the upper panel of the Figure~\ref{fig:snr_bias} the signal to noise ratio as a function of the maximum lensing scale considered. We see that for the noise levels considered, the polarization estimator brings the most information, but the temperature estimator cannot be neglected. The MAP estimator performs particularly better than the QE especially in the polarization channel. We see that for the polarization, the information gain saturates above the scale $L\sim1500$, while it still grows up to $L\sim3000$ for the temperature and minimum variance estimators. This is an important point as future lensing analyses, including cross-correlations with large scale structure, will be able to push the signal to noise to high significances, provided that systematic and modelling treatments are accurately handled.

We estimate the bias in the CMB lensing amplitude as in \cite{Amara:2007as}:
\begin{equation}
    \label{eq:bias_alens}
    b(A_L) =\frac{1}{\mathrm{SNR}^{2}} \sum_{L_{\rm min}}^{L_{\rm max}} \Delta A_L \, ,
\end{equation}
with
\begin{equation}
    \Delta A_L = \nlth  \mathbf{C}_{LL'}^{-1} C^{\phi\phi}_{L'} \ . 
\end{equation}

We show in the Table~\ref{tab:bias_alens} the bias on the lensing amplitude for the different estimators with $L_{\rm min}=0$, $L_{\rm max}=4000$, including both the LSS and post-Born effects in \nlth. We see that the MAP seems to be almost unbiased on temperature and polarization, but has an almost $1\sigma$ bias on the minimum variance combination.

We show in the lower panel of the Figure~\ref{fig:snr_bias} the bias $\Delta A_L$ as a function of scale. As it can be seen from this figure, the \nlth introduce a scale dependent bias. In some cases, the \nlth bias flip signs. This lowers the sum in Eq.~\ref{eq:bias_alens}, and thus gives a low bias despite a large absolute value for some given scales. This is the case for the QE in polarization, which has a negative bias for $L\in [300, 1200]$ and positive outside, and for MAP in temperature, which is negative for $L<1500$ and positive above. 
This scale dependent bias, which seems to be more important for the MAP than for the QE, could bias the measurements of cosmological parameters that are sensitive to the shape of the lensing spectrum. 
However, as the total bias almost cancels-out, the parameters combinations sensitive only to the amplitude of the lensing power spectrum should in principle be immune from shape changing effects.

\begin{table}
    \centering
        \begin{tabular}{ c c c c } 
            \hline
            \hline
             Bias  & TT & Pol & MV \\
            \hline
            QE & $-3.7 \sigma$ &  $0.07 \sigma$ &  $-2.13 \sigma$ \\
            MAP &$-0.36 \sigma$ & $-0.12 \sigma$ & $0.93 \sigma$ \\
            \hline
            \hline
        \end{tabular}
    \caption{Bias on the lensing power spectrum amplitude, estimated from the Fisher matrix, given in terms of number of sigmas, including both the LSS and post-Born effects in \nlth.}        
    \label{tab:bias_alens}
\end{table}

\subsection{Cosmological parameters}
\label{sec:params}

We now estimate the biases on cosmological parameters if one does not take into account the $\nlth$ bias in the analysis. We focus on the sum of the neutrino masses, as this is a key science goal of CMB-S4. We consider the temperature only (TT), the polarization only (Pol), and the combined (MV) estimators, for both QE and MAP.
Our fiducial cosmology is the Planck FFP10 cosmology\footnote{The CAMB parameter file used to generate the spectra can be found in \url{https://github.com/carronj/plancklens/blob/master/plancklens/data/cls/FFP10_wdipole_params.ini}}, with  one massive neutrino of $0.06 \, \rm eV$.

Our analysis combines the CMB lensing spectra, the primary (unlensed) CMB spectra and the BAO. We assume that all three sets of observables are independent, so we can sum their log-likelihood. 
For the BAO we consider a DESI configuration, following the recipes of \cite{Font-Ribera:2013rwa, DESI:2016fyo}. 
For the CMB likelihood, we consider a CMB-S4 experiment, with a beam $\theta_{\rm FWHM} = 1 \,\rm arcmin$, a temperature noise $\Delta T = 1 \, \mu \rm K{\text -}arcmin$, a polarization noise $\Delta P = \sqrt2 \, \mu \rm K{\text -}arcmin$ and a sky fraction $f_{\rm sky} = 0.4$.
For the primary CMB likelihood, we consider unlensed spectra $C_\ell^{TT}, C_\ell^{TE}$ and $C_\ell^{EE}$, between multipoles 30 and 3000, and assume that these spectra follow a Gaussian likelihood, with Gaussian covariance matrix.
We discuss in Appendix \ref{app:beck} the impact of using the unlensed CMB spectra or the lensed CMB spectra. We show there that the bias on the marginalized cosmological parameters are similar as long as we correctly model the non-Gaussian covariance of the lensed CMB and the correlations between the lensed CMB and the reconstructed lensing potential.


The Gaussian CMB lensing power spectrum likelihood is
\begin{equation}
    \label{eq:likecpp}
    -2 \ln \mathcal{L}(\theta) = \left(\hat C_L^{\rm \phi\phi} - C_L^{\rm th}(\theta)\right) \mathbf{C}_{LL'}^{-1}  \left(\hat C_{L'}^{\rm \phi\phi} - C_{L'}^{\rm th}(\theta)\right) \; ,
\end{equation}
where $\theta$ is the set of cosmological parameters being sampled. The covariance matrix is assumed to be diagonal
\begin{equation}
     \mathbf{C}_{LL'} =  \frac{2 \delta_{LL'}}{(2L+1) f_{\rm sky}} \left(C_{L}^{\rm \phi\phi} + N_L^{(0)} +  N_L^{(1)}\right)^2 \, ,
\end{equation}

where $C_{L}^{\rm \phi\phi}$, $ N_L^{(0)}$ and $N_L^{(1)}$ are evaluated in the fiducial cosmology. 
We consider lensing multipoles between 10 and 3000, and assume that the lensing field has been reconstructed with CMB multipoles between 10 and 4000 for both temperature and polarization channels.
For the different estimators and different configurations of \nlth bias (LSS, PB or Total ), we generate mock data vectors. These mock data vectors assume that the $N_L^{(0)}$ bias can be perfectly subtracted, and that the $N_L^{(1)}$ cosmology dependence is perfectly modelled. 
Our mock data vectors are given by 
\begin{equation}
\label{eq:cpp_datvec}
    \hat C_L^{\rm \phi\phi} = C_{L}^{\rm \phi\phi}(\theta^{\rm fid}) + N_L^{(1)}(\theta^{\rm fid}) + \frac{C_{L}^{\rm \phi\phi}(\theta^{\rm fid})}{C_{L}^{\rm \phi\phi, \rm sim}} N_L^{(3/2)} \, .
\end{equation}
where the $N_L^{(3/2)}$ bias has been estimated from simulations, as described in the previous sections, and the CMB lensing power spectrum and $N_L^{(1)}$ bias are taken at the fiducial cosmology. 
To cancel part of the realization variance of the simulation we divide \nlth by the lensing power spectrum of the simulation, bin this ratio in 19 multipole bins between 10 and 3810, and fit a spline to these points, weighting by the inverse variance of each bin.
We then multiply this spline by our fiducial lensing spectrum. This also allows to partially take into account the difference in the cosmology between the simulation used to estimate the \nlth (which do not have massive neutrinos, see Equation~\ref{eq:cosmodemuni}), and the  Planck FFP10 cosmology used to generate the data vector. We assume that higher order cosmology dependence of \nlth can be neglected.

To accelerate the computation of the theory vector $C_L^{\rm th}(\theta)$, we do not re-estimate the $N_L^{(1)}$ bias for the sampled cosmology. Instead we correct for the variations of the $N_L^{(1)}$ bias around the fiducial, at first order in $C_L^{\phi\phi}(\theta)$, following the procedure of \cite{plancklensing2018}. The theory vector is then 
\begin{equation}
\label{eq:cpp_nlone}
    C_L^{\rm th}(\theta) = C_L^{\phi\phi}(\theta) + N_L^{(1)}(\theta^{\rm fid}) + \frac{\partial N_L^{(1)}}{\partial C_{L'}^{\phi\phi}} \left[C_{L'}^{\phi\phi}(\theta) - C_{L'}^{\phi\phi}(\theta^{\rm fid})\right]
\end{equation}
where the $N_L^{(1)}$ correction matrix has been previously evaluated in the fiducial cosmology. 
We neglect the correction due to the variation of the response around the fiducial. We do not expect that this will impact the results. Indeed in the Gaussian case (i.e. without \nlth bias) we recover unbiased cosmological parameters.

In this analysis we do not model the reconstruction bias $N_L^{(0)}$ in the data vector, we assume it is perfectly subtracted. 
In a standard analysis one would use the realization dependent bias estimator $\text{RD-}N_L^{(0)}$, which makes the debiasing robust at first order to differences between the fiducial spectra assumed for the reconstruction, and the true CMB spectra of the maps. It was showed in \cite{legrand2022, legrand2023} that the realization dependent $\text{RD-}N_L^{(0)}$ allows for unbiased cosmological parameter estimates, for both the QE and the MAP estimators. However we note that this analysis was performed with a Gaussian lensing potential, contrary to the maps we are using here, but we assume that this will have a negligible impact on the $\text{RD-}N_L^{(0)}$ estimate.

We sample for seven cosmological parameters, namely $\ln(10^{10} A_{\rm s}), n_{\rm s}, \theta_{\rm MC}, \Omega_{\rm c} h^2, \Omega_{\rm b} h^2, \tau$ and $\sum m_\nu$. 
We include a strong Gaussian prior on $\tau$, the reionisation optical depth, assuming a cosmic variance limit of $\sigma_\tau= 0.002$.  

We rely on the CAMB cosmological Boltzmann code \cite{Lewis:1999bs,Howlett:2012mh}, and sample the posterior with adaptive, speed-hierarchy-aware MCMC sampler (adapted from CosmoMC) \cite{Lewis:2002ah,Lewis:2013hha}. We explore the posterior using the  Cobaya \cite{Torrado:2020dgo} and GetDist packages \cite{Lewis:2019xzd}.

\begin{figure}
    \centering
    \includegraphics[width=\columnwidth]{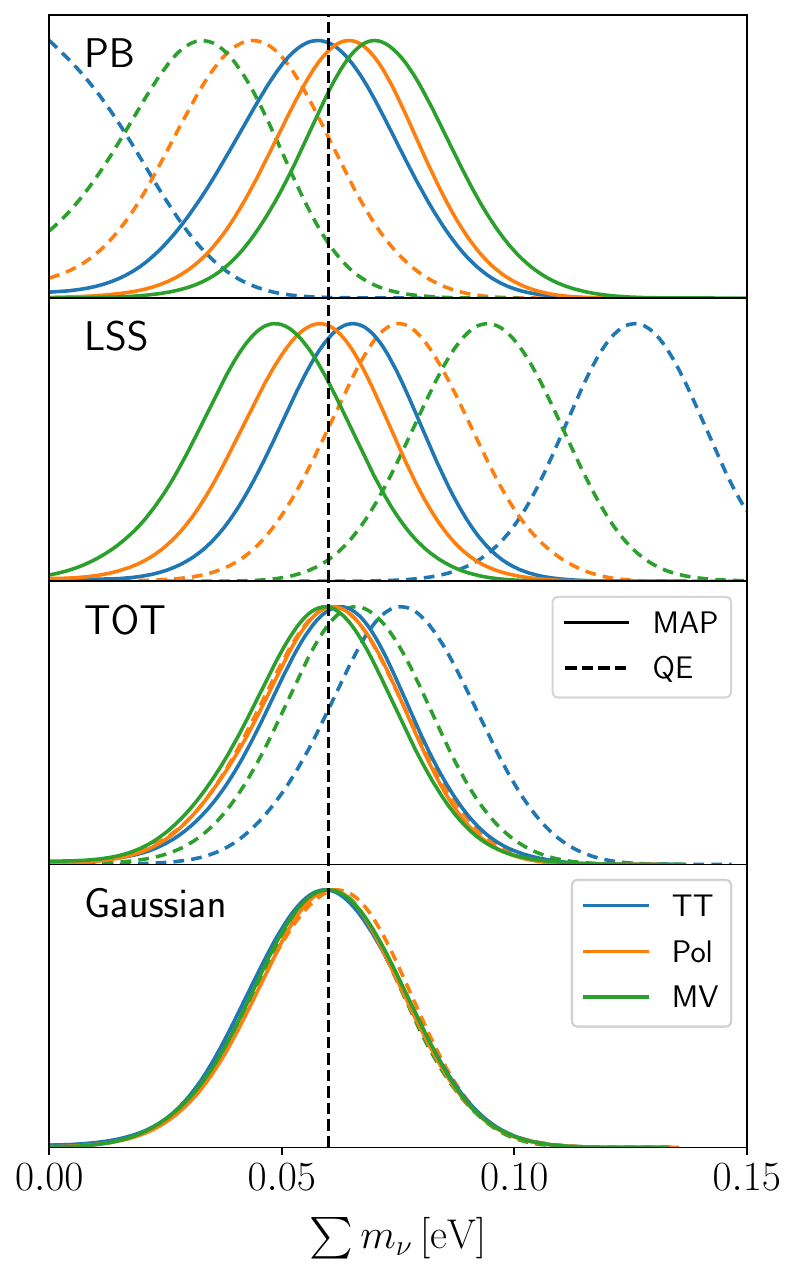}
    \caption{marginalized posteriors on the sum of the neutrino masses. Panels from top to bottom show respectively the impact of \nlth when considering the post-Born term only, the large scale structure term only, both terms together, or with no \nlth bias (fiducial, Gaussian case). The dashed lines are for the QE and the plain lines are for the MAP. Blue, orange and green lines show the temperature only, polarization only or minimum variance estimators. The fiducial cosmology assumes a sum of neutrino masses of 0.06 eV, showed as the vertical dashed line. We combine here the CMB-S4 likelihood with DESI-BAO, and we set a cosmic variance Gaussian prior on the optical depth to reionization.}
    \label{fig:mnu}
\end{figure}

The marginalized posterior distribution on the sum of the neutrino masses, for our different lensing estimators and source of \nlth bias, is shown in Figure~\ref{fig:mnu}. We show the impact of each term that contributes to the \nlth bias: the large scale structures non-Gaussianities (LSS) and the post-Born lensing. 
We confirm that taking into account only one of the two effects, either LSS or PB, will bias the neutrino mass estimates with the QE. We can see that in general, the MAP estimators are less impacted than the QE.
When looking at the total \nlth, we see that the MAP is unbiased, both for temperature and for polarization. The QE shows a $\sim 1\sigma$ bias with the temperature reconstruction, while it is unbiased in the polarization. 

We note also that as previously observed in the literature \cite{Allison:2015qca,legrand2022}, the marginalized constraints on the sum of the neutrino mass from the QE is not improved with the MAP. Indeed, even if, as we showed in the Figure~\ref{fig:snr_bias}, the SNR is increased with the MAP, there are some degeneracies between the cosmological parameters that prevent to reach the full statistical power of the MAP estimator on the marginalized constraints. The marginalized constraints also do not change much if we use the temperature, polarization or the minimum variance estimators, with a one sigma constraint of about $0.016 \, \rm eV$ for all cases.

\begin{figure}
    \centering
    \includegraphics[width=\columnwidth]{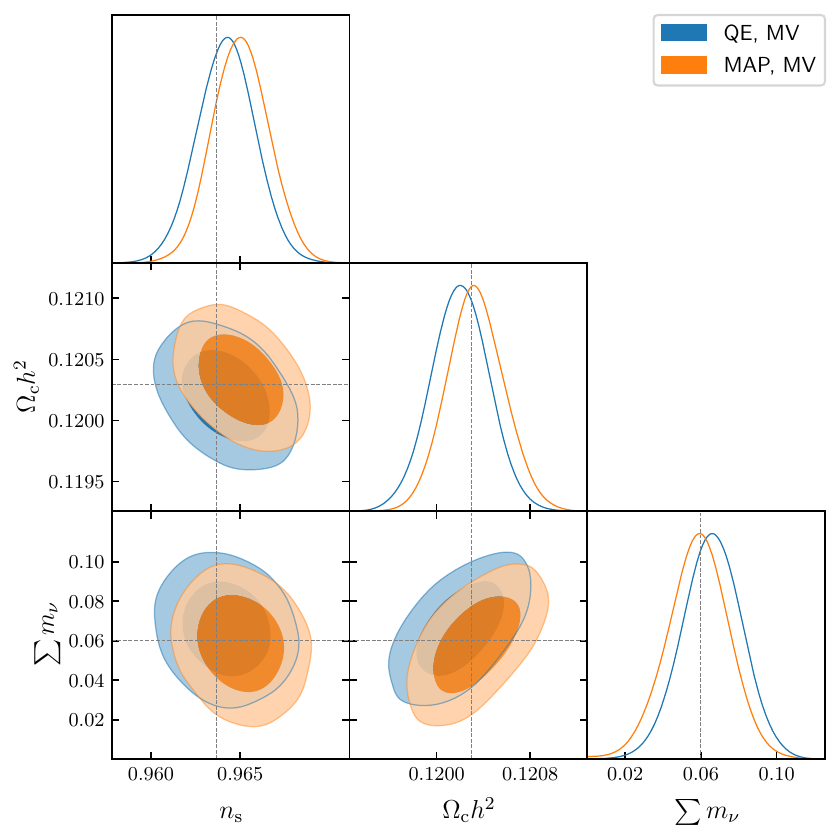}
    \caption{Posterior distribution for a subset of the sampled cosmological parameters. We show the reconstruction for the minimum variance estimator, for both QE (blue) and MAP (orange), considering the bias due to the total $\nlth$. The Figure~\ref{fig:trigmnu_full} in the Appendix shows the full posterior.}
    \label{fig:trigmnu}
\end{figure}

Figure~\ref{fig:trigmnu} shows the posterior distribution for the $n_s$, $\Omega_c h^2$ and $\sum m_{\nu}$ parameters, for the total \nlth bias with the minimum variance estimator. We see that even if the MAP is unbiased on the sum of the neutrino mass, it shows a slight $0.5 \sigma$ bias on the $n_s$ parameter. This bias could be sourced by the shape of the \nlth bias. Indeed, we showed in Figure~\ref{fig:snr_bias} that the \nlth creates a scale dependent bias. In particular, this could tilt the lensing power spectrum, and thus create a bias on the $n_s$ parameter. In comparison, the amplitude bias for the minimum variance QE is mostly negative, so it will not create a strong scale dependent bias, but rather an overall amplitude shift on the CMB lensing power spectrum.

In Appendix~\ref{app:beck}, we compare our results for the QE with the previous results from \cite{beckLensingReconstructionPostBorn2018a}. They obtained a larger bias on the sum of the neutrino masses. This higher bias comes from the fact that they consider lensed CMB spectra in the likelihood but neglect the non-Gaussian covariance due to lensing and the correlations between CMB and the QE, as described in \cite{Peloton:2016kbw}.
In Appendix~\ref{app:beck} we show that using lensed CMB spectra with the correct covariance results in a bias on the sum of the neutrino mass that is similar to the one obtained when considering unlensed spectra and neglecting the non diagonal correlations. 

\section{Conclusions}
Future CMB surveys will detect the CMB lensing auto-spectrum well over $100\sigma$ significance, and will enable powerful cosmological constraints on the structure formation history and on the sum of neutrino masses. 
Great care is required before interpreting these measurements at this high level of sensitivity. Standard quadratic estimators are only guaranteed to be unbiased for purely Gaussian fields and under idealized conditions.
When the Gaussianity assumptions fail, future CMB surveys such as CMB-S4 can deliver a biased lensing reconstruction induced by the bispectrum of the lensing potential which, in turn will affect constraints on the sum of neutrino masses~\citep{beckLensingReconstructionPostBorn2018a}.

In this work we studied the induced non-Gaussian biases in the auto-spectrum of the CMB lensing potential reconstructed through maximum a posteriori estimators. 
By using state-of-the art simulations that include non-Gaussian lensing deflections induced by nonlinear matter clustering as well as post-Born lensing corrections, we found that for scales relevant for a CMB-S4 like configuration the non-Gaussian induced bias is mitigated compared to the QE one. When including additional effects due to the post-Born rotation we do not find a significant difference.


When performing a $\Lambda \rm{CDM}$ cosmological forecast on the sum of neutrino masses through a full MCMC analysis, we found that contrary to the QE, the MAP estimator does not suffer from biases even if \nlth biases are not specifically accounted in the fitting of the data vector. 
This shows that cosmological analyses using a MAP reconstructed lensing potential should in principle be more robust than those based on the standard QE. 
However, if not correcting for the bias, caution should still be taken when investigating extensions to $\Lambda \rm{CDM}$, as the bias has not fully disappeared, but is only to weak to shift the maximum posterior point, and this could change in other models. 

We also tested simple modifications to the MAP estimator to take into account the non-Gaussian statistics of the deflection field. Using the prior of a lognormal field instead of a Gaussian field on the input lensing convergence further mitigates slightly the bias. 
We have also found that the bias itself found on our N-body simulations is reasonably well reproduced by much simpler lognormal simulations. Generally, differences between reconstructions using the Gaussian or lognormal prior are small, and the lensing reconstructions of similar quality, both on $N$-body and  lognormal simulated inputs. Lognormal simulations are effective in replicating non-Gaussian bias effects from large-scale structure non-linearity. They could be leveraged to predict and subtract biases with sufficient accuracy even when applied to more sophisticated models affecting the full probability density function of the projected matter distribution, such as baryonic effects or extended cosmologies. On the other hand we have found that the lognormal distribution is not a good approximation when including post-Born effects, suggesting that the lognormal approximation is insufficient for recovering the shape of the post-Born induced bispectrum (and same would be for highly non-linear fields, necessitating alternative approaches \citep{das2006, fabbian2018, barthelemy2020}). Nevertheless, using state-of-the art simulations one could learn realistic probability density functions of fields using machine learning techniques to generate new samples or even include alternative priors in our MAP formulation (e.g. \cite{rouhiainen2021normalizing}).

Our results allow us to conclude that the non-Gaussianity of CMB lensing deflections should not present a major challenge to CMB lensing science in the near future: the bias, small to start with, is mitigated, and the baseline MAP reconstructions with Gaussian prior remain very close to optimal.

In the era of precision-cosmology, additional information will come from the combination and cross-correlation of CMB lensing measurements and  multiple large-scale structure observables, such as cosmic shear and galaxy clustering, from DESI, Euclid, and Rubin. 
It is known that in this case, for QE-based reconstructions, the \nlth bias in the CMB lensing cross-correlations has a stronger impact depending on the specific tracer and redshift considered in the analysis \cite{fabbianCMBLensingReconstruction2019a}. This is left for future work.

\begin{acknowledgments}
We thank Antony Lewis for comments. OD, SB, LL and JC acknowledge support from a SNSF Eccellenza Professorial Fellowship (No. 186879).
This work was supported by a grant from the Swiss National Supercomputing Centre (CSCS) under project ID s1203. Part of the computations were performed at University of Geneva using Baobab HPC service. GF acknowledges the support of the European Research Council under the Marie Sk\l{}odowska Curie actions through the Individual Global Fellowship No.~892401 PiCOGAMBAS.
\end{acknowledgments}

\nocite{*}


\appendix

\section{Mean field impact \label{sec:meanfieldapp}}
\newcommand{\mbeq}{\overset{!}{=}}
The MAP estimator tries to find the best estimate of the CMB lensing potential by nulling the total gradient

\begin{equation}
    g^{\rm{tot}}(\phi) = g^{\rm{QD}}(\phi)+g^{\rm{MF}}(\phi)+g^{\rm{PR}}(\phi)
\end{equation}

Ignoring the mean field contribution $g^{\rm{MF}}$, as we often do for simplicity, one is effectively nulling instead

\begin{equation}
    g^{\rm{tot,noMF}}(\hat{\phi}) = g^{\rm{QD}}(\hat{\phi})+g^{\rm{PR}}(\hat{\phi}) \mbeq 0\ .
\end{equation}
The neglect of this gradient piece is equivalent to maximizing a slightly different, but still well-defined likelihood function. Hence the iterative procedure converges without problems. The actual total gradient at this point $\hat \phi$ is not zero but

\begin{equation}
   g^{\rm{tot}}(\hat{\phi}) = g^{\rm{MF}}(\hat{\phi}) \neq 0\ .
\end{equation}

Early indications from reconstructions from polarization suggested this term was small~\cite{Belkner:2023duz}. Physically, this is because this term accounts for the anisotropies in the noise maps induced by delensing, and these anisotropies trace for the most part the magnification part of the lensing signal, rather than the shear-like signal that the most powerful $EB$ polarized estimator is sensitive to. Delensing affects the local noise levels by changing areas according to the local magnification, given by $ 1 -2\kappa$ to first order, and the mean-field removes the contribution of this noise anisotropy to the quadratic gradient piece.

\begin{figure}[ht]
    \centering
\includegraphics{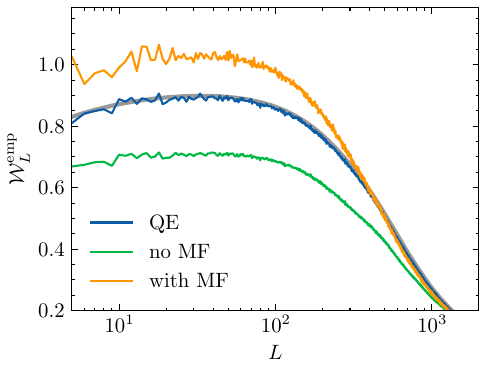}
    \caption{Comparison of the empirical normalization of output lensing maps, after a few iterations, for our temperature-based CMB-S4 like reconstructions. In blue we show the case of the QE estimator, in green for the MAP estimator if ignoring the mean field contribution, and in orange when including it, using an estimate from a finite number of Monte-Carlo at each step.}
    \label{fig:wf_mean_field}

\end{figure}

\begin{figure}[ht]
    \centering
    \includegraphics{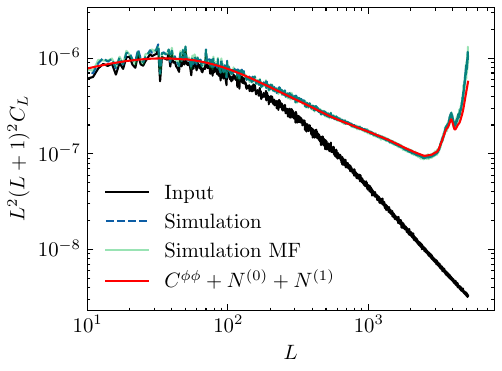}
    \caption{Lensing spectrum from a MAP temperature reconstruction, after dividing by the empirical normalization of the lensing map (blue). In red the naive theory prediction, obtained from a iterative $N^{(0)}_L$ and $N^{(1)}_L$ calculation. We get a similar result if we account for the mean field, justifying the usage of an empirical normalisation when not account for a mean-field.}
    \label{fig:normalized_recnoise}
\end{figure}

It turns out that that for reconstructions from temperature, that takes substantial contribution from the magnification-like lensing signals, the mean field contribution is much larger.
If the mean-field at the solution $\hat \kappa_{LM}$ were exactly proportional to $\hat \kappa_{LM}$ as suggested by the argument above, the mean-field contamination of $\hat \kappa$ acts simply as a rescaling of the output map. This would have no impact on the cross-correlation coefficient of the reconstruction with the input (the `quality' of the reconstruction), only on its normalization
\begin{equation}
	\mathcal W^{\rm emp}_L \equiv \frac{C_L^{\hat{\phi}\phi}}{C_L^{\phi\phi}}
\end{equation} (the empirical `Wiener-filter' of the reconstruction). The shift seen on temperature reconstruction for our CMB-S4-like configuration is shown on Fig.~\ref{fig:wf_mean_field}, and is substantial. Not accounting for the mean field lowers the Wiener-filter curve. This is because the magnification follows $\kappa$ but with a minus sign.
In Fig.~\ref{fig:normalized_recnoise} we show that after applying the empirical normalization, for the case of mean-field subtraction/no subtraction, we find a power spectrum in line with naive expectations,
\begin{equation}
C_L^{\hat \phi \hat \phi} \sim C_L^{\phi\phi} + N_L^{(0)}	+ N_L^{(1)}
\end{equation}
where the lensing biases are calculated using the partially delensed CMB spectra. The change in cross-correlation coefficient between the two maps is tiny.
This gives support to the idea that the bulk of the effect is only a rescaling. Hence, in this paper we proceed using the empirical normalization, and neglecting the delensed-noise mean-field altogether. A more careful study of the mean-field is ongoing.

\section{About the impact of non-Gaussian lensing correlations on the cosmological parameter biases}
\label{app:beck}

In the previous study of \cite{beckLensingReconstructionPostBorn2018a} (herefater Beck18), the likelihood was neglecting the correlations between the reconstructed lensing field and the CMB spectra. 
In the results we present in the Section~\ref{sec:params} above, assume we can neglect these correlations by using unlensed CMB spectra un the likelihood.
We will now make systematic study of this assumption, and evaluate the impact of using the lensed or unlensed CMB spectra, and the impact of the non-Guassian covariance on the final cosmological constraints.

We will consider four different likelihood configurations:
\begin{enumerate}
    \item The Beck18 likelihood
    \item The Beck18 likelihood but with unlensed CMB spectra
    \item The likelihood we introduced in Section~\ref{sec:params}, which uses unlensed CMB spectra and assumes no correlations between the lensing power spectrum and the CMB fields
    \item A complete likelihood including the full non-Gaussian correlations between the lensed CMB spectra and the lensing potential, following \cite{Peloton:2016kbw}
\end{enumerate}

The Beck18 CMB likelihood is
\begin{equation}
    \label{eq:becklike}
    - 2 \log \mathcal{L} (\theta | \mathbf{\hat C}) = \sum_\ell (2\ell+1) f_\mathrm{sky} \left(\ln \frac{|\mathbf{C}_\ell|}{|\mathbf{\hat C}_\ell|} + \mathbf{C}_\ell^{-1} \mathbf{\hat C}_\ell - 3 \right)
\end{equation}
where the theoretical covariance matrix $\mathbf{C}$ and the data covariance matrix $\mathbf{\hat C}$ are constructed with the lensed CMB spectra and are given by
\begin{equation}
    \label{eq:covclsbeck}
    \mathbf{C}_\ell = 
    \begin{pmatrix}
        C_\ell^{TT} + N_\ell^{TT} & C_\ell^{TE} & C_\ell^{T\phi} \\
        C_\ell^{TE} & C_\ell^{EE} + N_\ell^{EE} & 0 \\
        C_\ell^{T\phi}  & 0 & C_\ell^{\phi\phi} + N_\ell^{\phi\phi}
    \end{pmatrix}
\end{equation}
For the theoretical covariance $\mathbf{C}_\ell$, the CMB lensing noise is $ N_\ell^{\phi\phi} = N_\ell^{(0)} + N_\ell^{(1)}$. For the mock data covariance we include the $\nlth$ bias estimated from the simulations as
\begin{equation}
    N_L^{\phi\phi} = N_L^{(0)} + N_L^{(0)} + \frac{C_{L}^{\rm \phi\phi, fid}}{C_{L}^{\rm \phi\phi, \rm sim}} N_L^{(3/2), \rm sim} \; .
\end{equation} 
In this configuration, and in order to reproduce the results of Beck18, we do not take into account the cosmological dependence of the $N_L^{(0)}$ and $N_L^{(1)}$ biases, and we keep them fixed to their fiducial value. 

The Beck18 likelihood with unlensed spectra is the same as above but replaces the lensed CMB spectra by their unlensed version.


The likelihood which takes into account the non-Gaussian covariance introduced in \cite{Peloton:2016kbw} has the following data vector:
\begin{equation}
    \label{eq:datvecpeloton}
    \mathcal{C}_\ell = \left( C_\ell^{TT}, C_\ell^{EE}, C_\ell^{TE}, C_\ell^{\phi\phi} + N_\ell^{(1)} \right) \, . 
\end{equation}
For the theory vector $\mathcal{C}_\ell(\theta)$ we vary the theoretical $N_\ell^{(1)}$ at first order in the CMB lensing spectra when sampling the cosmology, like in Eq.~\ref{eq:cpp_nlone}. The mock data vector $\mathcal{\hat C}_\ell$ includes the \nlth bias, which is not modelled by the theory vector.
The full likelihood is then 
\begin{equation}
    \label{eq:full_like}
    -2 \log \mathcal{L}(\theta) = \left( \mathcal{\hat C}_\ell - \mathcal{C}_\ell(\theta) \right) \mathbf{C}_{\ell \ell'}^{-1} \left( \mathcal{\hat C}_{\ell '} - \mathcal{C}_{\ell '}(\theta) \right)
\end{equation}
The covariance matrix $\mathbf{C}_{\ell \ell'}$ includes the non-Gaussian terms described in \cite{Peloton:2016kbw}, such as the correlations between the CMB spectra due to lensing, and the correlation between the reconstructed lensing field and the CMB.

In all likelihood scenarios we generate the data vector using the same cosmology of Beck18, reproduced in Table~\ref{tab:beckcosmo}, contrary to the analysis in the Section~\ref{sec:params} where we used the FFP10 cosmology. Notably, we now consider massless neutrinos, instead of the minimal mass normal hierarchy considered before.
Moreover, contrary to the main analysis, the likelihoods used here do not include external BAO constraints. We follow again Beck18 and assume tight Gaussian priors on the parameters, except for the sum of neutrino mass which assumes a flat prior, given on Table~\ref{tab:beckcosmo}. 

\begin{table}
    \centering
        \begin{tabular}{ c c } 
            \hline
            \hline
            $\Omega_\mathrm{b} h^2$ & $ 0.02225 \pm 0.00016$ \\
            $\Omega_\mathrm{c} h^2$ &  $0.1198 \pm 0.0015$ \\
            $\tau$ & $0.058 \pm 0.012$ \\
             $ \log(10^{10} A_\mathrm{s})$ &$ 3.094 \pm 0.034$ \\
             $n_\mathrm{s}$ &$ 0.9645 \pm 0.0049$\\
             $100\theta_\mathrm{MC}$ & $1.04077 \pm 0.00032$\\
             $\sum m_\nu \; [\mathrm{eV}] $ & $[0, 300]$ \\
            \hline
            \hline
        \end{tabular}
    \label{tab:beckcosmo}
    \caption{Fiducial cosmological parameters used in this section, following Beck18, together with their $1\sigma$ Gaussian prior, or uniform parameter bound.}    
\end{table}

\begin{figure}
    \centering
    \includegraphics[width=\columnwidth]{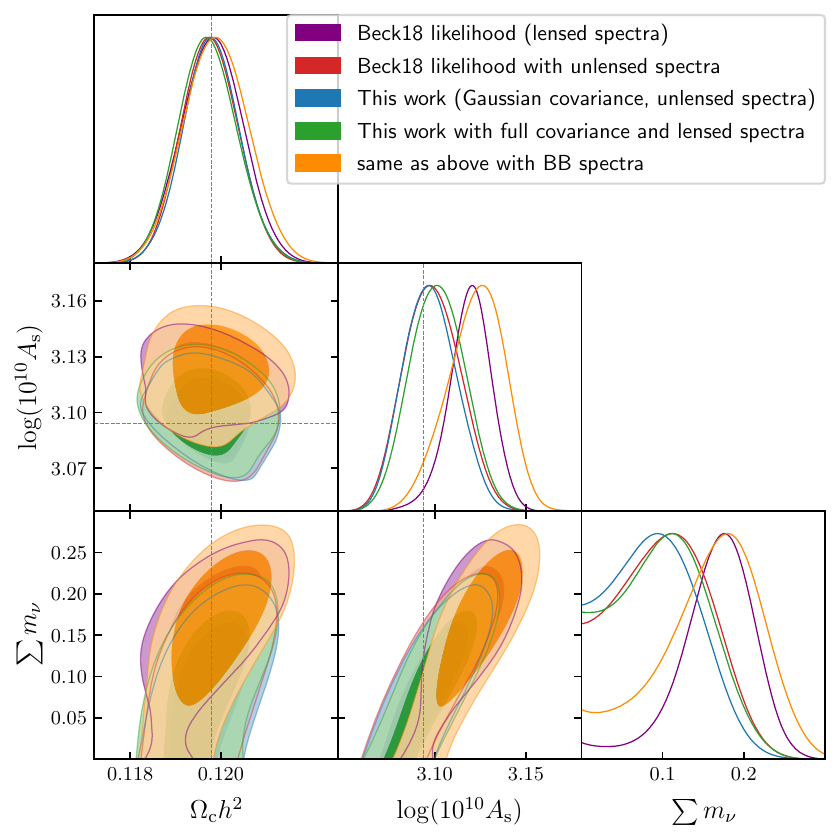}
    \caption{marginalized contours for three cosmological parameters, for the QE minimum variance, in the presence of the LSS only \nlth bias. We show the results for the Beck18 likelihood with the lensed CMB spectra (in purple) or with the unlensed CMB spectra (in red). We compare with our likelihood used for the main results in Section~\ref{sec:params} (in blue). We also show in green the likelihood which includes the full non-Gaussian covariance of the lensed CMB spectra, and the correlations between the reconstructed lensing spectrum and the CMB. We see that correctly taking into account these correlations gives almost the same posterior as considering unlensed CMB spectra and neglecting correlations. In orange we add the lensed BB spectrum in the data vector, taking into account its correlations with the lensed CMB and the reconstructed lensing potential.
    } 
    \label{fig:beck}
\end{figure}

We perform the MCMC samplings like in the main analysis, and we now discuss the results obtained for the QE minimum variance estimator, with the LSS only term in the $\nlth$ bias.
We show in the Figure~\ref{fig:beck} the constraints for three cosmological parameters $\Omega_c h^2$, $\ln(10^{10} A_s)$ and $\sum m_\nu$. The purple contours are the one obtained with the likelihood of Beck18, with the lensed CMB spectra. We retrieve the same results as in Beck18 (see their Figure 11), with a posterior estimate on the sum of neutrino masses peaking at $0.18\,\rm eV$. 
The red contours show the posterior with the Beck18 likelihood with unlensed CMB spectra in the data and theory vectors. 
The blue contours are for our likelihood, using unlensed CMB spectra, for Beck18 cosmology and priors. In practice, the main difference here with the Beck18 unlensed likelihood is that we take into account the variation of the $N_L^{(1)}$ bias when sampling the cosmology, while it is not the case for the Beck18 likelihood.
Finally, the green contours are for the likelihood with lensed CMB spectra and the \cite{Peloton:2016kbw} non-Gaussian covariance. 


It appears that the Beck18 likelihood, with the lensed CMB spectra, has the largest bias on the sum of the neutrino masses and on the amplitude of the matter power spectrum. 
We see that when considering unlensed CMB spectra the Beck18 likelihood obtains similar constraints as ours, whihc has unlensed spectra as well. 
We show that we obtain almost exactly the same posterior, with a reduced bias, if we use the lensed spectra but correctly taking into account the correlations due to lensing from \cite{Peloton:2016kbw}. This reduction of the bias when using the full non-Gaussian covariance can be understood as lowering the impact of the bias from the lensing power spectrum, by avoiding to ``double'' count the biased lensing power spectrum in the analysis. 
If we include as well the lensed $C_\ell^{BB}$ spectrum in the data vector, even if the correlations are properly taken into account, the bias is not reduced. We interpret that as the fact that the lensed  spectrum is another measurement of $C_L^{\phi\phi}$, as we can approximate for $\ell \ll 1000$ \cite{lewis2006}
\begin{equation}
    C_\ell^{BB} \sim \frac{1}{4\pi} \int \frac{d \ell'}{\ell'} \ell'^4 C_{\ell'}^{\phi\phi}\ell'^2 C_{\ell '}^{EE, \rm unl} \, .
\end{equation}
Thus there is a tension between the lensed BB spectrum, which measures an unbiased $C_L^{\phi\phi}$, and the estimate lensing power spectrum which is biased by \nlth. It appears that the biased $C_L^{\phi\phi}$ dominates and the likelihood is leaning towards the biased measurement of the neutrino mass. In practice, future surveys might not include the lensed BB spectrum in the analysis since the BB spectrum might be more prone to instrumental systematics and polarized foregrounds such as point sources. But comparing the marginalized constraints with or without the BB spectra in the likelihood could serve as additional robustness test to assess the presence of biases coming from non-Gaussian effects. 

We do not test the impact of the full non-Gaussian covariance for the MAP estimator. In that case, one would use the partially delensed CMB spectra in the likelihood, reducing the non-diagonal terms in the covariance matrix. Fisher forecasts using the covariance matrix computed with delensed spectra were published in \cite{Green:2016cjr, Hotinli:2021umk}. We leave for a future work a detailed comparison of our MAP reconstructed delensed spectra with this delensed analytical covariance.

\begin{figure}[h]
    \centering
    \includegraphics[width=0.8\columnwidth]{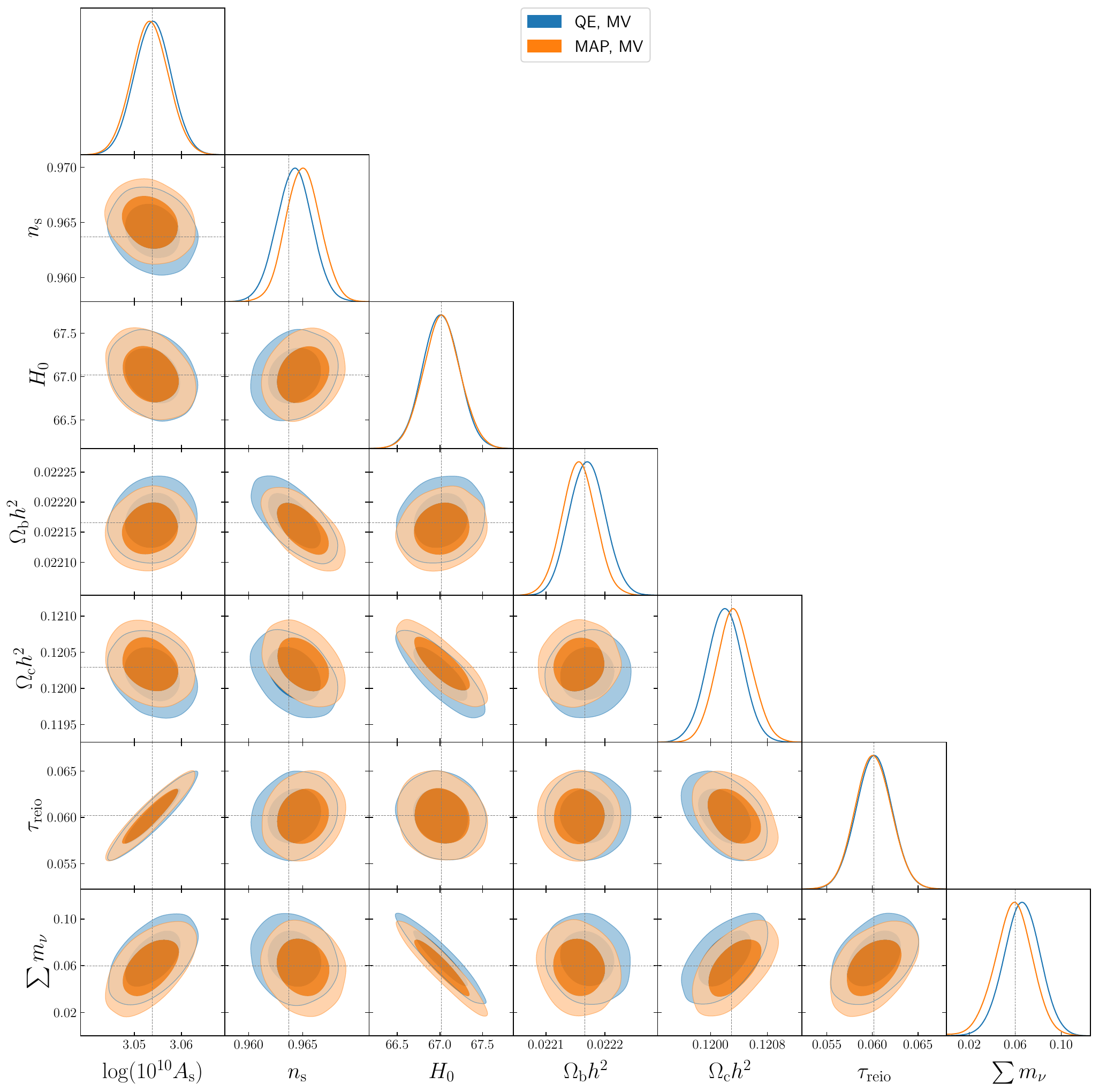}
    \caption{Full posterior distribution for the seven sampled cosmological parameters. We show the reconstruction for the minimum variance estimator, for both QE (blue) and MAP (orange), considering the bias due to the total $\nlth$.}
    \label{fig:trigmnu_full}
\end{figure}

We show in the Figure~\ref{fig:trigmnu_full} the full posterior of the likelihood for the QE and MAP reconstruction, considering the total \nlth bias, and seven cosmological parameters as described in the Section~\ref{sec:params}.

\onecolumngrid

\bibliography{n32}

\begin{thebibliography}{96}%
\makeatletter
\providecommand \@ifxundefined [1]{%
 \@ifx{#1\undefined}
}%
\providecommand \@ifnum [1]{%
 \ifnum #1\expandafter \@firstoftwo
 \else \expandafter \@secondoftwo
 \fi
}%
\providecommand \@ifx [1]{%
 \ifx #1\expandafter \@firstoftwo
 \else \expandafter \@secondoftwo
 \fi
}%
\providecommand \natexlab [1]{#1}%
\providecommand \enquote  [1]{``#1''}%
\providecommand \bibnamefont  [1]{#1}%
\providecommand \bibfnamefont [1]{#1}%
\providecommand \citenamefont [1]{#1}%
\providecommand \href@noop [0]{\@secondoftwo}%
\providecommand \href [0]{\begingroup \@sanitize@url \@href}%
\providecommand \@href[1]{\@@startlink{#1}\@@href}%
\providecommand \@@href[1]{\endgroup#1\@@endlink}%
\providecommand \@sanitize@url [0]{\catcode `\\12\catcode `\$12\catcode `\&12\catcode `\#12\catcode `\^12\catcode `\_12\catcode `\%12\relax}%
\providecommand \@@startlink[1]{}%
\providecommand \@@endlink[0]{}%
\providecommand \url  [0]{\begingroup\@sanitize@url \@url }%
\providecommand \@url [1]{\endgroup\@href {#1}{\urlprefix }}%
\providecommand \urlprefix  [0]{URL }%
\providecommand \Eprint [0]{\href }%
\providecommand \doibase [0]{https://doi.org/}%
\providecommand \selectlanguage [0]{\@gobble}%
\providecommand \bibinfo  [0]{\@secondoftwo}%
\providecommand \bibfield  [0]{\@secondoftwo}%
\providecommand \translation [1]{[#1]}%
\providecommand \BibitemOpen [0]{}%
\providecommand \bibitemStop [0]{}%
\providecommand \bibitemNoStop [0]{.\EOS\space}%
\providecommand \EOS [0]{\spacefactor3000\relax}%
\providecommand \BibitemShut  [1]{\csname bibitem#1\endcsname}%
\let\auto@bib@innerbib\@empty
\bibitem [{\citenamefont {Aguirre}\ \emph {et~al.}(2019)\citenamefont {Aguirre} \emph {et~al.}}]{so}%
  \BibitemOpen
  \bibfield  {author} {\bibinfo {author} {\bibfnamefont {J.}~\bibnamefont {Aguirre}} \emph {et~al.} (\bibinfo {collaboration} {Simons Observatory}),\ }\bibfield  {title} {\bibinfo {title} {{The Simons Observatory: Science goals and forecasts}},\ }\href {https://doi.org/10.1088/1475-7516/2019/02/056} {\bibfield  {journal} {\bibinfo  {journal} {\jcap}\ }\textbf {\bibinfo {volume} {1902}},\ \bibinfo {pages} {056} (\bibinfo {year} {2019})},\ \Eprint {https://arxiv.org/abs/1808.07445} {arXiv:1808.07445 [astro-ph.CO]} \BibitemShut {NoStop}%
\bibitem [{\citenamefont {{Abazajian}}\ \emph {et~al.}(2016)\citenamefont {{Abazajian}} \emph {et~al.}}]{cmbs4}%
  \BibitemOpen
  \bibfield  {author} {\bibinfo {author} {\bibfnamefont {K.~N.}\ \bibnamefont {{Abazajian}}} \emph {et~al.},\ }\bibfield  {title} {\bibinfo {title} {{CMB-S4 Science Book, First Edition}},\ }\href@noop {} {\  (\bibinfo {year} {2016})},\ \Eprint {https://arxiv.org/abs/1610.02743} {arXiv:1610.02743} \BibitemShut {NoStop}%
\bibitem [{\citenamefont {{Lewis}}\ and\ \citenamefont {{Challinor}}(2006)}]{lewis2006}%
  \BibitemOpen
  \bibfield  {author} {\bibinfo {author} {\bibfnamefont {A.}~\bibnamefont {{Lewis}}}\ and\ \bibinfo {author} {\bibfnamefont {A.}~\bibnamefont {{Challinor}}},\ }\bibfield  {title} {\bibinfo {title} {{Weak gravitational lensing of the CMB}},\ }\href {https://doi.org/10.1016/j.physrep.2006.03.002} {\bibfield  {journal} {\bibinfo  {journal} {\physrep}\ }\textbf {\bibinfo {volume} {429}},\ \bibinfo {pages} {1} (\bibinfo {year} {2006})},\ \Eprint {https://arxiv.org/abs/astro-ph/0601594} {arXiv:astro-ph/0601594} \BibitemShut {NoStop}%
\bibitem [{\citenamefont {{de Putter}}\ \emph {et~al.}(2009)\citenamefont {{de Putter}}, \citenamefont {{Zahn}},\ and\ \citenamefont {{Linder}}}]{2009PhRvD..79f5033D}%
  \BibitemOpen
  \bibfield  {author} {\bibinfo {author} {\bibfnamefont {R.}~\bibnamefont {{de Putter}}}, \bibinfo {author} {\bibfnamefont {O.}~\bibnamefont {{Zahn}}},\ and\ \bibinfo {author} {\bibfnamefont {E.~V.}\ \bibnamefont {{Linder}}},\ }\bibfield  {title} {\bibinfo {title} {{CMB lensing constraints on neutrinos and dark energy}},\ }\href {https://doi.org/10.1103/PhysRevD.79.065033} {\bibfield  {journal} {\bibinfo  {journal} {\prd}\ }\textbf {\bibinfo {volume} {79}},\ \bibinfo {eid} {065033} (\bibinfo {year} {2009})},\ \Eprint {https://arxiv.org/abs/0901.0916} {arXiv:0901.0916} \BibitemShut {NoStop}%
\bibitem [{\citenamefont {{Stompor}}\ and\ \citenamefont {{Efstathiou}}(1999)}]{1999MNRAS.302..735S}%
  \BibitemOpen
  \bibfield  {author} {\bibinfo {author} {\bibfnamefont {R.}~\bibnamefont {{Stompor}}}\ and\ \bibinfo {author} {\bibfnamefont {G.}~\bibnamefont {{Efstathiou}}},\ }\bibfield  {title} {\bibinfo {title} {{Gravitational lensing of cosmic microwave background anisotropies and cosmological parameter estimation}},\ }\href {https://doi.org/10.1046/j.1365-8711.1999.02174.x} {\bibfield  {journal} {\bibinfo  {journal} {\mnras}\ }\textbf {\bibinfo {volume} {302}},\ \bibinfo {pages} {735} (\bibinfo {year} {1999})},\ \Eprint {https://arxiv.org/abs/astro-ph/9805294} {arXiv:astro-ph/9805294 [astro-ph]} \BibitemShut {NoStop}%
\bibitem [{\citenamefont {Sailer}\ \emph {et~al.}(2021)\citenamefont {Sailer}, \citenamefont {Castorina}, \citenamefont {Ferraro},\ and\ \citenamefont {White}}]{Sailer_2021}%
  \BibitemOpen
  \bibfield  {author} {\bibinfo {author} {\bibfnamefont {N.}~\bibnamefont {Sailer}}, \bibinfo {author} {\bibfnamefont {E.}~\bibnamefont {Castorina}}, \bibinfo {author} {\bibfnamefont {S.}~\bibnamefont {Ferraro}},\ and\ \bibinfo {author} {\bibfnamefont {M.}~\bibnamefont {White}},\ }\bibfield  {title} {\bibinfo {title} {Cosmology at high redshift — a probe of fundamental physics},\ }\href {https://doi.org/10.1088/1475-7516/2021/12/049} {\bibfield  {journal} {\bibinfo  {journal} {Journal of Cosmology and Astroparticle Physics}\ }\textbf {\bibinfo {volume} {2021}}\bibinfo  {number} { (12)},\ \bibinfo {pages} {049}}\BibitemShut {NoStop}%
\bibitem [{\citenamefont {{Vallinotto}}(2012)}]{vallinotto2012}%
  \BibitemOpen
\bibfield  {number} {  }\bibfield  {author} {\bibinfo {author} {\bibfnamefont {A.}~\bibnamefont {{Vallinotto}}},\ }\bibfield  {title} {\bibinfo {title} {{Using Cosmic Microwave Background Lensing to Constrain the Multiplicative Bias of Cosmic Shear}},\ }\href {https://doi.org/10.1088/0004-637X/759/1/32} {\bibfield  {journal} {\bibinfo  {journal} {\apj}\ }\textbf {\bibinfo {volume} {759}},\ \bibinfo {eid} {32} (\bibinfo {year} {2012})},\ \Eprint {https://arxiv.org/abs/1110.5339} {arXiv:1110.5339} \BibitemShut {NoStop}%
\bibitem [{\citenamefont {{Schaan}}\ \emph {et~al.}(2017)\citenamefont {{Schaan}}, \citenamefont {{Krause}}, \citenamefont {{Eifler}}, \citenamefont {{Dor{\'e}}}, \citenamefont {{Miyatake}}, \citenamefont {{Rhodes}},\ and\ \citenamefont {{Spergel}}}]{2017PhRvD..95l3512S}%
  \BibitemOpen
  \bibfield  {author} {\bibinfo {author} {\bibfnamefont {E.}~\bibnamefont {{Schaan}}}, \bibinfo {author} {\bibfnamefont {E.}~\bibnamefont {{Krause}}}, \bibinfo {author} {\bibfnamefont {T.}~\bibnamefont {{Eifler}}}, \bibinfo {author} {\bibfnamefont {O.}~\bibnamefont {{Dor{\'e}}}}, \bibinfo {author} {\bibfnamefont {H.}~\bibnamefont {{Miyatake}}}, \bibinfo {author} {\bibfnamefont {J.}~\bibnamefont {{Rhodes}}},\ and\ \bibinfo {author} {\bibfnamefont {D.~N.}\ \bibnamefont {{Spergel}}},\ }\bibfield  {title} {\bibinfo {title} {{Looking through the same lens: Shear calibration for LSST, Euclid, and WFIRST with stage 4 CMB lensing}},\ }\href {https://doi.org/10.1103/PhysRevD.95.123512} {\bibfield  {journal} {\bibinfo  {journal} {\prd}\ }\textbf {\bibinfo {volume} {95}},\ \bibinfo {eid} {123512} (\bibinfo {year} {2017})},\ \Eprint {https://arxiv.org/abs/1607.01761} {arXiv:1607.01761} \BibitemShut {NoStop}%
\bibitem [{\citenamefont {{Cawthon}}(2020)}]{2020PhRvD.101f3509C}%
  \BibitemOpen
  \bibfield  {author} {\bibinfo {author} {\bibfnamefont {R.}~\bibnamefont {{Cawthon}}},\ }\bibfield  {title} {\bibinfo {title} {{Effects of redshift uncertainty on cross-correlations of CMB lensing and galaxy surveys}},\ }\href {https://doi.org/10.1103/PhysRevD.101.063509} {\bibfield  {journal} {\bibinfo  {journal} {\prd}\ }\textbf {\bibinfo {volume} {101}},\ \bibinfo {eid} {063509} (\bibinfo {year} {2020})},\ \Eprint {https://arxiv.org/abs/1809.09251} {arXiv:1809.09251} \BibitemShut {NoStop}%
\bibitem [{\citenamefont {{Vallinotto}}(2013)}]{vallinotto2013}%
  \BibitemOpen
  \bibfield  {author} {\bibinfo {author} {\bibfnamefont {A.}~\bibnamefont {{Vallinotto}}},\ }\bibfield  {title} {\bibinfo {title} {{The Synergy between the Dark Energy Survey and the South Pole Telescope}},\ }\href {https://doi.org/10.1088/0004-637X/778/2/108} {\bibfield  {journal} {\bibinfo  {journal} {\apj}\ }\textbf {\bibinfo {volume} {778}},\ \bibinfo {eid} {108} (\bibinfo {year} {2013})},\ \Eprint {https://arxiv.org/abs/1304.3474} {arXiv:1304.3474} \BibitemShut {NoStop}%
\bibitem [{\citenamefont {{Wenzl}}\ \emph {et~al.}(2022)\citenamefont {{Wenzl}}, \citenamefont {{Doux}}, \citenamefont {{Heinrich}}, \citenamefont {{Bean}}, \citenamefont {{Jain}}, \citenamefont {{Dor{\'e}}}, \citenamefont {{Eifler}},\ and\ \citenamefont {{Fang}}}]{roman}%
  \BibitemOpen
  \bibfield  {author} {\bibinfo {author} {\bibfnamefont {L.}~\bibnamefont {{Wenzl}}}, \bibinfo {author} {\bibfnamefont {C.}~\bibnamefont {{Doux}}}, \bibinfo {author} {\bibfnamefont {C.}~\bibnamefont {{Heinrich}}}, \bibinfo {author} {\bibfnamefont {R.}~\bibnamefont {{Bean}}}, \bibinfo {author} {\bibfnamefont {B.}~\bibnamefont {{Jain}}}, \bibinfo {author} {\bibfnamefont {O.}~\bibnamefont {{Dor{\'e}}}}, \bibinfo {author} {\bibfnamefont {T.}~\bibnamefont {{Eifler}}},\ and\ \bibinfo {author} {\bibfnamefont {X.}~\bibnamefont {{Fang}}},\ }\bibfield  {title} {\bibinfo {title} {{Cosmology with the Roman Space Telescope - Synergies with CMB lensing}},\ }\href {https://doi.org/10.1093/mnras/stac790} {\bibfield  {journal} {\bibinfo  {journal} {\mnras}\ }\textbf {\bibinfo {volume} {512}},\ \bibinfo {pages} {5311} (\bibinfo {year} {2022})},\ \Eprint {https://arxiv.org/abs/2112.07681} {arXiv:2112.07681} \BibitemShut {NoStop}%
\bibitem [{\citenamefont {{Euclid Collaboration}}\ \emph {et~al.}(2022)\citenamefont {{Euclid Collaboration}}, \citenamefont {{Ili{\'c}}} \emph {et~al.}}]{euclid}%
  \BibitemOpen
  \bibfield  {author} {\bibinfo {author} {\bibnamefont {{Euclid Collaboration}}}, \bibinfo {author} {\bibfnamefont {S.}~\bibnamefont {{Ili{\'c}}}}, \emph {et~al.},\ }\bibfield  {title} {\bibinfo {title} {{Euclid preparation. XV. Forecasting cosmological constraints for the Euclid and CMB joint analysis}},\ }\href {https://doi.org/10.1051/0004-6361/202141556} {\bibfield  {journal} {\bibinfo  {journal} {\aap}\ }\textbf {\bibinfo {volume} {657}},\ \bibinfo {eid} {A91} (\bibinfo {year} {2022})},\ \Eprint {https://arxiv.org/abs/2106.08346} {arXiv:2106.08346} \BibitemShut {NoStop}%
\bibitem [{\citenamefont {{Lewis}}\ and\ \citenamefont {{King}}(2006)}]{lewis-king}%
  \BibitemOpen
  \bibfield  {author} {\bibinfo {author} {\bibfnamefont {A.}~\bibnamefont {{Lewis}}}\ and\ \bibinfo {author} {\bibfnamefont {L.}~\bibnamefont {{King}}},\ }\bibfield  {title} {\bibinfo {title} {{Cluster masses from CMB and galaxy weak lensing}},\ }\href {https://doi.org/10.1103/PhysRevD.73.063006} {\bibfield  {journal} {\bibinfo  {journal} {\prd}\ }\textbf {\bibinfo {volume} {73}},\ \bibinfo {eid} {063006} (\bibinfo {year} {2006})},\ \Eprint {https://arxiv.org/abs/astro-ph/0512104} {arXiv:astro-ph/0512104 [astro-ph]} \BibitemShut {NoStop}%
\bibitem [{\citenamefont {{Hu}}\ \emph {et~al.}(2007)\citenamefont {{Hu}}, \citenamefont {{DeDeo}},\ and\ \citenamefont {{Vale}}}]{hu-dedeo}%
  \BibitemOpen
  \bibfield  {author} {\bibinfo {author} {\bibfnamefont {W.}~\bibnamefont {{Hu}}}, \bibinfo {author} {\bibfnamefont {S.}~\bibnamefont {{DeDeo}}},\ and\ \bibinfo {author} {\bibfnamefont {C.}~\bibnamefont {{Vale}}},\ }\bibfield  {title} {\bibinfo {title} {{Cluster mass estimators from CMB temperature and polarization lensing}},\ }\href {https://doi.org/10.1088/1367-2630/9/12/441} {\bibfield  {journal} {\bibinfo  {journal} {New Journal of Physics}\ }\textbf {\bibinfo {volume} {9}},\ \bibinfo {pages} {441} (\bibinfo {year} {2007})},\ \Eprint {https://arxiv.org/abs/astro-ph/0701276} {arXiv:astro-ph/0701276 [astro-ph]} \BibitemShut {NoStop}%
\bibitem [{\citenamefont {{Raghunathan}}\ \emph {et~al.}(2017)\citenamefont {{Raghunathan}}, \citenamefont {{Patil}}, \citenamefont {{Baxter}}, \citenamefont {{Bianchini}}, \citenamefont {{Bleem}}, \citenamefont {{Crawford}}, \citenamefont {{Holder}}, \citenamefont {{Manzotti}},\ and\ \citenamefont {{Reichardt}}}]{2017JCAP...08..030R}%
  \BibitemOpen
  \bibfield  {author} {\bibinfo {author} {\bibfnamefont {S.}~\bibnamefont {{Raghunathan}}}, \bibinfo {author} {\bibfnamefont {S.}~\bibnamefont {{Patil}}}, \bibinfo {author} {\bibfnamefont {E.~J.}\ \bibnamefont {{Baxter}}}, \bibinfo {author} {\bibfnamefont {F.}~\bibnamefont {{Bianchini}}}, \bibinfo {author} {\bibfnamefont {L.~E.}\ \bibnamefont {{Bleem}}}, \bibinfo {author} {\bibfnamefont {T.~M.}\ \bibnamefont {{Crawford}}}, \bibinfo {author} {\bibfnamefont {G.~P.}\ \bibnamefont {{Holder}}}, \bibinfo {author} {\bibfnamefont {A.}~\bibnamefont {{Manzotti}}},\ and\ \bibinfo {author} {\bibfnamefont {C.~L.}\ \bibnamefont {{Reichardt}}},\ }\bibfield  {title} {\bibinfo {title} {{Measuring galaxy cluster masses with CMB lensing using a Maximum Likelihood estimator: statistical and systematic error budgets for future experiments}},\ }\href {https://doi.org/10.1088/1475-7516/2017/08/030} {\bibfield  {journal} {\bibinfo  {journal} {\jcap}\ }\textbf {\bibinfo {volume} {2017}},\ \bibinfo {eid} {030} (\bibinfo {year}
  {2017})},\ \Eprint {https://arxiv.org/abs/1705.00411} {arXiv:1705.00411} \BibitemShut {NoStop}%
\bibitem [{\citenamefont {{Seljak}}\ and\ \citenamefont {{Hirata}}(2004)}]{2004PhRvD..69d3005S}%
  \BibitemOpen
  \bibfield  {author} {\bibinfo {author} {\bibfnamefont {U.}~\bibnamefont {{Seljak}}}\ and\ \bibinfo {author} {\bibfnamefont {C.~M.}\ \bibnamefont {{Hirata}}},\ }\bibfield  {title} {\bibinfo {title} {{Gravitational lensing as a contaminant of the gravity wave signal in the CMB}},\ }\href {https://doi.org/10.1103/PhysRevD.69.043005} {\bibfield  {journal} {\bibinfo  {journal} {\prd}\ }\textbf {\bibinfo {volume} {69}},\ \bibinfo {eid} {043005} (\bibinfo {year} {2004})},\ \Eprint {https://arxiv.org/abs/astro-ph/0310163} {arXiv:astro-ph/0310163 [astro-ph]} \BibitemShut {NoStop}%
\bibitem [{\citenamefont {{Smith}}\ \emph {et~al.}(2012)\citenamefont {{Smith}}, \citenamefont {{Hanson}}, \citenamefont {{LoVerde}}, \citenamefont {{Hirata}},\ and\ \citenamefont {{Zahn}}}]{smith2012}%
  \BibitemOpen
  \bibfield  {author} {\bibinfo {author} {\bibfnamefont {K.~M.}\ \bibnamefont {{Smith}}}, \bibinfo {author} {\bibfnamefont {D.}~\bibnamefont {{Hanson}}}, \bibinfo {author} {\bibfnamefont {M.}~\bibnamefont {{LoVerde}}}, \bibinfo {author} {\bibfnamefont {C.~M.}\ \bibnamefont {{Hirata}}},\ and\ \bibinfo {author} {\bibfnamefont {O.}~\bibnamefont {{Zahn}}},\ }\bibfield  {title} {\bibinfo {title} {{Delensing CMB polarization with external datasets}},\ }\href {https://doi.org/10.1088/1475-7516/2012/06/014} {\bibfield  {journal} {\bibinfo  {journal} {\jcap}\ }\textbf {\bibinfo {volume} {2012}},\ \bibinfo {eid} {014} (\bibinfo {year} {2012})},\ \Eprint {https://arxiv.org/abs/1010.0048} {arXiv:1010.0048} \BibitemShut {NoStop}%
\bibitem [{\citenamefont {{Hu}}\ and\ \citenamefont {{Okamoto}}(2002)}]{hu-okamoto}%
  \BibitemOpen
  \bibfield  {author} {\bibinfo {author} {\bibfnamefont {W.}~\bibnamefont {{Hu}}}\ and\ \bibinfo {author} {\bibfnamefont {T.}~\bibnamefont {{Okamoto}}},\ }\bibfield  {title} {\bibinfo {title} {{Mass Reconstruction with Cosmic Microwave Background Polarization}},\ }\href {https://doi.org/10.1086/341110} {\bibfield  {journal} {\bibinfo  {journal} {\apj}\ }\textbf {\bibinfo {volume} {574}},\ \bibinfo {pages} {566} (\bibinfo {year} {2002})},\ \Eprint {https://arxiv.org/abs/astro-ph/0111606} {arXiv:astro-ph/0111606} \BibitemShut {NoStop}%
\bibitem [{\citenamefont {Aghanim}\ \emph {et~al.}(2020)\citenamefont {Aghanim} \emph {et~al.}}]{plancklensing2018}%
  \BibitemOpen
  \bibfield  {author} {\bibinfo {author} {\bibfnamefont {N.}~\bibnamefont {Aghanim}} \emph {et~al.} (\bibinfo {collaboration} {Planck}),\ }\bibfield  {title} {\bibinfo {title} {{Planck 2018 results. VIII. Gravitational lensing}},\ }\href {https://doi.org/10.1051/0004-6361/201833886} {\bibfield  {journal} {\bibinfo  {journal} {\aap}\ }\textbf {\bibinfo {volume} {641}},\ \bibinfo {pages} {A8} (\bibinfo {year} {2020})},\ \Eprint {https://arxiv.org/abs/1807.06210} {arXiv:1807.06210 [astro-ph.CO]} \BibitemShut {NoStop}%
\bibitem [{\citenamefont {{Carron}}\ \emph {et~al.}(2022)\citenamefont {{Carron}}, \citenamefont {{Mirmelstein}},\ and\ \citenamefont {{Lewis}}}]{plancklensingdr4}%
  \BibitemOpen
  \bibfield  {author} {\bibinfo {author} {\bibfnamefont {J.}~\bibnamefont {{Carron}}}, \bibinfo {author} {\bibfnamefont {M.}~\bibnamefont {{Mirmelstein}}},\ and\ \bibinfo {author} {\bibfnamefont {A.}~\bibnamefont {{Lewis}}},\ }\bibfield  {title} {\bibinfo {title} {{CMB lensing from Planck PR4 maps}},\ }\href {https://doi.org/10.1088/1475-7516/2022/09/039} {\bibfield  {journal} {\bibinfo  {journal} {\jcap}\ }\textbf {\bibinfo {volume} {2022}},\ \bibinfo {eid} {039} (\bibinfo {year} {2022})},\ \Eprint {https://arxiv.org/abs/2206.07773} {arXiv:2206.07773} \BibitemShut {NoStop}%
\bibitem [{\citenamefont {{Madhavacheril}}\ \emph {et~al.}(2023)\citenamefont {{Madhavacheril}} \emph {et~al.}}]{actlensingdr6}%
  \BibitemOpen
  \bibfield  {author} {\bibinfo {author} {\bibfnamefont {M.~S.}\ \bibnamefont {{Madhavacheril}}} \emph {et~al.},\ }\bibfield  {title} {\bibinfo {title} {{The Atacama Cosmology Telescope: DR6 Gravitational Lensing Map and Cosmological Parameters}},\ }\href {https://doi.org/10.48550/arXiv.2304.05203} {\  (\bibinfo {year} {2023})},\ \Eprint {https://arxiv.org/abs/2304.05203} {arXiv:2304.05203} \BibitemShut {NoStop}%
\bibitem [{\citenamefont {{Wu}}\ \emph {et~al.}(2019)\citenamefont {{Wu}} \emph {et~al.}}]{sptpollensing}%
  \BibitemOpen
  \bibfield  {author} {\bibinfo {author} {\bibfnamefont {W.~L.~K.}\ \bibnamefont {{Wu}}} \emph {et~al.},\ }\bibfield  {title} {\bibinfo {title} {{A Measurement of the Cosmic Microwave Background Lensing Potential and Power Spectrum from 500 deg$^{2}$ of SPTpol Temperature and Polarization Data}},\ }\href {https://doi.org/10.3847/1538-4357/ab4186} {\bibfield  {journal} {\bibinfo  {journal} {\apj}\ }\textbf {\bibinfo {volume} {884}},\ \bibinfo {eid} {70} (\bibinfo {year} {2019})},\ \Eprint {https://arxiv.org/abs/1905.05777} {arXiv:1905.05777} \BibitemShut {NoStop}%
\bibitem [{\citenamefont {{Fa{\'u}ndez}}\ \emph {et~al.}(2020)\citenamefont {{Fa{\'u}ndez}} \emph {et~al.}}]{pblensing2020}%
  \BibitemOpen
  \bibfield  {author} {\bibinfo {author} {\bibfnamefont {M.~A.}\ \bibnamefont {{Fa{\'u}ndez}}} \emph {et~al.},\ }\bibfield  {title} {\bibinfo {title} {{Measurement of the Cosmic Microwave Background Polarization Lensing Power Spectrum from Two Years of POLARBEAR Data}},\ }\href {https://doi.org/10.3847/1538-4357/ab7e29} {\bibfield  {journal} {\bibinfo  {journal} {\apj}\ }\textbf {\bibinfo {volume} {893}},\ \bibinfo {eid} {85} (\bibinfo {year} {2020})},\ \Eprint {https://arxiv.org/abs/1911.10980} {arXiv:1911.10980} \BibitemShut {NoStop}%
\bibitem [{\citenamefont {{BICEP/Keck Collaboration}}\ \emph {et~al.}(2022)\citenamefont {{BICEP/Keck Collaboration}}, \citenamefont {{:}}, \citenamefont {{Ade}} \emph {et~al.}}]{biceplensing}%
  \BibitemOpen
  \bibfield  {author} {\bibinfo {author} {\bibnamefont {{BICEP/Keck Collaboration}}}, \bibinfo {author} {\bibnamefont {{:}}}, \bibinfo {author} {\bibfnamefont {P.~A.~R.}\ \bibnamefont {{Ade}}}, \emph {et~al.},\ }\bibfield  {title} {\bibinfo {title} {{BICEP / Keck XVII: Line of Sight Distortion Analysis: Estimates of Gravitational Lensing, Anisotropic Cosmic Birefringence, Patchy Reionization, and Systematic Errors}},\ }\href {https://doi.org/10.48550/arXiv.2210.08038} {\  (\bibinfo {year} {2022})},\ \Eprint {https://arxiv.org/abs/2210.08038} {arXiv:2210.08038} \BibitemShut {NoStop}%
\bibitem [{\citenamefont {Sailer}\ \emph {et~al.}(2023)\citenamefont {Sailer}, \citenamefont {Ferraro},\ and\ \citenamefont {Schaan}}]{Sailer:2022jwt}%
  \BibitemOpen
  \bibfield  {author} {\bibinfo {author} {\bibfnamefont {N.}~\bibnamefont {Sailer}}, \bibinfo {author} {\bibfnamefont {S.}~\bibnamefont {Ferraro}},\ and\ \bibinfo {author} {\bibfnamefont {E.}~\bibnamefont {Schaan}},\ }\bibfield  {title} {\bibinfo {title} {{Foreground-immune CMB lensing reconstruction with polarization}},\ }\href {https://doi.org/10.1103/PhysRevD.107.023504} {\bibfield  {journal} {\bibinfo  {journal} {\prl}\ }\textbf {\bibinfo {volume} {107}},\ \bibinfo {pages} {023504} (\bibinfo {year} {2023})},\ \Eprint {https://arxiv.org/abs/2211.03786} {arXiv:2211.03786} \BibitemShut {NoStop}%
\bibitem [{\citenamefont {Maniyar}\ \emph {et~al.}(2021)\citenamefont {Maniyar}, \citenamefont {Ali-Ha\"\i{}moud}, \citenamefont {Carron}, \citenamefont {Lewis},\ and\ \citenamefont {Madhavacheril}}]{Maniyar:2021msb}%
  \BibitemOpen
  \bibfield  {author} {\bibinfo {author} {\bibfnamefont {A.~S.}\ \bibnamefont {Maniyar}}, \bibinfo {author} {\bibfnamefont {Y.}~\bibnamefont {Ali-Ha\"\i{}moud}}, \bibinfo {author} {\bibfnamefont {J.}~\bibnamefont {Carron}}, \bibinfo {author} {\bibfnamefont {A.}~\bibnamefont {Lewis}},\ and\ \bibinfo {author} {\bibfnamefont {M.~S.}\ \bibnamefont {Madhavacheril}},\ }\bibfield  {title} {\bibinfo {title} {{Quadratic estimators for CMB weak lensing}},\ }\href {https://doi.org/10.1103/PhysRevD.103.083524} {\bibfield  {journal} {\bibinfo  {journal} {\prd}\ }\textbf {\bibinfo {volume} {103}},\ \bibinfo {pages} {083524} (\bibinfo {year} {2021})},\ \Eprint {https://arxiv.org/abs/2101.12193} {arXiv:2101.12193} \BibitemShut {NoStop}%
\bibitem [{\citenamefont {Carron}(2023)}]{Carron:2022edh}%
  \BibitemOpen
  \bibfield  {author} {\bibinfo {author} {\bibfnamefont {J.}~\bibnamefont {Carron}},\ }\bibfield  {title} {\bibinfo {title} {{Real-world CMB lensing quadratic estimator power spectrum response}},\ }\href {https://doi.org/10.1088/1475-7516/2023/02/057} {\bibfield  {journal} {\bibinfo  {journal} {\jcap}\ }\textbf {\bibinfo {volume} {02}},\ \bibinfo {pages} {057} (\bibinfo {year} {2023})}\BibitemShut {NoStop}%
\bibitem [{\citenamefont {Beck}\ \emph {et~al.}(2020)\citenamefont {Beck}, \citenamefont {Errard},\ and\ \citenamefont {Stompor}}]{Beck:2020dhe}%
  \BibitemOpen
  \bibfield  {author} {\bibinfo {author} {\bibfnamefont {D.}~\bibnamefont {Beck}}, \bibinfo {author} {\bibfnamefont {J.}~\bibnamefont {Errard}},\ and\ \bibinfo {author} {\bibfnamefont {R.}~\bibnamefont {Stompor}},\ }\bibfield  {title} {\bibinfo {title} {{Impact of Polarized Galactic Foreground Emission on CMB Lensing Reconstruction and Delensing of B-Modes}},\ }\href {https://doi.org/10.1088/1475-7516/2020/06/030} {\bibfield  {journal} {\bibinfo  {journal} {\jcap}\ }\textbf {\bibinfo {volume} {06}},\ \bibinfo {pages} {030} (\bibinfo {year} {2020})},\ \Eprint {https://arxiv.org/abs/2001.02641} {arXiv:2001.02641} \BibitemShut {NoStop}%
\bibitem [{\citenamefont {{Schaan}}\ and\ \citenamefont {{Ferraro}}(2019)}]{2019PhRvL.122r1301S}%
  \BibitemOpen
  \bibfield  {author} {\bibinfo {author} {\bibfnamefont {E.}~\bibnamefont {{Schaan}}}\ and\ \bibinfo {author} {\bibfnamefont {S.}~\bibnamefont {{Ferraro}}},\ }\bibfield  {title} {\bibinfo {title} {{Foreground-Immune Cosmic Microwave Background Lensing with Shear-Only Reconstruction}},\ }\href {https://doi.org/10.1103/PhysRevLett.122.181301} {\bibfield  {journal} {\bibinfo  {journal} {\prl}\ }\textbf {\bibinfo {volume} {122}},\ \bibinfo {eid} {181301} (\bibinfo {year} {2019})},\ \Eprint {https://arxiv.org/abs/1804.06403} {arXiv:1804.06403} \BibitemShut {NoStop}%
\bibitem [{\citenamefont {{Mirmelstein}}\ \emph {et~al.}(2021)\citenamefont {{Mirmelstein}}, \citenamefont {{Fabbian}}, \citenamefont {{Lewis}},\ and\ \citenamefont {{Peloton}}}]{mirmelstein2021}%
  \BibitemOpen
  \bibfield  {author} {\bibinfo {author} {\bibfnamefont {M.}~\bibnamefont {{Mirmelstein}}}, \bibinfo {author} {\bibfnamefont {G.}~\bibnamefont {{Fabbian}}}, \bibinfo {author} {\bibfnamefont {A.}~\bibnamefont {{Lewis}}},\ and\ \bibinfo {author} {\bibfnamefont {J.}~\bibnamefont {{Peloton}}},\ }\bibfield  {title} {\bibinfo {title} {{Instrumental systematics biases in CMB lensing reconstruction: A simulation-based assessment}},\ }\href {https://doi.org/10.1103/PhysRevD.103.123540} {\bibfield  {journal} {\bibinfo  {journal} {\prd}\ }\textbf {\bibinfo {volume} {103}},\ \bibinfo {eid} {123540} (\bibinfo {year} {2021})},\ \Eprint {https://arxiv.org/abs/2011.13910} {arXiv:2011.13910} \BibitemShut {NoStop}%
\bibitem [{\citenamefont {Darwish}\ \emph {et~al.}(2023)\citenamefont {Darwish}, \citenamefont {Sherwin}, \citenamefont {Sailer}, \citenamefont {Schaan},\ and\ \citenamefont {Ferraro}}]{darwish2023}%
  \BibitemOpen
  \bibfield  {author} {\bibinfo {author} {\bibfnamefont {O.}~\bibnamefont {Darwish}}, \bibinfo {author} {\bibfnamefont {B.~D.}\ \bibnamefont {Sherwin}}, \bibinfo {author} {\bibfnamefont {N.}~\bibnamefont {Sailer}}, \bibinfo {author} {\bibfnamefont {E.}~\bibnamefont {Schaan}},\ and\ \bibinfo {author} {\bibfnamefont {S.}~\bibnamefont {Ferraro}},\ }\bibfield  {title} {\bibinfo {title} {Optimizing foreground mitigation for cmb lensing with combined multifrequency and geometric methods},\ }\href {https://doi.org/10.1103/PhysRevD.107.043519} {\bibfield  {journal} {\bibinfo  {journal} {Phys. Rev. D}\ }\textbf {\bibinfo {volume} {107}},\ \bibinfo {pages} {043519} (\bibinfo {year} {2023})},\ \Eprint {https://arxiv.org/abs/2111.00462} {arXiv:2111.00462} \BibitemShut {NoStop}%
\bibitem [{\citenamefont {Millea}\ \emph {et~al.}(2019)\citenamefont {Millea}, \citenamefont {Anderes},\ and\ \citenamefont {Wandelt}}]{Millea:2017fyd}%
  \BibitemOpen
  \bibfield  {author} {\bibinfo {author} {\bibfnamefont {M.}~\bibnamefont {Millea}}, \bibinfo {author} {\bibfnamefont {E.}~\bibnamefont {Anderes}},\ and\ \bibinfo {author} {\bibfnamefont {B.~D.}\ \bibnamefont {Wandelt}},\ }\bibfield  {title} {\bibinfo {title} {{Bayesian delensing of CMB temperature and polarization}},\ }\href {https://doi.org/10.1103/PhysRevD.100.023509} {\bibfield  {journal} {\bibinfo  {journal} {\prd}\ }\textbf {\bibinfo {volume} {100}},\ \bibinfo {pages} {023509} (\bibinfo {year} {2019})}\BibitemShut {NoStop}%
\bibitem [{\citenamefont {Millea}\ \emph {et~al.}(2020)\citenamefont {Millea}, \citenamefont {Anderes},\ and\ \citenamefont {Wandelt}}]{Millea:2020cpw}%
  \BibitemOpen
  \bibfield  {author} {\bibinfo {author} {\bibfnamefont {M.}~\bibnamefont {Millea}}, \bibinfo {author} {\bibfnamefont {E.}~\bibnamefont {Anderes}},\ and\ \bibinfo {author} {\bibfnamefont {B.~D.}\ \bibnamefont {Wandelt}},\ }\bibfield  {title} {\bibinfo {title} {{Sampling-based inference of the primordial CMB and gravitational lensing}},\ }\href {https://doi.org/10.1103/PhysRevD.102.123542} {\bibfield  {journal} {\bibinfo  {journal} {\prd}\ }\textbf {\bibinfo {volume} {102}},\ \bibinfo {pages} {123542} (\bibinfo {year} {2020})}\BibitemShut {NoStop}%
\bibitem [{\citenamefont {Millea}\ and\ \citenamefont {Seljak}(2022)}]{Millea:2021had}%
  \BibitemOpen
  \bibfield  {author} {\bibinfo {author} {\bibfnamefont {M.}~\bibnamefont {Millea}}\ and\ \bibinfo {author} {\bibfnamefont {U.}~\bibnamefont {Seljak}},\ }\bibfield  {title} {\bibinfo {title} {{Marginal unbiased score expansion and application to CMB lensing}},\ }\href {https://doi.org/10.1103/PhysRevD.105.103531} {\bibfield  {journal} {\bibinfo  {journal} {\prd}\ }\textbf {\bibinfo {volume} {105}},\ \bibinfo {pages} {103531} (\bibinfo {year} {2022})}\BibitemShut {NoStop}%
\bibitem [{\citenamefont {Hirata}\ and\ \citenamefont {Seljak}(2003)}]{Hirata:2003ka}%
  \BibitemOpen
  \bibfield  {author} {\bibinfo {author} {\bibfnamefont {C.~M.}\ \bibnamefont {Hirata}}\ and\ \bibinfo {author} {\bibfnamefont {U.}~\bibnamefont {Seljak}},\ }\bibfield  {title} {\bibinfo {title} {{Reconstruction of lensing from the cosmic microwave background polarization}},\ }\href {https://doi.org/10.1103/PhysRevD.68.083002} {\bibfield  {journal} {\bibinfo  {journal} {\prd}\ }\textbf {\bibinfo {volume} {68}},\ \bibinfo {pages} {083002} (\bibinfo {year} {2003})}\BibitemShut {NoStop}%
\bibitem [{\citenamefont {Carron}\ and\ \citenamefont {Lewis}(2017)}]{maplensing}%
  \BibitemOpen
  \bibfield  {author} {\bibinfo {author} {\bibfnamefont {J.}~\bibnamefont {Carron}}\ and\ \bibinfo {author} {\bibfnamefont {A.}~\bibnamefont {Lewis}},\ }\bibfield  {title} {\bibinfo {title} {{Maximum a posteriori CMB lensing reconstruction}},\ }\href {https://doi.org/10.1103/PhysRevD.96.063510} {\bibfield  {journal} {\bibinfo  {journal} {\prd}\ }\textbf {\bibinfo {volume} {96}},\ \bibinfo {pages} {063510} (\bibinfo {year} {2017})}\BibitemShut {NoStop}%
\bibitem [{\citenamefont {{Adachi}}\ \emph {et~al.}(2020)\citenamefont {{Adachi}}, \citenamefont {others},\ and\ \citenamefont {{Polarbear Collaboration}}}]{pbdelens}%
  \BibitemOpen
  \bibfield  {author} {\bibinfo {author} {\bibfnamefont {S.}~\bibnamefont {{Adachi}}}, \bibinfo {author} {\bibnamefont {others}},\ and\ \bibinfo {author} {\bibnamefont {{Polarbear Collaboration}}},\ }\bibfield  {title} {\bibinfo {title} {{Internal Delensing of Cosmic Microwave Background Polarization B-Modes with the POLARBEAR Experiment}},\ }\href {https://doi.org/10.1103/PhysRevLett.124.131301} {\bibfield  {journal} {\bibinfo  {journal} {\prl}\ }\textbf {\bibinfo {volume} {124}},\ \bibinfo {eid} {131301} (\bibinfo {year} {2020})},\ \Eprint {https://arxiv.org/abs/1909.13832} {arXiv:1909.13832} \BibitemShut {NoStop}%
\bibitem [{\citenamefont {{Millea}}\ \emph {et~al.}(2021)\citenamefont {{Millea}} \emph {et~al.}}]{millea2021}%
  \BibitemOpen
  \bibfield  {author} {\bibinfo {author} {\bibfnamefont {M.}~\bibnamefont {{Millea}}} \emph {et~al.},\ }\bibfield  {title} {\bibinfo {title} {{Optimal Cosmic Microwave Background Lensing Reconstruction and Parameter Estimation with SPTpol Data}},\ }\href {https://doi.org/10.3847/1538-4357/ac02bb} {\bibfield  {journal} {\bibinfo  {journal} {\apj}\ }\textbf {\bibinfo {volume} {922}},\ \bibinfo {eid} {259} (\bibinfo {year} {2021})},\ \Eprint {https://arxiv.org/abs/2012.01709} {arXiv:2012.01709} \BibitemShut {NoStop}%
\bibitem [{\citenamefont {{Legrand}}\ and\ \citenamefont {{Carron}}(2022)}]{legrand2022}%
  \BibitemOpen
  \bibfield  {author} {\bibinfo {author} {\bibfnamefont {L.}~\bibnamefont {{Legrand}}}\ and\ \bibinfo {author} {\bibfnamefont {J.}~\bibnamefont {{Carron}}},\ }\bibfield  {title} {\bibinfo {title} {{Lensing power spectrum of the cosmic microwave background with deep polarization experiments}},\ }\href {https://doi.org/10.1103/PhysRevD.105.123519} {\bibfield  {journal} {\bibinfo  {journal} {\prd}\ }\textbf {\bibinfo {volume} {105}},\ \bibinfo {eid} {123519} (\bibinfo {year} {2022})},\ \Eprint {https://arxiv.org/abs/2112.05764} {arXiv:2112.05764} \BibitemShut {NoStop}%
\bibitem [{\citenamefont {Legrand}\ and\ \citenamefont {Carron}(2023)}]{legrand2023}%
  \BibitemOpen
  \bibfield  {author} {\bibinfo {author} {\bibfnamefont {L.}~\bibnamefont {Legrand}}\ and\ \bibinfo {author} {\bibfnamefont {J.}~\bibnamefont {Carron}},\ }\bibfield  {title} {\bibinfo {title} {{Robust and efficient CMB lensing power spectrum from polarization surveys}},\ }\href {https://doi.org/10.1103/PhysRevD.108.103516} {\bibfield  {journal} {\bibinfo  {journal} {Phys. Rev. D}\ }\textbf {\bibinfo {volume} {108}},\ \bibinfo {pages} {103516} (\bibinfo {year} {2023})},\ \Eprint {https://arxiv.org/abs/2304.02584} {arXiv:2304.02584 [astro-ph.CO]} \BibitemShut {NoStop}%
\bibitem [{\citenamefont {B{\"o}hm}\ \emph {et~al.}({\natexlab{a}})\citenamefont {B{\"o}hm}, \citenamefont {Schmittfull},\ and\ \citenamefont {Sherwin}}]{bohmBiasCMBLensing2016}%
  \BibitemOpen
  \bibfield  {author} {\bibinfo {author} {\bibfnamefont {V.}~\bibnamefont {B{\"o}hm}}, \bibinfo {author} {\bibfnamefont {M.}~\bibnamefont {Schmittfull}},\ and\ \bibinfo {author} {\bibfnamefont {B.~D.}\ \bibnamefont {Sherwin}},\ }\bibfield  {title} {\bibinfo {title} {A bias to {{CMB}} lensing measurements from the bispectrum of large-scale structure},\ }\href {https://doi.org/10.1103/PhysRevD.94.043519} {\ \textbf {\bibinfo {volume} {94}},\ \bibinfo {pages} {043519} ({\natexlab{a}})},\ \Eprint {https://arxiv.org/abs/1605.01392} {1605.01392} \BibitemShut {NoStop}%
\bibitem [{\citenamefont {Beck}\ \emph {et~al.}()\citenamefont {Beck}, \citenamefont {Fabbian},\ and\ \citenamefont {Errard}}]{beckLensingReconstructionPostBorn2018a}%
  \BibitemOpen
  \bibfield  {author} {\bibinfo {author} {\bibfnamefont {D.}~\bibnamefont {Beck}}, \bibinfo {author} {\bibfnamefont {G.}~\bibnamefont {Fabbian}},\ and\ \bibinfo {author} {\bibfnamefont {J.}~\bibnamefont {Errard}},\ }\bibfield  {title} {\bibinfo {title} {Lensing {{Reconstruction}} in {{Post-Born Cosmic Microwave Background Weak Lensing}}},\ }\href {https://doi.org/10.1103/PhysRevD.98.043512} {\ \textbf {\bibinfo {volume} {98}},\ \bibinfo {pages} {043512}},\ \Eprint {https://arxiv.org/abs/1806.01216} {1806.01216} \BibitemShut {NoStop}%
\bibitem [{\citenamefont {B{\"o}hm}\ \emph {et~al.}({\natexlab{b}})\citenamefont {B{\"o}hm}, \citenamefont {Sherwin}, \citenamefont {Liu}, \citenamefont {Hill}, \citenamefont {Schmittfull},\ and\ \citenamefont {Namikawa}}]{bohmEffectNonGaussianLensing2018}%
  \BibitemOpen
  \bibfield  {author} {\bibinfo {author} {\bibfnamefont {V.}~\bibnamefont {B{\"o}hm}}, \bibinfo {author} {\bibfnamefont {B.~D.}\ \bibnamefont {Sherwin}}, \bibinfo {author} {\bibfnamefont {J.}~\bibnamefont {Liu}}, \bibinfo {author} {\bibfnamefont {J.~C.}\ \bibnamefont {Hill}}, \bibinfo {author} {\bibfnamefont {M.}~\bibnamefont {Schmittfull}},\ and\ \bibinfo {author} {\bibfnamefont {T.}~\bibnamefont {Namikawa}},\ }\bibfield  {title} {\bibinfo {title} {On the effect of non-{{Gaussian}} lensing deflections on {{CMB}} lensing measurements},\ }\href {https://doi.org/10.1103/PhysRevD.98.123510} {\ \textbf {\bibinfo {volume} {98}},\ \bibinfo {pages} {123510} ({\natexlab{b}})},\ \Eprint {https://arxiv.org/abs/1806.01157} {1806.01157} \BibitemShut {NoStop}%
\bibitem [{\citenamefont {{Pratten}}\ and\ \citenamefont {{Lewis}}(2016)}]{pratten2016}%
  \BibitemOpen
  \bibfield  {author} {\bibinfo {author} {\bibfnamefont {G.}~\bibnamefont {{Pratten}}}\ and\ \bibinfo {author} {\bibfnamefont {A.}~\bibnamefont {{Lewis}}},\ }\bibfield  {title} {\bibinfo {title} {{Impact of post-Born lensing on the CMB}},\ }\href {https://doi.org/10.1088/1475-7516/2016/08/047} {\bibfield  {journal} {\bibinfo  {journal} {\jcap}\ }\textbf {\bibinfo {volume} {8}},\ \bibinfo {eid} {047} (\bibinfo {year} {2016})},\ \Eprint {https://arxiv.org/abs/1605.05662} {arXiv:1605.05662} \BibitemShut {NoStop}%
\bibitem [{\citenamefont {Fabbian}\ \emph {et~al.}()\citenamefont {Fabbian}, \citenamefont {Lewis},\ and\ \citenamefont {Beck}}]{fabbianCMBLensingReconstruction2019a}%
  \BibitemOpen
  \bibfield  {author} {\bibinfo {author} {\bibfnamefont {G.}~\bibnamefont {Fabbian}}, \bibinfo {author} {\bibfnamefont {A.}~\bibnamefont {Lewis}},\ and\ \bibinfo {author} {\bibfnamefont {D.}~\bibnamefont {Beck}},\ }\bibfield  {title} {\bibinfo {title} {{{CMB}} lensing reconstruction biases in cross-correlation with large-scale structure probes},\ }\href {https://doi.org/10.1088/1475-7516/2019/10/057} {\ \textbf {\bibinfo {volume} {2019}},\ \bibinfo {pages} {057}},\ \Eprint {https://arxiv.org/abs/1906.08760} {1906.08760} \BibitemShut {NoStop}%
\bibitem [{\citenamefont {Carron}\ and\ \citenamefont {Lewis}()}]{carronMaximumPosterioriCMB2017}%
  \BibitemOpen
  \bibfield  {author} {\bibinfo {author} {\bibfnamefont {J.}~\bibnamefont {Carron}}\ and\ \bibinfo {author} {\bibfnamefont {A.}~\bibnamefont {Lewis}},\ }\bibfield  {title} {\bibinfo {title} {Maximum a posteriori {{CMB}} lensing reconstruction},\ }\href {https://doi.org/10.1103/PhysRevD.96.063510} {\ \textbf {\bibinfo {volume} {96}},\ \bibinfo {pages} {063510}},\ \Eprint {https://arxiv.org/abs/1704.08230} {1704.08230} \BibitemShut {NoStop}%
\bibitem [{\citenamefont {Belkner}\ \emph {et~al.}(2023)\citenamefont {Belkner}, \citenamefont {Carron}, \citenamefont {Legrand}, \citenamefont {Umilt\`a}, \citenamefont {Pryke},\ and\ \citenamefont {Bischoff}}]{Belkner:2023duz}%
  \BibitemOpen
  \bibfield  {author} {\bibinfo {author} {\bibfnamefont {S.}~\bibnamefont {Belkner}}, \bibinfo {author} {\bibfnamefont {J.}~\bibnamefont {Carron}}, \bibinfo {author} {\bibfnamefont {L.}~\bibnamefont {Legrand}}, \bibinfo {author} {\bibfnamefont {C.}~\bibnamefont {Umilt\`a}}, \bibinfo {author} {\bibfnamefont {C.}~\bibnamefont {Pryke}},\ and\ \bibinfo {author} {\bibfnamefont {C.}~\bibnamefont {Bischoff}} (\bibinfo {collaboration} {CMB-S4}),\ }\bibfield  {title} {\bibinfo {title} {{CMB-S4: Iterative internal delensing and $r$ constraints}},\ }\href@noop {} {\  (\bibinfo {year} {2023})},\ \Eprint {https://arxiv.org/abs/2310.06729} {arXiv:2310.06729 [astro-ph.CO]} \BibitemShut {NoStop}%
\bibitem [{\citenamefont {Hirata}\ and\ \citenamefont {Seljak}({\natexlab{a}})}]{hirataAnalyzingWeakLensing2003}%
  \BibitemOpen
  \bibfield  {author} {\bibinfo {author} {\bibfnamefont {C.~M.}\ \bibnamefont {Hirata}}\ and\ \bibinfo {author} {\bibfnamefont {U.}~\bibnamefont {Seljak}},\ }\bibfield  {title} {\bibinfo {title} {Analyzing weak lensing of the cosmic microwave background using the likelihood function},\ }\href {https://doi.org/10.1103/PhysRevD.67.043001} {\ \textbf {\bibinfo {volume} {67}},\ \bibinfo {pages} {043001} ({\natexlab{a}})}\BibitemShut {NoStop}%
\bibitem [{\citenamefont {Hirata}\ and\ \citenamefont {Seljak}({\natexlab{b}})}]{hirataReconstructionLensingCosmic2003}%
  \BibitemOpen
  \bibfield  {author} {\bibinfo {author} {\bibfnamefont {C.~M.}\ \bibnamefont {Hirata}}\ and\ \bibinfo {author} {\bibfnamefont {U.}~\bibnamefont {Seljak}},\ }\bibfield  {title} {\bibinfo {title} {Reconstruction of lensing from the cosmic microwave background polarization},\ }\href {https://doi.org/10.1103/PhysRevD.68.083002} {\ \textbf {\bibinfo {volume} {68}},\ \bibinfo {pages} {083002} ({\natexlab{b}})}\BibitemShut {NoStop}%
\bibitem [{\citenamefont {Hanson}\ \emph {et~al.}(2010)\citenamefont {Hanson}, \citenamefont {Challinor},\ and\ \citenamefont {Lewis}}]{Hanson_2010}%
  \BibitemOpen
  \bibfield  {author} {\bibinfo {author} {\bibfnamefont {D.}~\bibnamefont {Hanson}}, \bibinfo {author} {\bibfnamefont {A.}~\bibnamefont {Challinor}},\ and\ \bibinfo {author} {\bibfnamefont {A.}~\bibnamefont {Lewis}},\ }\bibfield  {title} {\bibinfo {title} {Weak lensing of the {CMB}},\ }\href {https://doi.org/10.1007/s10714-010-1036-y} {\bibfield  {journal} {\bibinfo  {journal} {General Relativity and Gravitation}\ }\textbf {\bibinfo {volume} {42}},\ \bibinfo {pages} {2197} (\bibinfo {year} {2010})}\BibitemShut {NoStop}%
\bibitem [{\citenamefont {Kesden}\ \emph {et~al.}(2003)\citenamefont {Kesden}, \citenamefont {Cooray},\ and\ \citenamefont {Kamionkowski}}]{Kesden_2003}%
  \BibitemOpen
  \bibfield  {author} {\bibinfo {author} {\bibfnamefont {M.}~\bibnamefont {Kesden}}, \bibinfo {author} {\bibfnamefont {A.}~\bibnamefont {Cooray}},\ and\ \bibinfo {author} {\bibfnamefont {M.}~\bibnamefont {Kamionkowski}},\ }\bibfield  {title} {\bibinfo {title} {Lensing reconstruction with cmb temperature and polarization},\ }\bibfield  {journal} {\bibinfo  {journal} {Physical Review D}\ }\textbf {\bibinfo {volume} {67}},\ \href {https://doi.org/10.1103/physrevd.67.123507} {10.1103/physrevd.67.123507} (\bibinfo {year} {2003})\BibitemShut {NoStop}%
\bibitem [{\citenamefont {Namikawa}\ \emph {et~al.}(2013)\citenamefont {Namikawa}, \citenamefont {Hanson},\ and\ \citenamefont {Takahashi}}]{Namikawa_2013}%
  \BibitemOpen
  \bibfield  {author} {\bibinfo {author} {\bibfnamefont {T.}~\bibnamefont {Namikawa}}, \bibinfo {author} {\bibfnamefont {D.}~\bibnamefont {Hanson}},\ and\ \bibinfo {author} {\bibfnamefont {R.}~\bibnamefont {Takahashi}},\ }\bibfield  {title} {\bibinfo {title} {Bias-hardened cmb lensing},\ }\href {https://doi.org/10.1093/mnras/stt195} {\bibfield  {journal} {\bibinfo  {journal} {Monthly Notices of the Royal Astronomical Society}\ }\textbf {\bibinfo {volume} {431}},\ \bibinfo {pages} {609–620} (\bibinfo {year} {2013})}\BibitemShut {NoStop}%
\bibitem [{\citenamefont {Reinecke}\ \emph {et~al.}(2023)\citenamefont {Reinecke}, \citenamefont {Belkner},\ and\ \citenamefont {Carron}}]{reinecke2023improved}%
  \BibitemOpen
  \bibfield  {author} {\bibinfo {author} {\bibfnamefont {M.}~\bibnamefont {Reinecke}}, \bibinfo {author} {\bibfnamefont {S.}~\bibnamefont {Belkner}},\ and\ \bibinfo {author} {\bibfnamefont {J.}~\bibnamefont {Carron}},\ }\href@noop {} {\bibinfo {title} {Improved cosmic microwave background (de-)lensing using general spherical harmonic transforms}} (\bibinfo {year} {2023}),\ \Eprint {https://arxiv.org/abs/2304.10431} {arXiv:2304.10431 [astro-ph.CO]} \BibitemShut {NoStop}%
\bibitem [{\citenamefont {{Fabbian}}\ \emph {et~al.}(2018)\citenamefont {{Fabbian}}, \citenamefont {{Calabrese}},\ and\ \citenamefont {{Carbone}}}]{fabbian2018}%
  \BibitemOpen
  \bibfield  {author} {\bibinfo {author} {\bibfnamefont {G.}~\bibnamefont {{Fabbian}}}, \bibinfo {author} {\bibfnamefont {M.}~\bibnamefont {{Calabrese}}},\ and\ \bibinfo {author} {\bibfnamefont {C.}~\bibnamefont {{Carbone}}},\ }\bibfield  {title} {\bibinfo {title} {{CMB weak-lensing beyond the Born approximation: a numerical approach}},\ }\href {https://doi.org/10.1088/1475-7516/2018/02/050} {\bibfield  {journal} {\bibinfo  {journal} {\jcap}\ }\textbf {\bibinfo {volume} {2}},\ \bibinfo {eid} {050} (\bibinfo {year} {2018})},\ \Eprint {https://arxiv.org/abs/1702.03317} {arXiv:1702.03317} \BibitemShut {NoStop}%
\bibitem [{\citenamefont {{Hilbert}}\ \emph {et~al.}(2009)\citenamefont {{Hilbert}}, \citenamefont {{Hartlap}}, \citenamefont {{White}},\ and\ \citenamefont {{Schneider}}}]{hilbert2009}%
  \BibitemOpen
  \bibfield  {author} {\bibinfo {author} {\bibfnamefont {S.}~\bibnamefont {{Hilbert}}}, \bibinfo {author} {\bibfnamefont {J.}~\bibnamefont {{Hartlap}}}, \bibinfo {author} {\bibfnamefont {S.~D.~M.}\ \bibnamefont {{White}}},\ and\ \bibinfo {author} {\bibfnamefont {P.}~\bibnamefont {{Schneider}}},\ }\bibfield  {title} {\bibinfo {title} {{Ray-tracing through the Millennium Simulation: Born corrections and lens-lens coupling in cosmic shear and galaxy-galaxy lensing}},\ }\href {https://doi.org/10.1051/0004-6361/200811054} {\bibfield  {journal} {\bibinfo  {journal} {\aap}\ }\textbf {\bibinfo {volume} {499}},\ \bibinfo {pages} {31} (\bibinfo {year} {2009})},\ \Eprint {https://arxiv.org/abs/0809.5035} {arXiv:0809.5035} \BibitemShut {NoStop}%
\bibitem [{\citenamefont {{Carbone}}\ \emph {et~al.}(2016)\citenamefont {{Carbone}}, \citenamefont {{Petkova}},\ and\ \citenamefont {{Dolag}}}]{carbone2016}%
  \BibitemOpen
  \bibfield  {author} {\bibinfo {author} {\bibfnamefont {C.}~\bibnamefont {{Carbone}}}, \bibinfo {author} {\bibfnamefont {M.}~\bibnamefont {{Petkova}}},\ and\ \bibinfo {author} {\bibfnamefont {K.}~\bibnamefont {{Dolag}}},\ }\bibfield  {title} {\bibinfo {title} {{DEMNUni: ISW, Rees-Sciama, and weak-lensing in the presence of massive neutrinos}},\ }\href {https://doi.org/10.1088/1475-7516/2016/07/034} {\bibfield  {journal} {\bibinfo  {journal} {\jcap}\ }\textbf {\bibinfo {volume} {7}},\ \bibinfo {eid} {034} (\bibinfo {year} {2016})},\ \Eprint {https://arxiv.org/abs/1605.02024} {arXiv:1605.02024} \BibitemShut {NoStop}%
\bibitem [{\citenamefont {{Castorina}}\ \emph {et~al.}(2015)\citenamefont {{Castorina}}, \citenamefont {{Carbone}}, \citenamefont {{Bel}}, \citenamefont {{Sefusatti}},\ and\ \citenamefont {{Dolag}}}]{castorina2015}%
  \BibitemOpen
  \bibfield  {author} {\bibinfo {author} {\bibfnamefont {E.}~\bibnamefont {{Castorina}}}, \bibinfo {author} {\bibfnamefont {C.}~\bibnamefont {{Carbone}}}, \bibinfo {author} {\bibfnamefont {J.}~\bibnamefont {{Bel}}}, \bibinfo {author} {\bibfnamefont {E.}~\bibnamefont {{Sefusatti}}},\ and\ \bibinfo {author} {\bibfnamefont {K.}~\bibnamefont {{Dolag}}},\ }\bibfield  {title} {\bibinfo {title} {{DEMNUni: the clustering of large-scale structures in the presence of massive neutrinos}},\ }\href {https://doi.org/10.1088/1475-7516/2015/07/043} {\bibfield  {journal} {\bibinfo  {journal} {\jcap}\ }\textbf {\bibinfo {volume} {7}},\ \bibinfo {eid} {043} (\bibinfo {year} {2015})},\ \Eprint {https://arxiv.org/abs/1505.07148} {arXiv:1505.07148} \BibitemShut {NoStop}%
\bibitem [{\citenamefont {{Carbone}}\ \emph {et~al.}(2008)\citenamefont {{Carbone}}, \citenamefont {{Springel}}, \citenamefont {{Baccigalupi}}, \citenamefont {{Bartelmann}},\ and\ \citenamefont {{Matarrese}}}]{carbone2008}%
  \BibitemOpen
  \bibfield  {author} {\bibinfo {author} {\bibfnamefont {C.}~\bibnamefont {{Carbone}}}, \bibinfo {author} {\bibfnamefont {V.}~\bibnamefont {{Springel}}}, \bibinfo {author} {\bibfnamefont {C.}~\bibnamefont {{Baccigalupi}}}, \bibinfo {author} {\bibfnamefont {M.}~\bibnamefont {{Bartelmann}}},\ and\ \bibinfo {author} {\bibfnamefont {S.}~\bibnamefont {{Matarrese}}},\ }\bibfield  {title} {\bibinfo {title} {{Full-sky maps for gravitational lensing of the cosmic microwave background}},\ }\href {https://doi.org/10.1111/j.1365-2966.2008.13544.x} {\bibfield  {journal} {\bibinfo  {journal} {\mnras}\ }\textbf {\bibinfo {volume} {388}},\ \bibinfo {pages} {1618} (\bibinfo {year} {2008})},\ \Eprint {https://arxiv.org/abs/0711.2655} {arXiv:0711.2655} \BibitemShut {NoStop}%
\bibitem [{\citenamefont {{Calabrese}}\ \emph {et~al.}(2015)\citenamefont {{Calabrese}}, \citenamefont {{Carbone}}, \citenamefont {{Fabbian}}, \citenamefont {{Baldi}},\ and\ \citenamefont {{Baccigalupi}}}]{calabrese2015}%
  \BibitemOpen
  \bibfield  {author} {\bibinfo {author} {\bibfnamefont {M.}~\bibnamefont {{Calabrese}}}, \bibinfo {author} {\bibfnamefont {C.}~\bibnamefont {{Carbone}}}, \bibinfo {author} {\bibfnamefont {G.}~\bibnamefont {{Fabbian}}}, \bibinfo {author} {\bibfnamefont {M.}~\bibnamefont {{Baldi}}},\ and\ \bibinfo {author} {\bibfnamefont {C.}~\bibnamefont {{Baccigalupi}}},\ }\bibfield  {title} {\bibinfo {title} {{Multiple lensing of the cosmic microwave background anisotropies}},\ }\href {https://doi.org/10.1088/1475-7516/2015/03/049} {\bibfield  {journal} {\bibinfo  {journal} {\jcap}\ }\textbf {\bibinfo {volume} {3}},\ \bibinfo {eid} {049} (\bibinfo {year} {2015})},\ \Eprint {https://arxiv.org/abs/1409.7680} {arXiv:1409.7680} \BibitemShut {NoStop}%
\bibitem [{\citenamefont {{Fosalba}}\ \emph {et~al.}(2008)\citenamefont {{Fosalba}}, \citenamefont {{Gazta{\~n}aga}}, \citenamefont {{Castander}},\ and\ \citenamefont {{Manera}}}]{fosalba2008}%
  \BibitemOpen
  \bibfield  {author} {\bibinfo {author} {\bibfnamefont {P.}~\bibnamefont {{Fosalba}}}, \bibinfo {author} {\bibfnamefont {E.}~\bibnamefont {{Gazta{\~n}aga}}}, \bibinfo {author} {\bibfnamefont {F.~J.}\ \bibnamefont {{Castander}}},\ and\ \bibinfo {author} {\bibfnamefont {M.}~\bibnamefont {{Manera}}},\ }\bibfield  {title} {\bibinfo {title} {{The onion universe: all sky lightcone simulations in spherical shells}},\ }\href {https://doi.org/10.1111/j.1365-2966.2008.13910.x} {\bibfield  {journal} {\bibinfo  {journal} {\mnras}\ }\textbf {\bibinfo {volume} {391}},\ \bibinfo {pages} {435} (\bibinfo {year} {2008})},\ \Eprint {https://arxiv.org/abs/0711.1540} {arXiv:0711.1540} \BibitemShut {NoStop}%
\bibitem [{\citenamefont {{Fabbian}}\ and\ \citenamefont {{Stompor}}(2013)}]{fabbian2013}%
  \BibitemOpen
  \bibfield  {author} {\bibinfo {author} {\bibfnamefont {G.}~\bibnamefont {{Fabbian}}}\ and\ \bibinfo {author} {\bibfnamefont {R.}~\bibnamefont {{Stompor}}},\ }\bibfield  {title} {\bibinfo {title} {{High-precision simulations of the weak lensing effect on cosmic microwave background polarization}},\ }\href {https://doi.org/10.1051/0004-6361/201321575} {\bibfield  {journal} {\bibinfo  {journal} {\aap}\ }\textbf {\bibinfo {volume} {556}},\ \bibinfo {eid} {A109} (\bibinfo {year} {2013})},\ \Eprint {https://arxiv.org/abs/1303.6550} {arXiv:1303.6550 [astro-ph.CO]} \BibitemShut {NoStop}%
\bibitem [{\citenamefont {Xavier}\ \emph {et~al.}()\citenamefont {Xavier}, \citenamefont {Abdalla},\ and\ \citenamefont {Joachimi}}]{xavierImprovingLognormalModels2016a}%
  \BibitemOpen
  \bibfield  {author} {\bibinfo {author} {\bibfnamefont {H.~S.}\ \bibnamefont {Xavier}}, \bibinfo {author} {\bibfnamefont {F.~B.}\ \bibnamefont {Abdalla}},\ and\ \bibinfo {author} {\bibfnamefont {B.}~\bibnamefont {Joachimi}},\ }\bibfield  {title} {\bibinfo {title} {Improving lognormal models for cosmological fields},\ }\href {https://doi.org/10.1093/mnras/stw874} {\ \textbf {\bibinfo {volume} {459}},\ \bibinfo {pages} {3693}},\ \Eprint {https://arxiv.org/abs/1602.08503} {1602.08503} \BibitemShut {NoStop}%
\bibitem [{\citenamefont {{G{\'o}rski}}\ \emph {et~al.}(2005)\citenamefont {{G{\'o}rski}}, \citenamefont {{Hivon}}, \citenamefont {{Banday}}, \citenamefont {{Wandelt}}, \citenamefont {{Hansen}}, \citenamefont {{Reinecke}},\ and\ \citenamefont {{Bartelmann}}}]{2005ApJ...622..759G}%
  \BibitemOpen
  \bibfield  {author} {\bibinfo {author} {\bibfnamefont {K.~M.}\ \bibnamefont {{G{\'o}rski}}}, \bibinfo {author} {\bibfnamefont {E.}~\bibnamefont {{Hivon}}}, \bibinfo {author} {\bibfnamefont {A.~J.}\ \bibnamefont {{Banday}}}, \bibinfo {author} {\bibfnamefont {B.~D.}\ \bibnamefont {{Wandelt}}}, \bibinfo {author} {\bibfnamefont {F.~K.}\ \bibnamefont {{Hansen}}}, \bibinfo {author} {\bibfnamefont {M.}~\bibnamefont {{Reinecke}}},\ and\ \bibinfo {author} {\bibfnamefont {M.}~\bibnamefont {{Bartelmann}}},\ }\bibfield  {title} {\bibinfo {title} {{HEALPix: A Framework for High-Resolution Discretization and Fast Analysis of Data Distributed on the Sphere}},\ }\href {https://doi.org/10.1086/427976} {\bibfield  {journal} {\bibinfo  {journal} {\apj}\ }\textbf {\bibinfo {volume} {622}},\ \bibinfo {pages} {759} (\bibinfo {year} {2005})},\ \Eprint {https://arxiv.org/abs/astro-ph/0409513} {arXiv:astro-ph/0409513 [astro-ph]} \BibitemShut {NoStop}%
\bibitem [{\citenamefont {Zonca}\ \emph {et~al.}(2019)\citenamefont {Zonca}, \citenamefont {Singer}, \citenamefont {Lenz}, \citenamefont {Reinecke}, \citenamefont {Rosset}, \citenamefont {Hivon},\ and\ \citenamefont {Gorski}}]{Zonca2019}%
  \BibitemOpen
  \bibfield  {author} {\bibinfo {author} {\bibfnamefont {A.}~\bibnamefont {Zonca}}, \bibinfo {author} {\bibfnamefont {L.~P.}\ \bibnamefont {Singer}}, \bibinfo {author} {\bibfnamefont {D.}~\bibnamefont {Lenz}}, \bibinfo {author} {\bibfnamefont {M.}~\bibnamefont {Reinecke}}, \bibinfo {author} {\bibfnamefont {C.}~\bibnamefont {Rosset}}, \bibinfo {author} {\bibfnamefont {E.}~\bibnamefont {Hivon}},\ and\ \bibinfo {author} {\bibfnamefont {K.~M.}\ \bibnamefont {Gorski}},\ }\bibfield  {title} {\bibinfo {title} {healpy: equal area pixelization and spherical harmonics transforms for data on the sphere in python},\ }\href {https://doi.org/10.21105/joss.01298} {\bibfield  {journal} {\bibinfo  {journal} {Journal of Open Source Software}\ }\textbf {\bibinfo {volume} {4}},\ \bibinfo {pages} {1298} (\bibinfo {year} {2019})}\BibitemShut {NoStop}%
\bibitem [{\citenamefont {{Hall}}\ and\ \citenamefont {{Taylor}}(2022)}]{hall2022}%
  \BibitemOpen
  \bibfield  {author} {\bibinfo {author} {\bibfnamefont {A.}~\bibnamefont {{Hall}}}\ and\ \bibinfo {author} {\bibfnamefont {A.}~\bibnamefont {{Taylor}}},\ }\bibfield  {title} {\bibinfo {title} {{Non-Gaussian likelihood of weak lensing power spectra}},\ }\href {https://doi.org/10.1103/PhysRevD.105.123527} {\bibfield  {journal} {\bibinfo  {journal} {\prd}\ }\textbf {\bibinfo {volume} {105}},\ \bibinfo {eid} {123527} (\bibinfo {year} {2022})},\ \Eprint {https://arxiv.org/abs/2202.04095} {arXiv:2202.04095 [astro-ph.CO]} \BibitemShut {NoStop}%
\bibitem [{\citenamefont {Gil-Mar{\'\i}n}\ \emph {et~al.}()\citenamefont {Gil-Mar{\'\i}n}, \citenamefont {Wagner}, \citenamefont {Fragkoudi}, \citenamefont {Jimenez},\ and\ \citenamefont {Verde}}]{gil-marinImprovedFittingFormula2012}%
  \BibitemOpen
  \bibfield  {author} {\bibinfo {author} {\bibfnamefont {H.}~\bibnamefont {Gil-Mar{\'\i}n}}, \bibinfo {author} {\bibfnamefont {C.}~\bibnamefont {Wagner}}, \bibinfo {author} {\bibfnamefont {F.}~\bibnamefont {Fragkoudi}}, \bibinfo {author} {\bibfnamefont {R.}~\bibnamefont {Jimenez}},\ and\ \bibinfo {author} {\bibfnamefont {L.}~\bibnamefont {Verde}},\ }\bibfield  {title} {\bibinfo {title} {An improved fitting formula for the dark matter bispectrum},\ }\href {https://doi.org/10.1088/1475-7516/2012/02/047} {\ \textbf {\bibinfo {volume} {2012}},\ \bibinfo {pages} {047}},\ \Eprint {https://arxiv.org/abs/1111.4477} {1111.4477} \BibitemShut {NoStop}%
\bibitem [{\citenamefont {Namikawa}\ \emph {et~al.}()\citenamefont {Namikawa}, \citenamefont {Bose}, \citenamefont {Bouchet}, \citenamefont {Takahashi},\ and\ \citenamefont {Taruya}}]{namikawaCMBLensingBispectrum2019}%
  \BibitemOpen
  \bibfield  {author} {\bibinfo {author} {\bibfnamefont {T.}~\bibnamefont {Namikawa}}, \bibinfo {author} {\bibfnamefont {B.}~\bibnamefont {Bose}}, \bibinfo {author} {\bibfnamefont {F.~R.}\ \bibnamefont {Bouchet}}, \bibinfo {author} {\bibfnamefont {R.}~\bibnamefont {Takahashi}},\ and\ \bibinfo {author} {\bibfnamefont {A.}~\bibnamefont {Taruya}},\ }\bibfield  {title} {\bibinfo {title} {{{CMB}} lensing bi-spectrum: Assessing analytical predictions against full-sky lensing simulations},\ }\href {https://doi.org/10.1103/PhysRevD.99.063511} {\ \textbf {\bibinfo {volume} {99}},\ \bibinfo {pages} {063511}},\ \Eprint {https://arxiv.org/abs/1812.10635} {1812.10635} \BibitemShut {NoStop}%
\bibitem [{\citenamefont {{Das}}\ and\ \citenamefont {{Ostriker}}(2006)}]{das2006}%
  \BibitemOpen
  \bibfield  {author} {\bibinfo {author} {\bibfnamefont {S.}~\bibnamefont {{Das}}}\ and\ \bibinfo {author} {\bibfnamefont {J.~P.}\ \bibnamefont {{Ostriker}}},\ }\bibfield  {title} {\bibinfo {title} {{Testing a New Analytic Model for Gravitational Lensing Probabilities}},\ }\href {https://doi.org/10.1086/504032} {\bibfield  {journal} {\bibinfo  {journal} {\apj}\ }\textbf {\bibinfo {volume} {645}},\ \bibinfo {pages} {1} (\bibinfo {year} {2006})},\ \Eprint {https://arxiv.org/abs/astro-ph/0512644} {arXiv:astro-ph/0512644 [astro-ph]} \BibitemShut {NoStop}%
\bibitem [{\citenamefont {{Barthelemy}}\ \emph {et~al.}(2020)\citenamefont {{Barthelemy}}, \citenamefont {{Codis}},\ and\ \citenamefont {{Bernardeau}}}]{barthelemy2020}%
  \BibitemOpen
  \bibfield  {author} {\bibinfo {author} {\bibfnamefont {A.}~\bibnamefont {{Barthelemy}}}, \bibinfo {author} {\bibfnamefont {S.}~\bibnamefont {{Codis}}},\ and\ \bibinfo {author} {\bibfnamefont {F.}~\bibnamefont {{Bernardeau}}},\ }\bibfield  {title} {\bibinfo {title} {{Post-Born corrections to the one-point statistics of (CMB) lensing convergence obtained via large deviation theory}},\ }\href {https://doi.org/10.1093/mnras/staa931} {\bibfield  {journal} {\bibinfo  {journal} {\mnras}\ }\textbf {\bibinfo {volume} {494}},\ \bibinfo {pages} {3368} (\bibinfo {year} {2020})},\ \Eprint {https://arxiv.org/abs/2002.03625} {arXiv:2002.03625 [astro-ph.CO]} \BibitemShut {NoStop}%
\bibitem [{\citenamefont {Osborne}\ \emph {et~al.}(2014)\citenamefont {Osborne}, \citenamefont {Hanson},\ and\ \citenamefont {Doré}}]{Osborne_2014}%
  \BibitemOpen
  \bibfield  {author} {\bibinfo {author} {\bibfnamefont {S.~J.}\ \bibnamefont {Osborne}}, \bibinfo {author} {\bibfnamefont {D.}~\bibnamefont {Hanson}},\ and\ \bibinfo {author} {\bibfnamefont {O.}~\bibnamefont {Doré}},\ }\bibfield  {title} {\bibinfo {title} {Extragalactic foreground contamination in temperature-based cmb lens reconstruction},\ }\href {https://doi.org/10.1088/1475-7516/2014/03/024} {\bibfield  {journal} {\bibinfo  {journal} {Journal of Cosmology and Astroparticle Physics}\ }\textbf {\bibinfo {volume} {2014}}\bibinfo  {number} { (03)},\ \bibinfo {pages} {024–024}}\BibitemShut {NoStop}%
\bibitem [{\citenamefont {Sailer}\ \emph {et~al.}(2020)\citenamefont {Sailer}, \citenamefont {Schaan},\ and\ \citenamefont {Ferraro}}]{Sailer_2020}%
  \BibitemOpen
\bibfield  {number} {  }\bibfield  {author} {\bibinfo {author} {\bibfnamefont {N.}~\bibnamefont {Sailer}}, \bibinfo {author} {\bibfnamefont {E.}~\bibnamefont {Schaan}},\ and\ \bibinfo {author} {\bibfnamefont {S.}~\bibnamefont {Ferraro}},\ }\bibfield  {title} {\bibinfo {title} {Lower bias, lower noise cmb lensing with foreground-hardened estimators},\ }\bibfield  {journal} {\bibinfo  {journal} {Physical Review D}\ }\textbf {\bibinfo {volume} {102}},\ \href {https://doi.org/10.1103/physrevd.102.063517} {10.1103/physrevd.102.063517} (\bibinfo {year} {2020})\BibitemShut {NoStop}%
\bibitem [{\citenamefont {Amara}\ and\ \citenamefont {Refregier}(2008)}]{Amara:2007as}%
  \BibitemOpen
  \bibfield  {author} {\bibinfo {author} {\bibfnamefont {A.}~\bibnamefont {Amara}}\ and\ \bibinfo {author} {\bibfnamefont {A.}~\bibnamefont {Refregier}},\ }\bibfield  {title} {\bibinfo {title} {{Systematic Bias in Cosmic Shear: Beyond the Fisher Matrix}},\ }\href {https://doi.org/10.1111/j.1365-2966.2008.13880.x} {\bibfield  {journal} {\bibinfo  {journal} {Mon. Not. Roy. Astron. Soc.}\ }\textbf {\bibinfo {volume} {391}},\ \bibinfo {pages} {228} (\bibinfo {year} {2008})},\ \Eprint {https://arxiv.org/abs/0710.5171} {arXiv:0710.5171 [astro-ph]} \BibitemShut {NoStop}%
\bibitem [{\citenamefont {Font-Ribera}\ \emph {et~al.}(2014)\citenamefont {Font-Ribera}, \citenamefont {McDonald}, \citenamefont {Mostek}, \citenamefont {Reid}, \citenamefont {Seo},\ and\ \citenamefont {Slosar}}]{Font-Ribera:2013rwa}%
  \BibitemOpen
  \bibfield  {author} {\bibinfo {author} {\bibfnamefont {A.}~\bibnamefont {Font-Ribera}}, \bibinfo {author} {\bibfnamefont {P.}~\bibnamefont {McDonald}}, \bibinfo {author} {\bibfnamefont {N.}~\bibnamefont {Mostek}}, \bibinfo {author} {\bibfnamefont {B.~A.}\ \bibnamefont {Reid}}, \bibinfo {author} {\bibfnamefont {H.-J.}\ \bibnamefont {Seo}},\ and\ \bibinfo {author} {\bibfnamefont {A.}~\bibnamefont {Slosar}},\ }\bibfield  {title} {\bibinfo {title} {{DESI and other dark energy experiments in the era of neutrino mass measurements}},\ }\href {https://doi.org/10.1088/1475-7516/2014/05/023} {\bibfield  {journal} {\bibinfo  {journal} {JCAP}\ }\textbf {\bibinfo {volume} {05}},\ \bibinfo {pages} {023}},\ \Eprint {https://arxiv.org/abs/1308.4164} {arXiv:1308.4164 [astro-ph.CO]} \BibitemShut {NoStop}%
\bibitem [{\citenamefont {Aghamousa}\ \emph {et~al.}(2016)\citenamefont {Aghamousa} \emph {et~al.}}]{DESI:2016fyo}%
  \BibitemOpen
  \bibfield  {author} {\bibinfo {author} {\bibfnamefont {A.}~\bibnamefont {Aghamousa}} \emph {et~al.} (\bibinfo {collaboration} {DESI}),\ }\bibfield  {title} {\bibinfo {title} {{The DESI Experiment Part I: Science,Targeting, and Survey Design}},\ }\href@noop {} {\  (\bibinfo {year} {2016})},\ \Eprint {https://arxiv.org/abs/1611.00036} {arXiv:1611.00036 [astro-ph.IM]} \BibitemShut {NoStop}%
\bibitem [{\citenamefont {Lewis}\ \emph {et~al.}(2000)\citenamefont {Lewis}, \citenamefont {Challinor},\ and\ \citenamefont {Lasenby}}]{Lewis:1999bs}%
  \BibitemOpen
  \bibfield  {author} {\bibinfo {author} {\bibfnamefont {A.}~\bibnamefont {Lewis}}, \bibinfo {author} {\bibfnamefont {A.}~\bibnamefont {Challinor}},\ and\ \bibinfo {author} {\bibfnamefont {A.}~\bibnamefont {Lasenby}},\ }\bibfield  {title} {\bibinfo {title} {{Efficient computation of CMB anisotropies in closed FRW models}},\ }\href {https://doi.org/10.1086/309179} {\bibfield  {journal} {\bibinfo  {journal} {Astrophys. J.}\ }\textbf {\bibinfo {volume} {538}},\ \bibinfo {pages} {473} (\bibinfo {year} {2000})},\ \Eprint {https://arxiv.org/abs/astro-ph/9911177} {arXiv:astro-ph/9911177 [astro-ph]} \BibitemShut {NoStop}%
\bibitem [{\citenamefont {Howlett}\ \emph {et~al.}(2012)\citenamefont {Howlett}, \citenamefont {Lewis}, \citenamefont {Hall},\ and\ \citenamefont {Challinor}}]{Howlett:2012mh}%
  \BibitemOpen
  \bibfield  {author} {\bibinfo {author} {\bibfnamefont {C.}~\bibnamefont {Howlett}}, \bibinfo {author} {\bibfnamefont {A.}~\bibnamefont {Lewis}}, \bibinfo {author} {\bibfnamefont {A.}~\bibnamefont {Hall}},\ and\ \bibinfo {author} {\bibfnamefont {A.}~\bibnamefont {Challinor}},\ }\bibfield  {title} {\bibinfo {title} {{CMB power spectrum parameter degeneracies in the era of precision cosmology}},\ }\href {https://doi.org/10.1088/1475-7516/2012/04/027} {\bibfield  {journal} {\bibinfo  {journal} {JCAP}\ }\textbf {\bibinfo {volume} {1204}},\ \bibinfo {pages} {027}},\ \Eprint {https://arxiv.org/abs/1201.3654} {arXiv:1201.3654 [astro-ph.CO]} \BibitemShut {NoStop}%
\bibitem [{\citenamefont {Lewis}\ and\ \citenamefont {Bridle}(2002)}]{Lewis:2002ah}%
  \BibitemOpen
  \bibfield  {author} {\bibinfo {author} {\bibfnamefont {A.}~\bibnamefont {Lewis}}\ and\ \bibinfo {author} {\bibfnamefont {S.}~\bibnamefont {Bridle}},\ }\bibfield  {title} {\bibinfo {title} {{Cosmological parameters from CMB and other data: A Monte Carlo approach}},\ }\href {https://doi.org/10.1103/PhysRevD.66.103511} {\bibfield  {journal} {\bibinfo  {journal} {Phys. Rev.}\ }\textbf {\bibinfo {volume} {D66}},\ \bibinfo {pages} {103511} (\bibinfo {year} {2002})},\ \Eprint {https://arxiv.org/abs/astro-ph/0205436} {arXiv:astro-ph/0205436 [astro-ph]} \BibitemShut {NoStop}%
\bibitem [{\citenamefont {Lewis}(2013)}]{Lewis:2013hha}%
  \BibitemOpen
  \bibfield  {author} {\bibinfo {author} {\bibfnamefont {A.}~\bibnamefont {Lewis}},\ }\bibfield  {title} {\bibinfo {title} {{Efficient sampling of fast and slow cosmological parameters}},\ }\href {https://doi.org/10.1103/PhysRevD.87.103529} {\bibfield  {journal} {\bibinfo  {journal} {Phys. Rev.}\ }\textbf {\bibinfo {volume} {D87}},\ \bibinfo {pages} {103529} (\bibinfo {year} {2013})},\ \Eprint {https://arxiv.org/abs/1304.4473} {arXiv:1304.4473 [astro-ph.CO]} \BibitemShut {NoStop}%
\bibitem [{\citenamefont {Torrado}\ and\ \citenamefont {Lewis}(2021)}]{Torrado:2020dgo}%
  \BibitemOpen
  \bibfield  {author} {\bibinfo {author} {\bibfnamefont {J.}~\bibnamefont {Torrado}}\ and\ \bibinfo {author} {\bibfnamefont {A.}~\bibnamefont {Lewis}},\ }\bibfield  {title} {\bibinfo {title} {{Cobaya: Code for Bayesian Analysis of hierarchical physical models}},\ }\href {https://doi.org/10.1088/1475-7516/2021/05/057} {\bibfield  {journal} {\bibinfo  {journal} {JCAP}\ }\textbf {\bibinfo {volume} {05}},\ \bibinfo {pages} {057}},\ \Eprint {https://arxiv.org/abs/2005.05290} {arXiv:2005.05290 [astro-ph.IM]} \BibitemShut {NoStop}%
\bibitem [{\citenamefont {Lewis}(2019)}]{Lewis:2019xzd}%
  \BibitemOpen
  \bibfield  {author} {\bibinfo {author} {\bibfnamefont {A.}~\bibnamefont {Lewis}},\ }\bibfield  {title} {\bibinfo {title} {{GetDist: a Python package for analysing Monte Carlo samples}},\ }\href {https://getdist.readthedocs.io} {\  (\bibinfo {year} {2019})},\ \Eprint {https://arxiv.org/abs/1910.13970} {arXiv:1910.13970 [astro-ph.IM]} \BibitemShut {NoStop}%
\bibitem [{\citenamefont {Allison}\ \emph {et~al.}(2015)\citenamefont {Allison}, \citenamefont {Caucal}, \citenamefont {Calabrese}, \citenamefont {Dunkley},\ and\ \citenamefont {Louis}}]{Allison:2015qca}%
  \BibitemOpen
  \bibfield  {author} {\bibinfo {author} {\bibfnamefont {R.}~\bibnamefont {Allison}}, \bibinfo {author} {\bibfnamefont {P.}~\bibnamefont {Caucal}}, \bibinfo {author} {\bibfnamefont {E.}~\bibnamefont {Calabrese}}, \bibinfo {author} {\bibfnamefont {J.}~\bibnamefont {Dunkley}},\ and\ \bibinfo {author} {\bibfnamefont {T.}~\bibnamefont {Louis}},\ }\bibfield  {title} {\bibinfo {title} {{Towards a cosmological neutrino mass detection}},\ }\href {https://doi.org/10.1103/PhysRevD.92.123535} {\bibfield  {journal} {\bibinfo  {journal} {Phys. Rev. D}\ }\textbf {\bibinfo {volume} {92}},\ \bibinfo {pages} {123535} (\bibinfo {year} {2015})},\ \Eprint {https://arxiv.org/abs/1509.07471} {arXiv:1509.07471 [astro-ph.CO]} \BibitemShut {NoStop}%
\bibitem [{\citenamefont {Peloton}\ \emph {et~al.}(2017)\citenamefont {Peloton}, \citenamefont {Schmittfull}, \citenamefont {Lewis}, \citenamefont {Carron},\ and\ \citenamefont {Zahn}}]{Peloton:2016kbw}%
  \BibitemOpen
  \bibfield  {author} {\bibinfo {author} {\bibfnamefont {J.}~\bibnamefont {Peloton}}, \bibinfo {author} {\bibfnamefont {M.}~\bibnamefont {Schmittfull}}, \bibinfo {author} {\bibfnamefont {A.}~\bibnamefont {Lewis}}, \bibinfo {author} {\bibfnamefont {J.}~\bibnamefont {Carron}},\ and\ \bibinfo {author} {\bibfnamefont {O.}~\bibnamefont {Zahn}},\ }\bibfield  {title} {\bibinfo {title} {{Full covariance of CMB and lensing reconstruction power spectra}},\ }\href {https://doi.org/10.1103/PhysRevD.95.043508} {\bibfield  {journal} {\bibinfo  {journal} {Phys. Rev. D}\ }\textbf {\bibinfo {volume} {95}},\ \bibinfo {pages} {043508} (\bibinfo {year} {2017})},\ \Eprint {https://arxiv.org/abs/1611.01446} {arXiv:1611.01446 [astro-ph.CO]} \BibitemShut {NoStop}%
\bibitem [{\citenamefont {Rouhiainen}\ \emph {et~al.}(2021)\citenamefont {Rouhiainen}, \citenamefont {Giri},\ and\ \citenamefont {Münchmeyer}}]{rouhiainen2021normalizing}%
  \BibitemOpen
  \bibfield  {author} {\bibinfo {author} {\bibfnamefont {A.}~\bibnamefont {Rouhiainen}}, \bibinfo {author} {\bibfnamefont {U.}~\bibnamefont {Giri}},\ and\ \bibinfo {author} {\bibfnamefont {M.}~\bibnamefont {Münchmeyer}},\ }\href@noop {} {\bibinfo {title} {Normalizing flows for random fields in cosmology}} (\bibinfo {year} {2021}),\ \Eprint {https://arxiv.org/abs/2105.12024} {arXiv:2105.12024} \BibitemShut {NoStop}%
\bibitem [{\citenamefont {{Neal}}(2005)}]{Neal:2005}%
  \BibitemOpen
  \bibfield  {author} {\bibinfo {author} {\bibfnamefont {R.~M.}\ \bibnamefont {{Neal}}},\ }\bibfield  {title} {\bibinfo {title} {{Taking Bigger Metropolis Steps by Dragging Fast Variables}},\ }\href {https://arxiv.org/abs/math/0502099} {\bibfield  {journal} {\bibinfo  {journal} {ArXiv Mathematics e-prints}\ } (\bibinfo {year} {2005})},\ \Eprint {https://arxiv.org/abs/math/0502099} {math/0502099} \BibitemShut {NoStop}%
\bibitem [{\citenamefont {B{\"o}hm}\ \emph {et~al.}({\natexlab{c}})\citenamefont {B{\"o}hm}, \citenamefont {Modi},\ and\ \citenamefont {Castorina}}]{bohmLensingCorrectionsGalaxylensing2020}%
  \BibitemOpen
  \bibfield  {author} {\bibinfo {author} {\bibfnamefont {V.}~\bibnamefont {B{\"o}hm}}, \bibinfo {author} {\bibfnamefont {C.}~\bibnamefont {Modi}},\ and\ \bibinfo {author} {\bibfnamefont {E.}~\bibnamefont {Castorina}},\ }\bibfield  {title} {\bibinfo {title} {Lensing corrections on galaxy-lensing cross correlations and galaxy-galaxy auto correlations},\ }\href {https://doi.org/10.1088/1475-7516/2020/03/045} {\ \textbf {\bibinfo {volume} {2020}},\ \bibinfo {pages} {045} ({\natexlab{c}})},\ \Eprint {https://arxiv.org/abs/1910.06722} {1910.06722} \BibitemShut {NoStop}%
\bibitem [{\citenamefont {Boruah}\ \emph {et~al.}()\citenamefont {Boruah}, \citenamefont {Rozo},\ and\ \citenamefont {Fiedorowicz}}]{boruahMapbasedCosmologyInference2022}%
  \BibitemOpen
  \bibfield  {author} {\bibinfo {author} {\bibfnamefont {S.~S.}\ \bibnamefont {Boruah}}, \bibinfo {author} {\bibfnamefont {E.}~\bibnamefont {Rozo}},\ and\ \bibinfo {author} {\bibfnamefont {P.}~\bibnamefont {Fiedorowicz}},\ }\href {http://arxiv.org/abs/2204.13216} {\bibinfo {title} {Map-based cosmology inference with lognormal cosmic shear maps}},\ \Eprint {https://arxiv.org/abs/2204.13216} {2204.13216} \BibitemShut {NoStop}%
\bibitem [{\citenamefont {Buchalter}\ \emph {et~al.}()\citenamefont {Buchalter}, \citenamefont {Kamionkowski},\ and\ \citenamefont {Jaffe}}]{buchalterAngularThreePointCorrelation2000}%
  \BibitemOpen
  \bibfield  {author} {\bibinfo {author} {\bibfnamefont {A.}~\bibnamefont {Buchalter}}, \bibinfo {author} {\bibfnamefont {M.}~\bibnamefont {Kamionkowski}},\ and\ \bibinfo {author} {\bibfnamefont {A.~H.}\ \bibnamefont {Jaffe}},\ }\bibfield  {title} {\bibinfo {title} {The {{Angular Three-Point Correlation Function}} in the {{Quasilinear Regime}}},\ }\href {https://doi.org/10.1086/308339} {\ \textbf {\bibinfo {volume} {530}},\ \bibinfo {pages} {36}},\ \Eprint {https://arxiv.org/abs/astro-ph/9903486} {astro-ph/9903486} \BibitemShut {NoStop}%
\bibitem [{\citenamefont {Coles}\ and\ \citenamefont {Jones}()}]{colesLognormalModelCosmological1991}%
  \BibitemOpen
  \bibfield  {author} {\bibinfo {author} {\bibfnamefont {P.}~\bibnamefont {Coles}}\ and\ \bibinfo {author} {\bibfnamefont {B.}~\bibnamefont {Jones}},\ }\bibfield  {title} {\bibinfo {title} {A lognormal model for the cosmological mass distribution},\ }\href {https://doi.org/10.1093/mnras/248.1.1} {\ \textbf {\bibinfo {volume} {248}},\ \bibinfo {pages} {1}}\BibitemShut {NoStop}%
\bibitem [{\citenamefont {Hilbert}\ \emph {et~al.}()\citenamefont {Hilbert}, \citenamefont {Hartlap},\ and\ \citenamefont {Schneider}}]{hilbertCosmicShearCovariance2011}%
  \BibitemOpen
  \bibfield  {author} {\bibinfo {author} {\bibfnamefont {S.}~\bibnamefont {Hilbert}}, \bibinfo {author} {\bibfnamefont {J.}~\bibnamefont {Hartlap}},\ and\ \bibinfo {author} {\bibfnamefont {P.}~\bibnamefont {Schneider}},\ }\bibfield  {title} {\bibinfo {title} {Cosmic shear covariance: The log-normal approximation},\ }\href {https://doi.org/10.1051/0004-6361/201117294} {\ \textbf {\bibinfo {volume} {536}},\ \bibinfo {pages} {A85}}\BibitemShut {NoStop}%
\bibitem [{\citenamefont {{Lewis}}\ and\ \citenamefont {{Pratten}}(2016)}]{lewis2016}%
  \BibitemOpen
  \bibfield  {author} {\bibinfo {author} {\bibfnamefont {A.}~\bibnamefont {{Lewis}}}\ and\ \bibinfo {author} {\bibfnamefont {G.}~\bibnamefont {{Pratten}}},\ }\bibfield  {title} {\bibinfo {title} {{Effect of lensing non-Gaussianity on the CMB power spectra}},\ }\href {https://doi.org/10.1088/1475-7516/2016/12/003} {\bibfield  {journal} {\bibinfo  {journal} {\jcap}\ }\textbf {\bibinfo {volume} {12}},\ \bibinfo {eid} {003} (\bibinfo {year} {2016})},\ \Eprint {https://arxiv.org/abs/1608.01263} {arXiv:1608.01263} \BibitemShut {NoStop}%
\bibitem [{\citenamefont {{Carbone}}\ \emph {et~al.}(2009)\citenamefont {{Carbone}}, \citenamefont {{Baccigalupi}}, \citenamefont {{Bartelmann}}, \citenamefont {{Matarrese}},\ and\ \citenamefont {{Springel}}}]{carbone2009}%
  \BibitemOpen
  \bibfield  {author} {\bibinfo {author} {\bibfnamefont {C.}~\bibnamefont {{Carbone}}}, \bibinfo {author} {\bibfnamefont {C.}~\bibnamefont {{Baccigalupi}}}, \bibinfo {author} {\bibfnamefont {M.}~\bibnamefont {{Bartelmann}}}, \bibinfo {author} {\bibfnamefont {S.}~\bibnamefont {{Matarrese}}},\ and\ \bibinfo {author} {\bibfnamefont {V.}~\bibnamefont {{Springel}}},\ }\bibfield  {title} {\bibinfo {title} {{Lensed CMB temperature and polarization maps from the Millennium Simulation}},\ }\href {https://doi.org/10.1111/j.1365-2966.2009.14746.x} {\bibfield  {journal} {\bibinfo  {journal} {\mnras}\ }\textbf {\bibinfo {volume} {396}},\ \bibinfo {pages} {668} (\bibinfo {year} {2009})},\ \Eprint {https://arxiv.org/abs/0810.4145} {arXiv:0810.4145} \BibitemShut {NoStop}%
\bibitem [{\citenamefont {{Becker}}(2013)}]{becker2013}%
  \BibitemOpen
  \bibfield  {author} {\bibinfo {author} {\bibfnamefont {M.~R.}\ \bibnamefont {{Becker}}},\ }\bibfield  {title} {\bibinfo {title} {{CALCLENS: weak lensing simulations for large-area sky surveys and second-order effects in cosmic shear power spectra}},\ }\href {https://doi.org/10.1093/mnras/stt1352} {\bibfield  {journal} {\bibinfo  {journal} {\mnras}\ }\textbf {\bibinfo {volume} {435}},\ \bibinfo {pages} {115} (\bibinfo {year} {2013})},\ \Eprint {https://arxiv.org/abs/arXiv:1210.3069} {arXiv:arXiv:1210.3069} \BibitemShut {NoStop}%
\bibitem [{\citenamefont {Hotinli}\ \emph {et~al.}(2022)\citenamefont {Hotinli}, \citenamefont {Meyers}, \citenamefont {Trendafilova}, \citenamefont {Green},\ and\ \citenamefont {van Engelen}}]{Hotinli:2021umk}%
  \BibitemOpen
  \bibfield  {author} {\bibinfo {author} {\bibfnamefont {S.~C.}\ \bibnamefont {Hotinli}}, \bibinfo {author} {\bibfnamefont {J.}~\bibnamefont {Meyers}}, \bibinfo {author} {\bibfnamefont {C.}~\bibnamefont {Trendafilova}}, \bibinfo {author} {\bibfnamefont {D.}~\bibnamefont {Green}},\ and\ \bibinfo {author} {\bibfnamefont {A.}~\bibnamefont {van Engelen}},\ }\bibfield  {title} {\bibinfo {title} {{The benefits of CMB delensing}},\ }\href {https://doi.org/10.1088/1475-7516/2022/04/020} {\bibfield  {journal} {\bibinfo  {journal} {JCAP}\ }\textbf {\bibinfo {volume} {04}}\bibfield  {number} {\bibinfo  {number} { (04)},\ \bibinfo {pages} {020}},\ }\Eprint {https://arxiv.org/abs/2111.15036} {arXiv:2111.15036 [astro-ph.CO]} \BibitemShut {NoStop}%
\bibitem [{\citenamefont {Green}\ \emph {et~al.}(2017)\citenamefont {Green}, \citenamefont {Meyers},\ and\ \citenamefont {van Engelen}}]{Green:2016cjr}%
  \BibitemOpen
  \bibfield  {author} {\bibinfo {author} {\bibfnamefont {D.}~\bibnamefont {Green}}, \bibinfo {author} {\bibfnamefont {J.}~\bibnamefont {Meyers}},\ and\ \bibinfo {author} {\bibfnamefont {A.}~\bibnamefont {van Engelen}},\ }\bibfield  {title} {\bibinfo {title} {{CMB Delensing Beyond the B Modes}},\ }\href {https://doi.org/10.1088/1475-7516/2017/12/005} {\bibfield  {journal} {\bibinfo  {journal} {JCAP}\ }\textbf {\bibinfo {volume} {12}},\ \bibinfo {pages} {005}},\ \Eprint {https://arxiv.org/abs/1609.08143} {arXiv:1609.08143 [astro-ph.CO]} \BibitemShut {NoStop}%
\bibitem [{\citenamefont {Fantaye}\ \emph {et~al.}(2012)\citenamefont {Fantaye}, \citenamefont {Baccigalupi}, \citenamefont {Leach},\ and\ \citenamefont {Yadav}}]{Fantaye:2012ha}%
  \BibitemOpen
  \bibfield  {author} {\bibinfo {author} {\bibfnamefont {Y.}~\bibnamefont {Fantaye}}, \bibinfo {author} {\bibfnamefont {C.}~\bibnamefont {Baccigalupi}}, \bibinfo {author} {\bibfnamefont {S.}~\bibnamefont {Leach}},\ and\ \bibinfo {author} {\bibfnamefont {A.~P.~S.}\ \bibnamefont {Yadav}},\ }\bibfield  {title} {\bibinfo {title} {{CMB lensing reconstruction in the presence of diffuse polarized foregrounds}},\ }\href {https://doi.org/10.1088/1475-7516/2012/12/017} {\bibfield  {journal} {\bibinfo  {journal} {\jcap}\ }\textbf {\bibinfo {volume} {12}},\ \bibinfo {pages} {017} (\bibinfo {year} {2012})},\ \Eprint {https://arxiv.org/abs/1207.0508} {arXiv:1207.0508} \BibitemShut {NoStop}%
\bibitem [{\citenamefont {{Schaan}}\ \emph {et~al.}(2020)\citenamefont {{Schaan}}, \citenamefont {{Ferraro}},\ and\ \citenamefont {{Seljak}}}]{2020JCAP...12..001S}%
  \BibitemOpen
  \bibfield  {author} {\bibinfo {author} {\bibfnamefont {E.}~\bibnamefont {{Schaan}}}, \bibinfo {author} {\bibfnamefont {S.}~\bibnamefont {{Ferraro}}},\ and\ \bibinfo {author} {\bibfnamefont {U.}~\bibnamefont {{Seljak}}},\ }\bibfield  {title} {\bibinfo {title} {{Photo-z outlier self-calibration in weak lensing surveys}},\ }\href {https://doi.org/10.1088/1475-7516/2020/12/001} {\bibfield  {journal} {\bibinfo  {journal} {\jcap}\ }\textbf {\bibinfo {volume} {2020}},\ \bibinfo {eid} {001} (\bibinfo {year} {2020})},\ \Eprint {https://arxiv.org/abs/2007.12795} {arXiv:2007.12795} \BibitemShut {NoStop}%
\end{thebibliography}%

\end{document}